%% file: OR_FINAL.tex
\documentclass[11pt]{article}

\usepackage[numbers]{natbib}
\renewcommand{\baselinestretch}{1.5}

\usepackage[left=2.54cm,bottom=2.54cm,right=2.54cm,top=2.54cm]{geometry}

\usepackage{float}

\usepackage[small,bf]{caption}
\usepackage{authblk}

\usepackage{amsmath, amsthm, amssymb,multirow,hyperref}

\usepackage{epsfig,psfrag,url,verbatim}
\usepackage{graphics}
\usepackage{graphicx, algorithm}
\usepackage[parfill]{parskip}
\usepackage{subcaption}

\usepackage{titlesec}
\usepackage{lipsum}

\titlespacing{\paragraph}{%
  0pt}{
  0.0\baselineskip}{
  .3em}

\usepackage{amsfonts,paralist}

\usepackage{color}

\newtheorem{cor}{Corollary}
\newtheorem{thm}{Theorem}
\newtheorem{lem}{Lemma}
\newtheorem{prop}{Proposition}
\newtheorem{mydef}{Definition}
\newtheorem{rem}{Remark}

\newcommand{\B}{\boldsymbol}
\newcommand{\M}{\mathbf}
\newcommand{\sgn}{\operatorname{sgn}}

\newcommand{\sbt}{\mathrm{s.t.}}
\newcommand{\rnk}{\mathrm{rank}}

\newcommand{\s}{\mathrm S}

\newcommand{\plotA}{Pdf}

\newcommand{\mm}{{\mathcal M}}

\renewcommand{\c}{\psi}

\DeclareMathOperator*{\argmin}{arg\,min}

\DeclareMathOperator*{\mini}{minimize}

\newenvironment{myarray}[2][1]
  {\def\arraystretch{#1}\array{#2}}
  {\endarray}

\newcommand{\mmk}{k}

\usepackage{array,graphicx}
\usepackage{booktabs}
\usepackage{pifont}

\newcommand{\bepsilon}{\boldsymbol \epsilon}

\newcommand{\bbeta}{\boldsymbol \beta}

\newcommand{\bD}{\boldsymbol \Delta}

\newcommand{\bY}{{\bf y}}

\newcommand{\bX}{{\bf X}}

\newcommand{\be}{{\bf e}}

\newcommand{\bv}{{\bf v}}

\newcommand{\bff}{{\bf f}}
\newcommand{\bu}{{\bf u}}
\def\RR{\mathbb{R}}

\newcommand{\plotB}{Pdf}

\begin{document}

\title{Subset Selection with Shrinkage: Sparse Linear Modeling when the SNR is low}

\author[1]{Rahul Mazumder\thanks{Rahul Mazumder's research was partially supported by the Office of Naval Research (N000141512342, N000141812298 -- Young Investigator Award) and the National Science Foundation (NSF-IIS-1718258).
}}
\author[2]{Peter Radchenko
}
\author[1]{Antoine Dedieu\thanks{Now at Vicarious AI; performed a major part of his work while a graduate student at MIT.
}}


\affil[1]{Massachusetts Institute of Technology}
\affil[2]{University of Sydney}

\date{December, 2021
\footnote{This is a major revision of an earlier manuscript dated August 2017.}
}

\maketitle

\begin{abstract}

We study a seemingly unexpected and relatively less understood overfitting aspect of a fundamental tool in sparse linear modeling --
best subset selection, which minimizes the residual sum of squares subject to a constraint on the number of nonzero coefficients.
While the best subset selection procedure is often perceived as the ``gold standard''
in sparse learning when the signal to noise ratio (SNR) is high,
its predictive performance deteriorates when the SNR is low.  In particular, it is outperformed by continuous shrinkage methods, such as ridge regression and the Lasso. We investigate the behavior of best subset selection in the high-noise regimes and propose an alternative approach based on a regularized version of the least-squares criterion.
Our proposed estimators (a) mitigate, to a large extent, the poor predictive performance of best subset selection in the high-noise regimes; and (b) perform favorably, while generally delivering substantially sparser models, relative to the best predictive models available via ridge regression and the Lasso. We  conduct  an extensive  theoretical  analysis  of  the  predictive  properties of the proposed approach and provide justification for  its  superior  predictive  performance  relative to  best subset selection when the noise-level is high.
Our estimators can be expressed as solutions to mixed integer second order conic optimization problems and, hence, are amenable to  modern computational tools from mathematical optimization.
\end{abstract}




\section{Introduction}\label{sec:intro}

We consider the usual linear regression framework, with
response $\M{y} \in \RR^{n}$,
model matrix $\M{X}\in  \mathbb{R}^{n \times p}$ and regression coefficients $\B\beta \in  \mathbb{R}^{p}$.
We assume that columns of $\M{X}$ have been standardized to have zero means and unit $\ell_{2}$-norms.
In many classical and modern statistical applications it is desirable to obtain a parsimonious model with good data-fidelity.
Towards this end, a natural candidate is the well-known \textit{best-subsets} estimator \citep{miller2002subset}, given by the following combinatorial optimization problem:
\begin{equation}\label{subset-1}
\hat{\B\beta}_{\ell_0} ~~\in ~~ \argmin ~~  \| \M{y} - \M{X} \B\beta \|_{2}^2 ~~~~~ \sbt ~~~~~ \| \B\beta \|_{0} \leq k.
\end{equation}
Problem~\eqref{subset-1} has a simple interpretation: it seeks to obtain the best least squares fit with at most~$k$ nonzero regression coefficients.
There is a rich body of theoretical work studying the statistical properties of this estimator -- see, for example, \cite{GR2004,greenshtein2006best,raskutti2011minimax,zhang2012general} and the references therein.
The caveat, however, is that Problem~\eqref{subset-1} is often perceived as computationally \emph{infeasible}~\citep{natarajan1995sparse} --
the popular {\texttt{R}}-package \textit{leaps}, for example,
is unable to obtain solutions to~\eqref{subset-1} when $p>30$.
Inability to compute the best-subsets estimator has perhaps contributed towards an aura of mystery around its operational characteristics
on problem-instances that arise in practice.
Recently, \cite{bertsimas2015best} demonstrated that  Problem~\eqref{subset-1} can be solved to certifiable global optimality via mixed integer optimization (MIO) techniques~\citep{nemhauser1988integer,bertsimas2005optimization_new}, leveraging the impressive advances in MIO over the past ten or so years -- see \cite{bertsimas2015best,mazumder2015discrete,hazimeh2021grouped} and the references therein.  
From a practical viewpoint, this line of research has made it possible to use subset selection procedures on real and synthetic datasets and gather insights regarding their operating characteristics, previously unseen due to the perceived computational limits. This paper investigates one such insight.


\noindent {\textbf{Does best subset selection overfit?}} Suppose that the data are generated from a linear model $\M{y} = \M{X}\B\beta^* + \B\epsilon$, where matrix~$\M{X}$ is deterministic and the elements of $\B\epsilon\in\RR^n$ are independent
$N(0, \sigma^2)$.  We focus on the case where~$\B\beta^*$ is sparse, with few nonzero elements.
It is well known that if the noise-level, measured by~$\sigma$, is small relative to the signal-level, measured by $\|\M{X}\B\beta^*\|_2$, for example, then the best-subsets estimator leads to models with excellent
statistical properties \citep{raskutti2011minimax,zhang2012general,buhlmann2011statistics} in terms of prediction, estimation and variable selection (minor additional assumptions are required for the latter two metrics). However, the situation is different when the noise level is high -- this was observed in \cite{breiman1996heuristics}, which highlighted the instability of best subset selection.
Deterioration of the predictive performance in high-noise regimes is a significant drawback of best-subsets that, to our knowledge, has received limited attention in the literature thus far. It is important to note that SNR alone does not control the difficulty of the underlying statistical problem; model parameters~$p$, $n$, $k^*$, $\bbeta^*$, and~$\M{X}$, also affect the performance of the estimator.  In our theoretical analysis, presented in Section~\ref{sec:theory}, we use ratios~$\|\bbeta^*\|_1/\sigma$ and~$\|\bbeta^*\|_2/\sigma$ to characterize the relevant noise-level regimes.

The best-subsets estimator given by Problem~\eqref{subset-1} focuses on two goals: (a) searching for the best
subset, ${\mathcal I}$, containing~$k$ features; and (b) estimating $\hat{\B\beta}_{\ell_0}$ by implementing the unconstrained least-squares method on the selected features~${\mathcal I}$.
Even if best-subsets selects ${\mathcal I}$ to be the support of $\B\beta^*$, the un-regularized fit on features ${\mathcal I}$ can be improved by shrinking the coefficients when~$\sigma$ is large.
For a simple illustration of this, consider the setting where $n>p$ and $k=p$.   Here, estimator~$\hat{\B\beta}_{\ell_0}$ is the usual least-squares solution, which benefits from additional shrinkage \citep{james1961estimation} to achieve a better bias-variance trade-off in the presence of noise. Further problems arise when the SNR is low
due to the variability associated with the choice of ${\mathcal I}$.
See for example, the works of~\cite{wainwright2009sharp,comminges2012tight,david2017high} discussing the impossibility of variable selection when the signal is weak.


\begin{figure}[h!]
\centering
\resizebox{\textwidth}{0.25\textheight}{\begin{tabular}{l c c c}
& \scriptsize{$\rho=0$, SNR=0.5} & \scriptsize{$\rho=0$, SNR=2} &  \scriptsize{$\rho=0$, SNR=7} \vspace{-.5em}\\
\rotatebox{90}{\sf {\scriptsize{~~~~~~~~~~~~~Prediction Error}}}&
\includegraphics[width=0.33\textwidth,height=0.2\textheight,  trim =1.1cm 1.6cm 1.cm 1.7cm, clip = true ]{"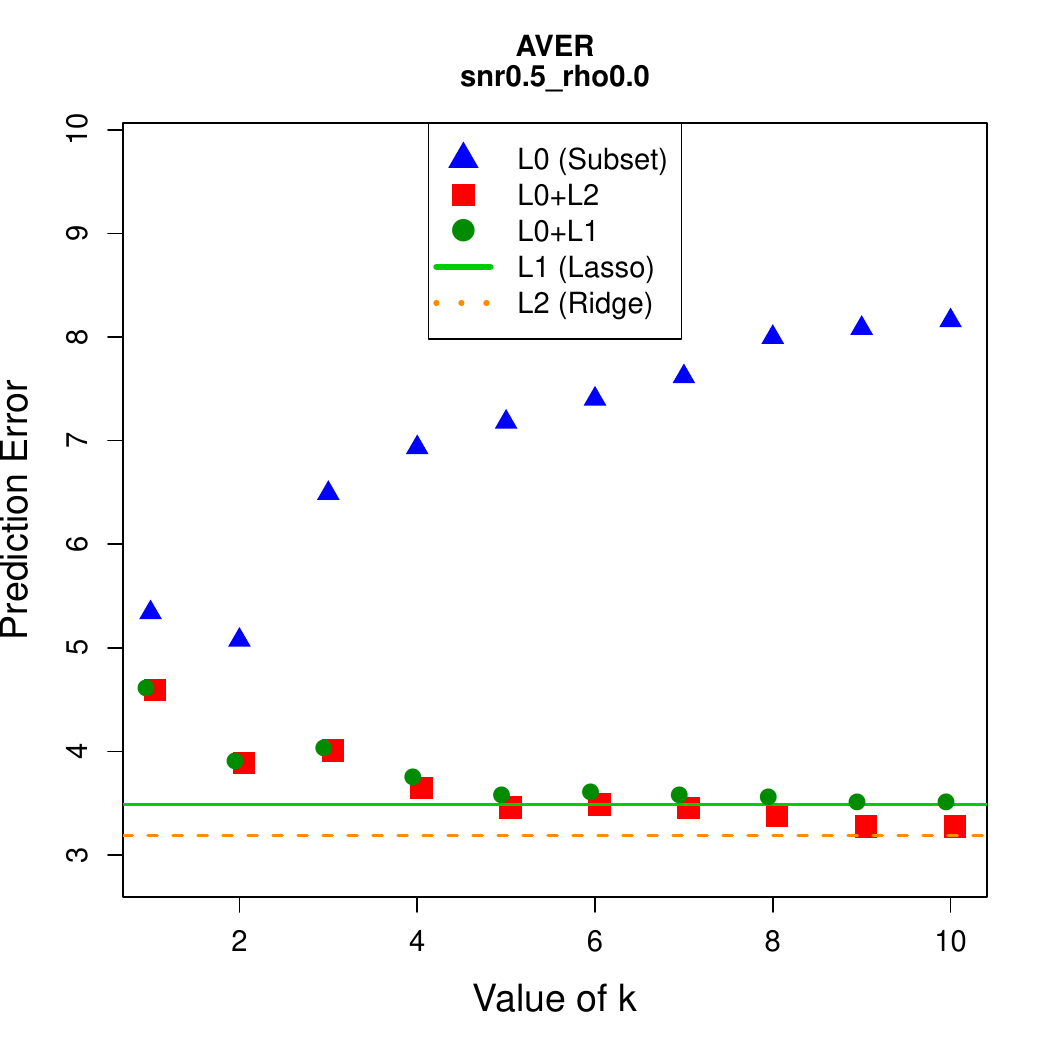"}&
\includegraphics[width=0.33\textwidth,height=0.2\textheight,  trim =1.1cm 1.6cm 1.cm 1.7cm, clip = true ]{"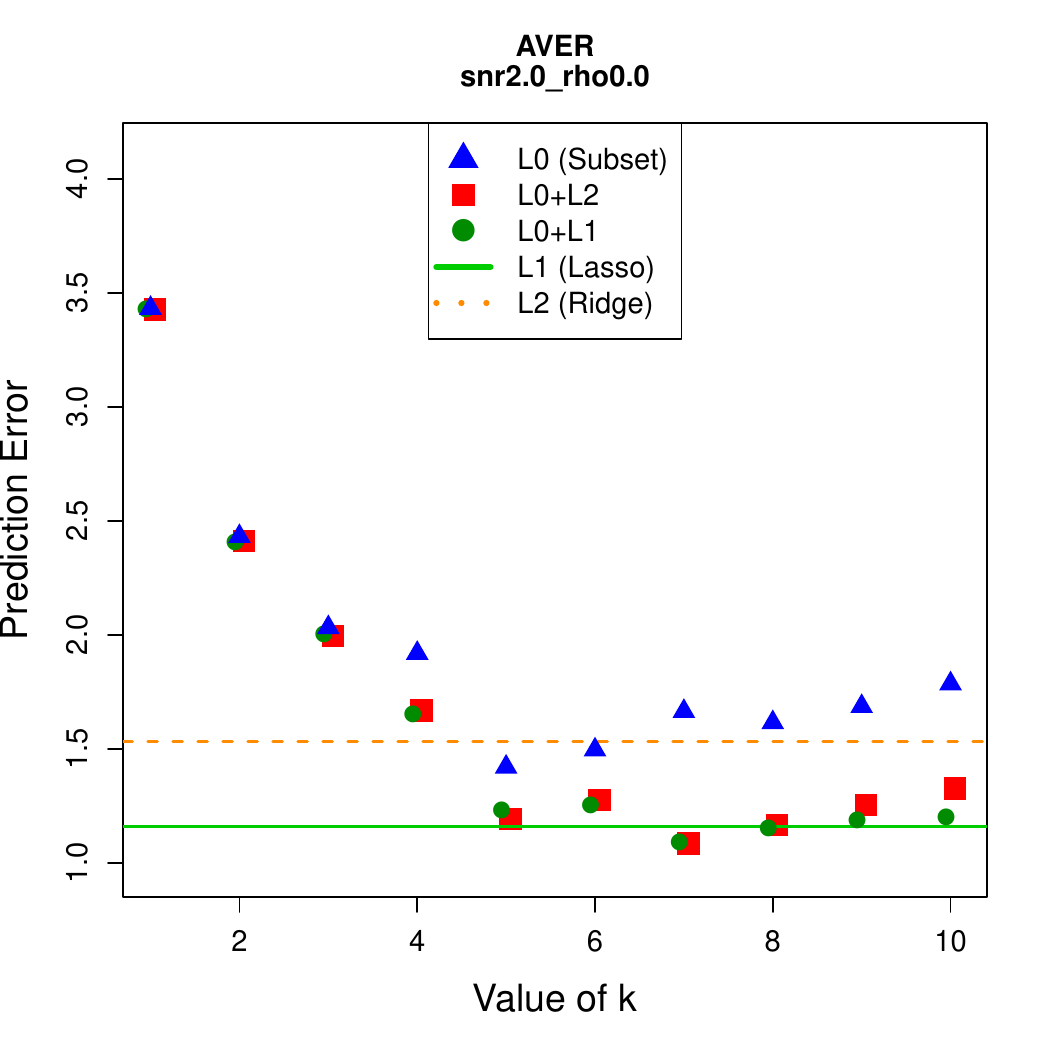"}&
\includegraphics[width=0.33\textwidth,height=0.2\textheight,  trim =1.1cm 1.6cm 1.cm 1.7cm, clip = true ]{"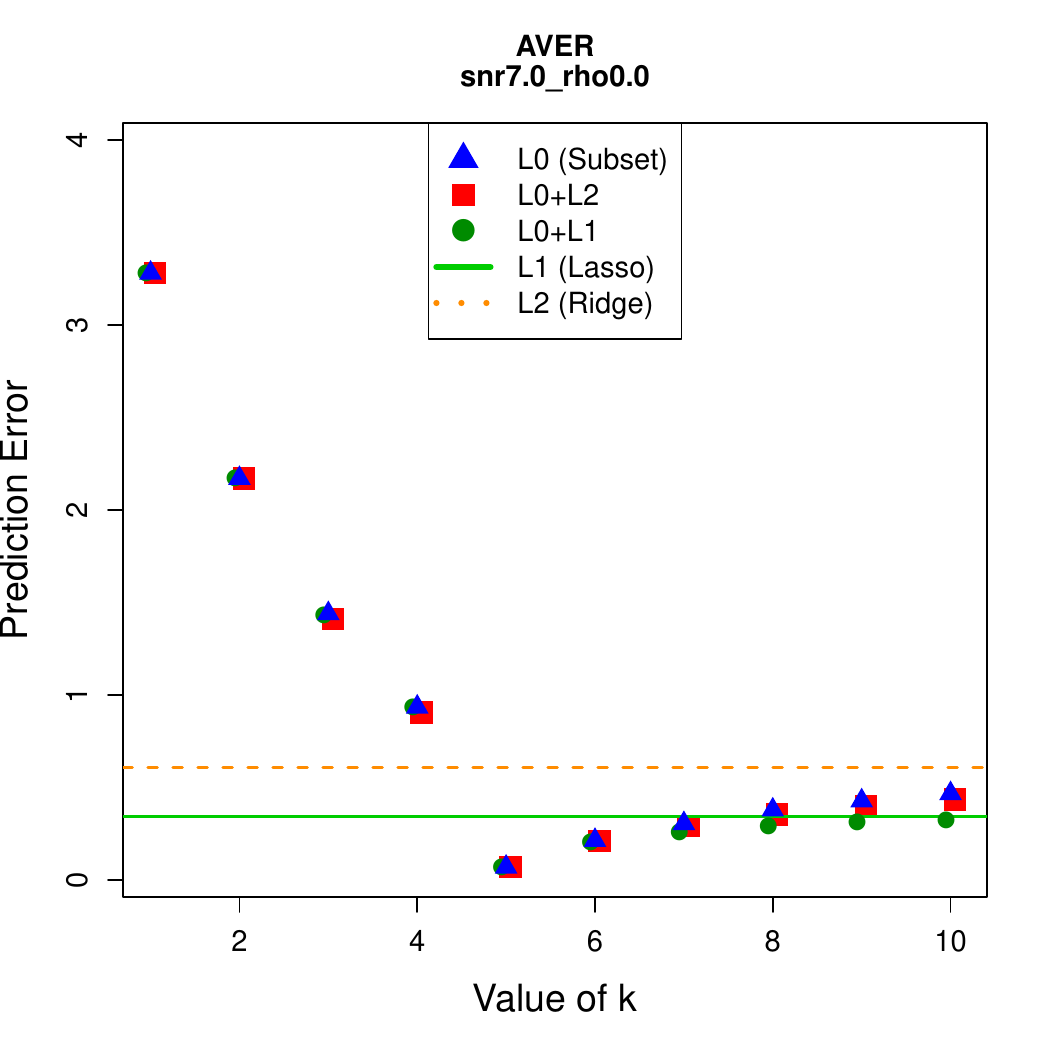"} \vspace{-.5em}\\
& \scriptsize{model size budget ($k$)} &  \scriptsize{model size budget ($k$)}  &   \scriptsize{model size budget ($k$)} \vspace{.5em}  \\
& \scriptsize{$\rho=0.8$, SNR=0.5} & \scriptsize{$\rho=0.8$, SNR=2} &  \scriptsize{$\rho=0.8$, SNR=7} \vspace{-.5em} \\
\rotatebox{90}{\sf {\scriptsize{~~~~~~~~~~~~~Prediction Error}}}&
\includegraphics[width=0.33\textwidth,height=0.2\textheight,  trim =1.1cm 1.5cm 1.cm 1.7cm, clip = true ]{"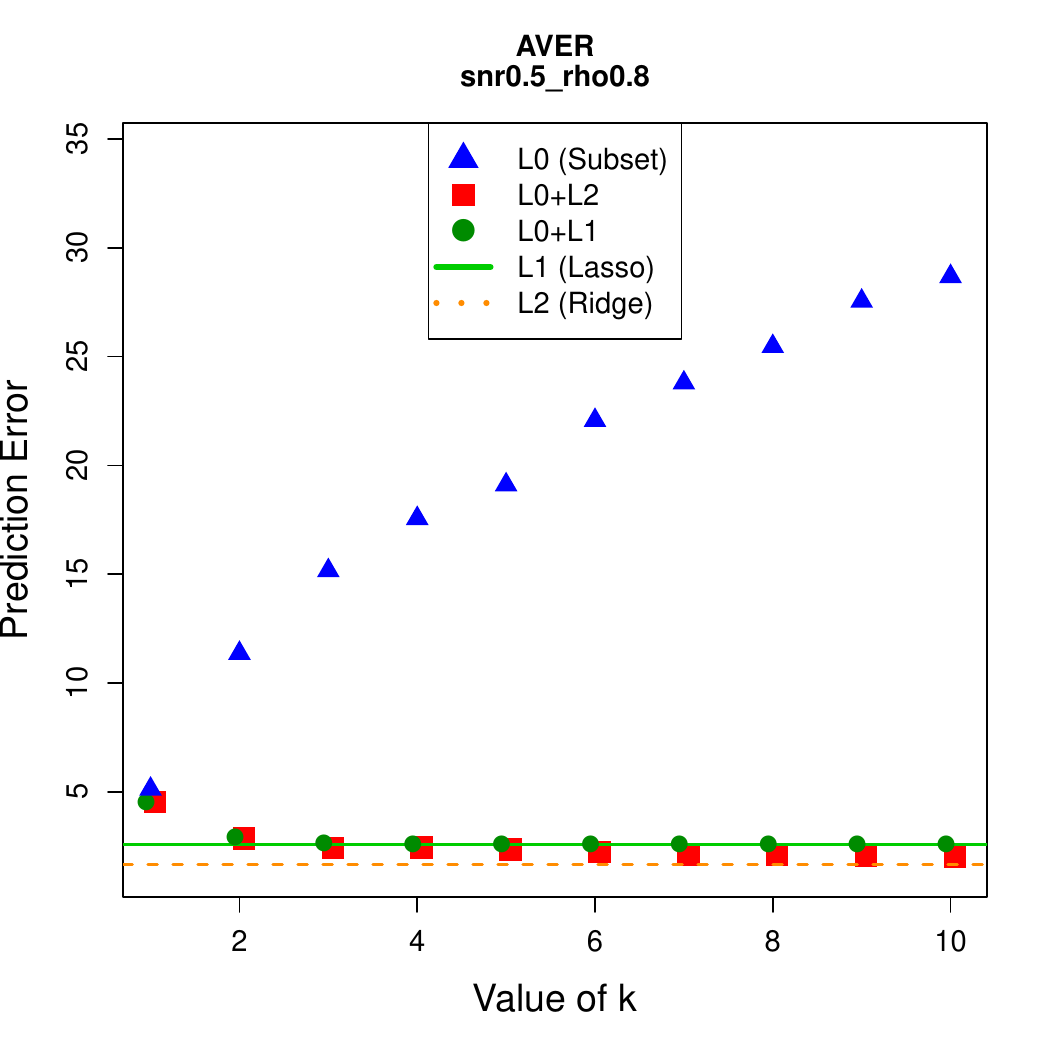"}&
\includegraphics[width=0.33\textwidth,height=0.2\textheight,  trim =1.1cm 1.5cm 1.cm 1.7cm, clip = true ]{"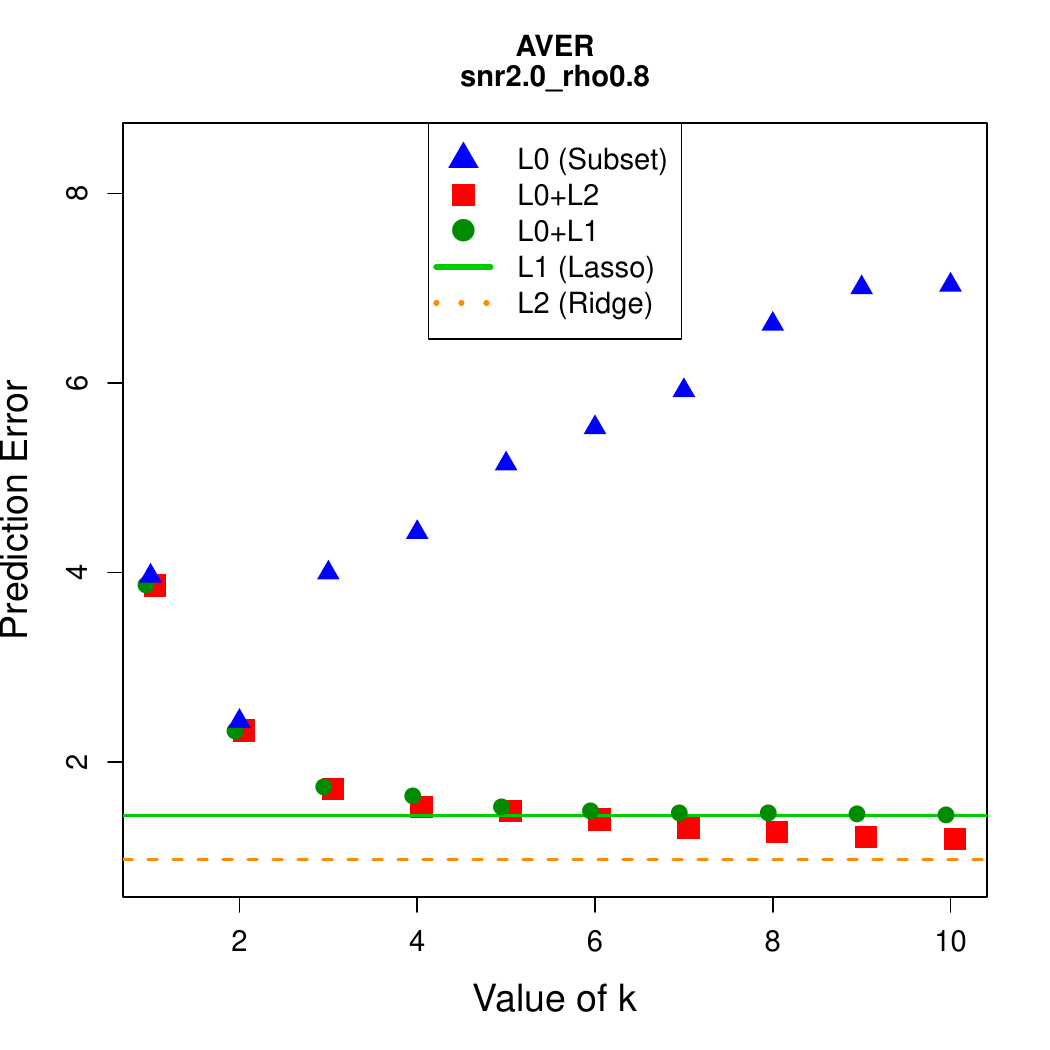"}&
\includegraphics[width=0.33\textwidth,height=0.2\textheight,  trim =1.1cm 1.5cm 1.cm 1.7cm, clip = true ]{"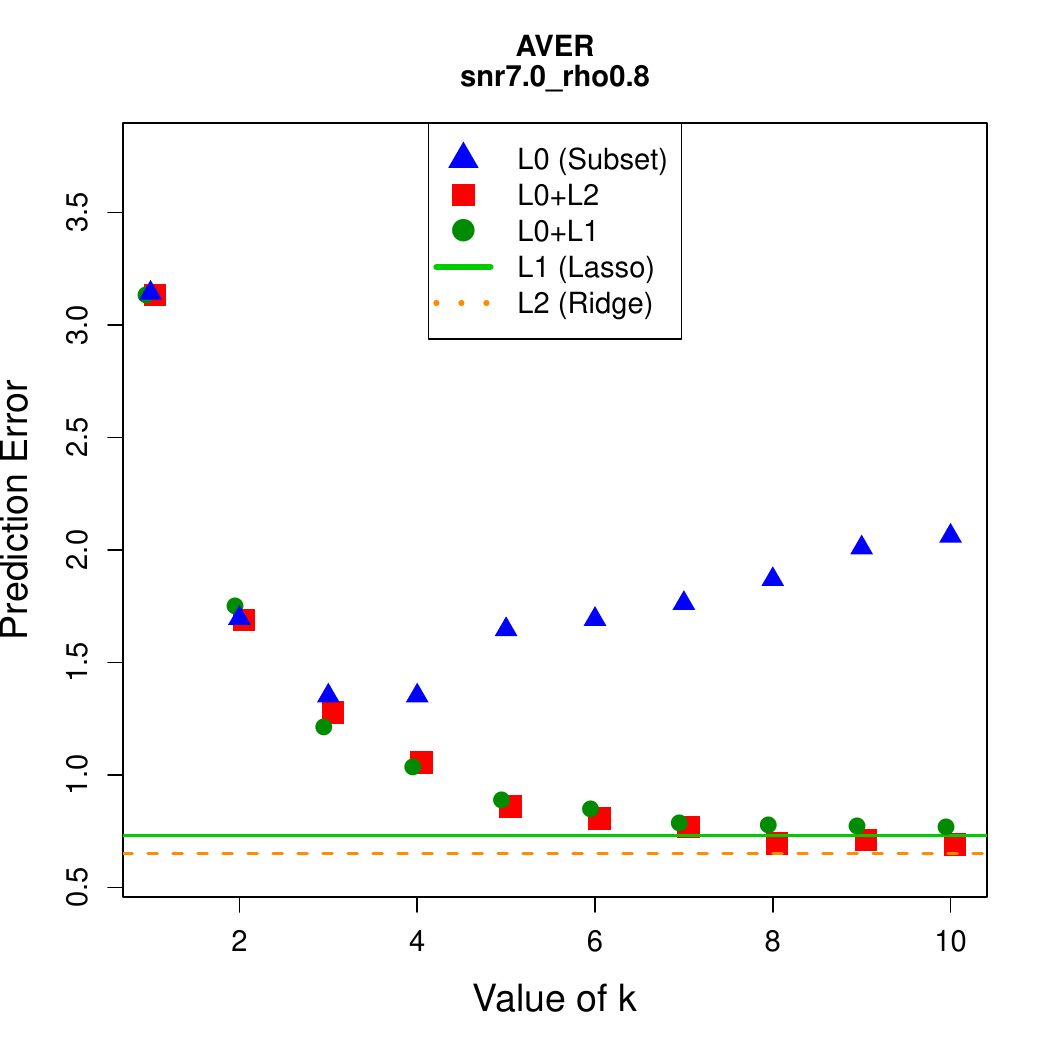"} \vspace{-.5em}\\
& \scriptsize{model size budget ($k$)} &  \scriptsize{model size budget ($k$)}  &   \scriptsize{model size budget ($k$)}  \\
\end{tabular}}\caption{ \small{Prediction error $\| \M{X}(\hat{\B\beta} - \B\beta^*)\|_{2}^2$, averaged over the simulated datasets described in the text, for the Lasso (L1),  ridge regression (L2), best-subsets (L0), and the estimators proposed in Problem~\eqref{best-subset-penalized}: L0+L1 ($q=1$) and L0+L2 ($q=2$).  Given the model size parameter~$k$ (irrelevant to L1 and L2), the average prediction error of best predictive model across~$\lambda$ (irrelevant to L0) is plotted for each method.  The best L1 models have average sizes  11.4, 17.8, 18.4 [top panel] and 8.3,  12.0, 16.6 [bottom panel], while the L2 models are completely dense.
}}
\label{fig-one}
\end{figure}
%

The discussion above suggests that the best-subsets estimator is not the right approach when the noise-level is high.    Figure~\ref{fig-one}  presents a concrete example illustrating this point.
The data are generated from a linear model with $n=40$, $p=60$, five true coefficients equal to one, and the rest equal to zero.
The rows of~$\M{X}$ are drawn from a multivariate Gaussian distribution with the mean equal to zero and all the pairwise correlations equal to~$\rho$. The features are standardized to have unit $\ell_{2}$-norm, and $\sigma^2$ is set to match specific values of $\text{SNR} = \| \M{X}\B\beta^*\|_{2}^2/\|\B\epsilon\|_{2}^2$.  Figure~\ref{fig-one} illustrates the performance of the best-subsets estimator, computed using the framework of~\cite{bertsimas2015best} for different values of~$k$; the results are averaged over ten different replications of $(\M{X},\B\epsilon)$.
As expected, the predictive accuracy of best-subsets deteriorates as the SNR decreases -- it is outperformed by continuous shrinkage methods such as ridge regression~\citep{HK70} and the Lasso~\citep{Ti96}.
The overfitting behavior of best-subsets can be attributed to its aggressive search for the best feature subset~${\mathcal I}$ and not performing any shrinkage on the selected coefficients.

We contend that the classical best-subsets estimator~\eqref{subset-1}  is not designed to be used in high-noise regimes.  Our theoretical and empirical investigations in Sections~\ref{sec:theory} and~\ref{sec:expts}  highlight the shortcomings of best-subsets when contrasted with shrinkage methods.  A natural question to ask at this point is: how might we \emph{fix} this problem? Addressing this question with an associated methodological development is the main focus of this paper. We rule out the ambitious goal of correct variable selection, as this may be not be statistically possible when the noise-level is high.
Instead, we focus on improving the predictive performance of the best-subsets approach, with an explicit control of the model size -- we also wish to devise an estimator that is based on a simple and easy-to-interpret optimization criterion.

In Section~\ref{sec:methodology} we formulate the optimization problem for our proposed estimator and describe how to compute the corresponding solutions using modern computational tools from mathematical optimization.
In Section~\ref{sec:theory} we study the theoretical properties of our proposed approach.  First, we establish non-asymptotic error bounds for the new estimators.  Second, we derive novel lower-bounds on the prediction error for the best-subsets estimator in settings where the noise-level is high,  and then contrast the predictive performance of best-subsets with that of our estimators.  In Section~\ref{sec:broader} we discuss the connections between our proposal and existing work, and in Section~\ref{sec:expts} we evaluate the performance of the proposed estimators empirically.  Theoretical proofs and some computational details are provided in the Supplementary Material.

\section{Methodological Framework}
\label{sec:methodology}

Continuous shrinkage methods that solve optimization problems of the form
\begin{equation}\label{subset-lq}
\mini \limits_{\B{\beta}}~~~~ \ \frac12 \| \M{y} - \M{X}\B\beta \|_{2}^2 + \lambda \| \B\beta \|^q_{q},
\end{equation}
such as ridge regression ($q=2$) and the Lasso ($q=1$), are generally recognized for producing estimators with excellent predictive performance, however, their estimated models are denser than those produced by best-subsets (see Figure~\ref{fig-one}).
Similarly to the best-subsets approach, the Lasso \emph{searches} for a subset of features, however, unlike best-subsets,
it then regularizes the least-squares regression performed on the selected features. The superior predictive performance of the Lasso can be attributed in part to the shrinkage effect of the $\ell_{1}$-penalty.
Perhaps even more compelling is the example of ridge regression -- there is no searching here per se, as all the estimated coefficients are generally nonzero.  The excellent predictive performance of ridge regression can be attributed fully to the shrinkage induced by the $\ell_2$-penalty.


\subsection{The proposed estimator}
The above discussion suggests the possibility of obtaining a \emph{sparse} linear model with predictive performance better than best-subsets and comparable to, or even better than, ridge regression and the Lasso.
In terms of sparsity, we desire an estimator with fewer nonzero coefficients than the Lasso, for example.
We propose the following regularized best-subsets estimator{\footnote{Estimator~\eqref{best-subset-penalized}
is inspired by \textit{regularized SVD} estimators (involving a nuclear norm penalty and a rank constraint) commonly used in collaborative filtering~\citep{koren2009matrix} and matrix completion~\citep{hastie2015matrix}.}}:
\begin{equation}\label{best-subset-penalized}
\mini \limits_{\B{\beta}}~~~~ \ \frac12 \|\mathbf{y} - \mathbf{X} \B{\beta} \|_2^2 + ~\underbrace{\lambda \| \B\beta \|_{q}}_{\text{Shrinkage}} ~~~~ \sbt ~~~~~ \underbrace{\|\B{\beta}\|_0 \le k}_{\text{Sparsity}}.
\end{equation}
Above, the cardinality constraint on $\B\beta$ directly controls the model size, and the $\ell_{q}$-penalty{\footnote{Note that Problem~\eqref{best-subset-penalized} uses the $\ell_{q}$ rather than the~$\ell_{q}^q$ penalization, to be consistent with the theoretical results in Section~\ref{sec:theory}. However, our computational framework can handle both versions of the problem.}} with $q \in \{1, 2 \}$ shrinks the regression coefficients towards zero using $\lambda>0$ as the shrinkage parameter.
Informally speaking\footnote{When $q=1$, the shrinkage penalty may induce further sparsity.}, Problem~\eqref{best-subset-penalized} separates out the effects of shrinkage (via  $\lambda \| \B\beta \|_{q}$) and sparsity (via $\|\B{\beta}\|_0 \le k$) -- this may be contrasted with the Lasso, where the penalty simultaneously controls both shrinkage and sparsity, and best subset selection, which only selects but does not shrink.
The family of estimators~\eqref{best-subset-penalized} contains as special cases the best-subsets estimator given by Problem~\ref{subset-1} ($\lambda=0$), the Lasso family ($k=p$, $q=1$) and the ridge regression\footnote{The coefficient path for Problem~\eqref{best-subset-penalized} contains the ridge regression coefficient path.} family ($k=p$, $q=2$) of estimators. For other values of~$\lambda$ and~$k$, Problem~\eqref{best-subset-penalized} combines the best of both worlds: best-subsets (Problem~\ref{subset-1}) and continuous shrinkage methods (Problem~\ref{subset-lq}).

Figure~\ref{fig-one} shows that
when $k > \| \B\beta^*\|_0$, continuous shrinkage regulates the overfitting behavior of best-subsets: as~$k$ increases, estimator~\eqref{best-subset-penalized} overfits more slowly when compared to best-subsets.
This observation is also supported by our theory in Section~\ref{sec:theory}.
When the SNR is low, shrinkage imparted via $\ell_{q}$-regularization becomes critical -- estimator~\eqref{best-subset-penalized} prefers to choose a strictly positive value of~$\lambda$ to produce a good predictive model.
The $\ell_{1}$-penalty in estimator~\eqref{best-subset-penalized} with $q=1$ can also act as an additional sparsification tool when $k$ is large -- this partially explains its (marginally) superior predictive accuracy over $q=2$
for larger SNR values. Overall, Figure~\ref{fig-one} illustrates that estimator~\eqref{best-subset-penalized} produces sparser models than the Lasso, while its predictive performance is consistently as good as or better than that of the continuous shrinkage methods.

Problem~\eqref{best-subset-penalized} is a nonconvex optimization problem. However, as we show in Section~\ref{sec:formulations}, it can be expressed as a mixed integer second order conic optimization (MISOCO) problem and solved (in practice) to certifiable optimality by leveraging advances in modern integer optimization techniques, using standard  solvers like Cplex, Gurobi, Knitro, Mosek, Glpk, Scip~\citep{linderoth2010milp,vielma2016extended}.
To obtain high-quality solutions to Problem~\eqref{best-subset-penalized} at low computational cost, we develop specialized discrete first order methods~\citep{nesterov2004introductorynew} in Section~\ref{dfo}, by extending the framework in~\cite{bertsimas2015best,mazumder2015discrete}.  When these algorithms are used with our proposed continuation schemes across $(\lambda, k)$ and randomized local search heuristics~\citep{aarts1997local,mladenovic1997variable}, a family of (near optimal) feasible solutions to Problem~\eqref{best-subset-penalized} can be computed within minutes.
These algorithms, however, do not certify the quality of the solutions in terms of lower-bounds on the objective function. For this we need the power of MIO techniques. When our heuristic algorithms are used in conjunction with MISOCO  solvers for Problem~\eqref{best-subset-penalized}, they lead to improved computational performance -- see, for example,~\cite{bertsimas2015best,mazumder2015discrete} for similar observations on related problems.

\subsection{Mixed Integer Optimization formulations}\label{sec:formulations}
Here we present the MIO formulation for Problem~\eqref{best-subset-penalized}. Denoting $\{1,\ldots,p\}$ by $[p]$ and assuming, without loss of generality,\footnote{Note that every solution to~\eqref{best-subset-penalized} is bounded when $\lambda>0$, because the level sets of the objective function are
bounded. The case for $\lambda=0$ has been addressed in~\cite{bertsimas2015best}.} that $\B\beta \in [-\mathcal M, \mathcal M]^{p}$, we can rewrite~\eqref{best-subset-penalized} as follows:
\begin{equation}\label{mio-form-1}
\mini~~\frac12\| \M{y} - \M{X}\B\beta \|_{2}^2 + \lambda\| \B\beta \|_{q}~~\sbt~~ -{\mathcal M} z_{j} \leq  \beta_{j} \leq {\mathcal M} z_{j}, j \in [p]; ~
 \M{z} \in \{ 0 , 1 \}^p;~\sum\limits_{j} z_{j} = k.
\end{equation}
Here, $\B\beta$ and~$\M{z}$ are the optimization variables and ${\mathcal M} < \infty$ is a BigM parameter~\citep{bertsimas2005optimization_new,bertsimas2015best}, which is sufficiently large, so that a solution to Problem~\eqref{mio-form-1} is also a solution to Problem~\eqref{best-subset-penalized}.
 The binary variable $z_{j}$ controls whether $\beta_{j}$ is zero or not: $z_{j}=1$ implies that $\beta_{j}$ is \emph{free} to vary in $[- {\mathcal M}, {\mathcal M}]$, and $z_{j}=0$ implies $\beta_{j}=0$. The constraint
 $\sum_{j} z_{j} = k$ allows at most~$k$ regression coefficients to be nonzero. The nonconvexity in~\eqref{mio-form-1} stems from the binary variables in~$\M{z}$.
Problem~\eqref{mio-form-1} can be reformulated as a
MISOCO, i.e., a second order conic optimization problem~\cite{BV2004} where a subset of the variables is binary. Thanks to the impressive advances in MIO, these problems can be solved in practice using state-of-the-art MIO solvers~\citep[see, for example, the recent work of][]{vielma2016extended}.
To this end, we note that~\eqref{mio-form-1} can be written as follows:
\begin{equation}\label{mio-form-2}
\mini~~{u}/{2}  + \lambda v~~\sbt~~\| \M{y} - \M{X} \B\beta\|_{2}^2 \leq u,~\| \B\beta \|_{q} \leq v, ~ (\B\beta, \M{z}) \in {\mathcal C},
\end{equation}
where the optimization variables are $(u,v, \B\beta, \M{z}) \subset  \mathbb{R} \times \mathbb{R} \times \mathbb{R}^{p} \times \{0,1\}^{p}$, and~$\mathcal C$ denotes the mixed integral polyhedral constraint in~\eqref{mio-form-1}.  The first term in the constraint can be expressed as a second order cone~\citep{BV2004},
$$ \left\{ (\B\beta, u) :   \|\M{y} - \M{X} \B\beta\|_{2}^2 \leq u, u \geq 0 \right\}~~ \equiv~~\left \{ (\B\beta, u ) : \left\| \big( [\M{y} - \M{X} \B\beta]^\top , [u-1]/2  \big) \right\|_{2} \leq (u+1)/2, u \geq 0 \right\}. $$
For $q=1$, the term $\| \B\beta \|_{q} \leq v$ in the constraint can be
expressed via linear inequalities using auxiliary continuous variables $\{\bar{\beta}\}_{1}^{p}$:
\begin{equation}\label{epi-L1-norm}
\{(\B\beta, v) : \| \B\beta \|_{1} \leq v, v \geq 0 \} ~~ \equiv ~~\{(\B\beta, v) :\exists ~ \bar{\B\beta}~\M{\geq}~ \M{0} ~~\text{s.t.}~~ - \bar{\beta}_{j} \leq \beta_{j}  \leq \bar{\beta}_{j}, \; \sum_{j} \bar{\beta}_{j} \leq v, v \geq 0 \},
\end{equation}
thereby leading to a MISOCO formulation for~\eqref{mio-form-2} when $q=1$.
When $q=2$, the epigraph version of
$\| \B\beta \|_{q} \leq v $ is already a second order cone, so~\eqref{mio-form-2} admits a MISOCO formulation.

\noindent {\textbf{Other Formulations.}} Computational performance of MISOCO solvers (Gurobi, for example) is found to improve by adding structural implied inequalities, or cuts, to the basic formulation~\eqref{mio-form-2} -- see Section~\ref{strong-formu-1} of the Supplementary Material.
Computation of problem-specific BigM parameters and other bounds is discussed in Section~\ref{bigM-compute1}.

Problem~\eqref{mio-form-1} with $q=1$ can also be expressed as a mixed integer quadratic optimization (MIQO) problem.
Note that if we replaced the $\ell_{2}$-penalty in~\eqref{mio-form-2} with the squared-$\ell_{2}$-penalty, then the resulting problem would be readily expressed as MIQO as well -- both problems leading to the same family of solutions\footnote{If we denote the solution to the modified problem by $\hat{\B\beta}_{\ell_{2}^2}(\lambda',k)$, then, for every fixed $k$, the solution path $\{\hat{\B\beta}_{\ell_{2}^2}(\lambda',k)\}_{\lambda' \geq 0}$ recovers the corresponding path for the original Problem~\eqref{mio-form-1} with~$q=2$.}.
In what follows, we will focus on the MISOCO formulation  presented above to be consistent with our theoretical results in Section~\ref{sec:theory}.


\subsection{Discrete First Order Algorithms}\label{dfo}
Inspired by proximal gradient methods~\citep{nesterov2004introductorynew,nesterov2013gradient}, popularly used in convex optimization, we present discrete first order (DFO) methods  to obtain good upper bounds for~\eqref{best-subset-penalized}.
The DFO methods
have a low iteration complexity and can nicely exploit warm-start information across the $(\lambda, k)$-space: Using a combination of neighborhood continuation schemes and local combinatorial search methods proposed here, they lead to near-optimal{\footnote{In our experiments, we observed that the solutions obtained by our elaborate heuristics are often close to the
optimal solutions returned by the MIO solvers in the neighborhood of the optimal $(\lambda, k)$ choice, made by minimizing the prediction error on a separate validation set.}}
solutions to~\eqref{best-subset-penalized}. We note that the DFO methods are heuristics--
they do not certify solution quality (i.e., global optimality) via dual-bounds. For the latter,
we critically rely on MIO technology.
The MIO solvers accept warm-starts available from the DFO algorithm, then subsequently improve the solution and certify optimality, at the cost of additional (but still reasonable) computation times.
		
	

We describe a DFO method for the following problem (in composite form~\cite{nesterov2013gradient}):
\begin{equation}\label{gen-form-1}
\mini ~~~F(\B\beta):=~~f(\B\beta) + \lambda \| \B\beta \|_{q} ~~~ \sbt ~~~ \| \B\beta \|_{0} \leq k,
\end{equation}
where
$f(\B\beta)$ is a $L_0$-smooth convex function, i.e., it satisfies
\begin{equation}\label{lip-grad-1}
\|\nabla f(\B\beta) - \nabla f(\B\alpha) \|_{2} \leq L_0 \| \B\beta - \B\alpha\|_{2}~~~\forall~ \B\beta, \B\alpha \in \mathbb{R}^{p}.
\end{equation}
For $f(\B\beta) = \frac12 \| \M{y} - \M{X} \B\beta \|_{2}^2$, we can use $L_0= \sigma_{\max}(\M{X})^2$, where $\sigma_{\max}(\cdot)$ is the maximum singular value of $\M{X}$.
As a consequence of~\eqref{lip-grad-1}, for any $L \geq L_0$, we have the following bound~\citep{nesterov2004introductorynew} in place:
\begin{equation}\label{ubound-1}
f(\B{\beta}) \le f(\B{\alpha}) + \langle \nabla f(\B{\alpha}), \B{\beta}-\B{\alpha} \rangle+ \frac{L}{2} \| \B{\beta} -\B{\alpha} \|_2^2 := Q_L(\B{\beta};\B\alpha),~~\forall \B\alpha, \B\beta \in \mathbb{R}^p.
\end{equation}
Given a current solution $\B\alpha$, our algorithm minimizes an upper bound to $F(\B\beta)$ around $\B\alpha$:
\begin{equation}\label{upper-bound-cont1}
 \mini_{\|\B\beta\|_{0} \leq k}~~  Q_L(\B{\beta};\B\alpha)  + \lambda \| \B\beta \|_{q} \iff   \mini_{\|\B\beta\|_{0} \leq k}~~  \frac{L}{2} \left\| \B\beta - \left( \B\alpha - \frac{1}{L} \nabla f( \B\alpha)  \right) \right\|_{2}^2 + \lambda \| \B\beta \|_{q}.
 \end{equation}
A key ingredient in solving the above is the thresholding operator,
\begin{equation}\label{thresh-op-1}
\s(\M{u} ; k; \lambda \ell_{q} ) := \argmin_{\B\beta: \| \B\beta \|_{0} \leq k}~~ \frac{1}{2}  \left\| \B\beta - \M{u} \right\|_{2}^2 + \lambda \| \B\beta \|_{q},
\end{equation}
where $\s(\M{u} ; k; \lambda \ell_{q})$ denotes the set of optimal solutions to Problem~\eqref{thresh-op-1}.
We note that $ \s(\M{u}; k; \lambda \ell_{q} )$ may be set-valued -- the non-uniqueness of an optimal solution to Problem~\eqref{thresh-op-1} arises from the fact that the ordering of~$|u_{j}|$ for $j \in [p]$ may have ties.

\smallskip

\begin{prop}
Let $(1), \ldots, (p)$ be a permutation of the indices $1,\ldots, p$, such that the entries in $\M{u}$ are sorted as:
$|u_{(1)}| \geq |u_{(2)}| \geq \ldots \geq | u_{(p)}|$. Then, the thresholding operator~\eqref{thresh-op-1} has the following form:
\begin{itemize}
\item[(a)] For the $\ell_{1}$-regularizer (with $q=1$) any $\hat{\B\beta} \in \s(\M{u}; k; \lambda \ell_{q})$ is given by:
\begin{equation}\label{soft-hard-1}
\hat{\beta}_{i} = \begin{cases} \sgn(u_{i}) \max \{ | u_{i}| - \lambda, 0 \} & i \in \{ (1), (2), \ldots, (k) \} \\
0 & \text{otherwise}.
\end{cases}
\end{equation}
\item[(a)] For the $\ell_{2}$-regularizer (with $q=2$) any $\hat{\B\beta} \in \s(\M{u}; k; \lambda \ell_{q})$ is given by:
\begin{equation}\label{soft-hard-2}
\hat{\beta}_{i} = \begin{cases} \frac{u_{i}}{\tau_{u}} \max \{\tau_{u} - \lambda, 0 \} & i \in \{ (1), (2), \ldots, (k) \} \\
0 & \text{otherwise},
\end{cases}
\end{equation}
where $\tau_{u} =  \sqrt{\sum_{i=1}^{k} u_{(i)}^2}$ is the $\ell_{2}$-norm of the~$k$  largest (in magnitude) entries of $\M{u}$.
\end{itemize}
\end{prop}

The DFO algorithm performs the following updates (for $m \geq 1$)
\begin{equation}\label{update-seq-1}
\B\beta^{(m+1)}  \in  \s_{} \left(  \B\beta^{(m)} - \tfrac{1}{L} \nabla f( \B\beta^{(m)}) ; k; \tfrac{\lambda}{L} \ell_{q} \right),
\end{equation}
till some convergence criterion is met. The algorithm is summarized below for convenience.

\begin{itemize}
\item[] ~~~~~~~~~~~~~~~~~~~~ \emph{ \underline{Discrete First Order Algorithm (DFO)}}
\item[1.] Fix $L \geq {L}_0$ and a convergence threshold $\tau >0$.  Initialize with $\B\beta^{(1)}$ that is $k$-sparse. Repeat update~\eqref{update-seq-1}
until $\| \B\beta^{(m+1)}  - \B\beta^{(m)} \|^2_{2} \leq \tau.$
\item[2.] Let $\texttt{I}({\widetilde{\B\beta}})$ denote the support of the $\widetilde{\B\beta}$ obtained from Step 1,
i.e., $\texttt{I}({\widetilde{\B\beta}}) = \{ i : \widetilde{\beta}_{j} \neq 0, j \in [p] \}$.  Solve the convex problem~\eqref{gen-form-1} restricted to the support $\texttt{I}({\widetilde{\B\beta}})$:
$\min ~ F(\B\beta) ~ \text{s.t.}~ \beta_{j} = 0, j \notin \texttt{I}({\widetilde{\B\beta}})$.
\end{itemize}

For the sake of completeness, we establish convergence properties of the sequence $\{\B\beta^{(m)}\}_{m \geq 1}$ in terms of reaching a first order stationary point. Our work adapts the framework proposed in~\cite{bertsimas2015best} to the composite form. Towards this end, we need the following definition.

\smallskip

\begin{mydef}\label{def-1} We say that $\B\eta$ is a
 first order stationary point of Problem~\eqref{gen-form-1} if
  $\B\eta \in \s( \B\eta - \frac{1}{L} \nabla f(\B\eta); k; \frac{\lambda}{L} \ell_{q} )$. We say that~$\B\eta$ is an $\epsilon$-accurate
 first order stationary point if  $\| \B\eta \|_{0} \leq k$ and $ \| \B\eta -  \s( \B\eta - \frac{1}{L} \nabla g(\B\eta); k; \frac{\lambda}{L} \ell_{q} )\|^2_{2} \leq \epsilon$.
\end{mydef}

The following result presents convergence properties of the sequence $\{\B\beta^{(m)}\}_{m \geq 1}$ in terms of reaching a  first order stationary point (see Section~\ref{proof-prop-dfo-1}, Supplementary Material for the proof).

\smallskip

 \begin{prop}\label{prop-conv-1}
Let $\{\B\beta^{(m)}\}$  denote a sequence generated by the DFO algorithm. Then,
\begin{itemize}
\item[(a)] for $L \geq {L}_0$, the sequence $F(\B\beta^{(m)})$ is decreasing, and it converges to some~$F^* \geq 0$;
\item[(b)] for $L > {L}_0$, we have the following finite-time convergence rate:
$$   \min_{1 \leq m \leq M} \| \B\beta^{(m+1)}  - \B\beta^{(m)} \|_{2}^2 \leq \frac{2(F(\B\beta^{(1)} ) - F^*)}{M(L- L_0)}.$$
\end{itemize}
\end{prop}

Proposition~\ref{prop-conv-1} suggests that the DFO algorithm applied to Problem~\eqref{gen-form-1} leads to a decreasing sequence of objective values, which eventually converges.
When $L>L_0$ the algorithm reaches an $\epsilon$-accurate first order stationary point (Definition~\ref{def-1}) in $O(\epsilon^{-1})$ iterations. We note that the proposition makes no assumption on the data at hand -- improved convergence rates
may be achievable by making further assumptions on the problem data (see, for example, \cite{bertsimas2015best} and the discussion therein).
In practice however, the DFO algorithm converges much faster (especially when using warm-start continuation) than the sublinear rate suggested by Proposition~\ref{prop-conv-1}.

\subsection{Neighborhood continuation and local search heuristics}\label{nbhd-conti-1}
Due to the nonconvexity of Problem~\eqref{best-subset-penalized}, the DFO algorithm is sensitive to the initialization $\B\beta^{(1)}$. The effect of initialization becomes particularly pronounced when $n$ is relatively small compared to $p$, the pairwise (sample) correlations among the features are high; and the SNR is low.
These solutions can be improved, often substantially (in terms of the objective value), using continuation schemes and randomized local search-heuristics, as we discuss below.
The continuation scheme, which makes use of the warm-starting capabilities of the DFO algorithm, is quite efficient. Note that these algorithms serve as stand-alone methods to obtain good feasible solutions for~\eqref{best-subset-penalized}, for a family of tuning parameters $(\lambda, k)$ -- this makes them practically appealing.
Furthermore, these methods can be used to obtain a good estimate of an optimal tuning parameter (for example, based on validation set tuning) with relatively low computational cost.

\noindent {\textbf{Neighborhood Continuation.}} Let $\hat{\B\beta}(\lambda, k)$ denote a solution delivered by the DFO algorithm for~\eqref{best-subset-penalized} (we drop the dependence on $q$ for notational convenience).
We let $F(\lambda, k)$ denote the corresponding objective value.
We consider a 2D grid of tuning parameters in $\Lambda \times K = \{\lambda_{1}, \ldots, \lambda_{N} \} \times \{k_{1}, \ldots, k_{r} \}$ with $\lambda_{i} > \lambda_{i+1}$ and $k_{i} > k_{i+1}$ for all $i$.
We set $k_{1} = p, k_{r} = 1$. We set $\lambda_{1} = \|\M{X}^\top\M{y}\|_{\bar{q}}$ with $\bar{q} = \infty$ if $q = 1$ and $\bar{q} = 2$ if $q=2$ -- the rationale being that if
$\lambda = \lambda_{1}$, then an optimal solution to Problem~\eqref{best-subset-penalized} is zero.

\begin{compactitem}
\item[]~~~~~~~~~~~~~~~~~~~~~~~\emph{\underline{Algorithm~1: Neighborhood Continuation}}
\item[(i)] Initialize $\hat{\B\beta}(\lambda_{i}; k_{j}) \leftarrow \M{0}$ for every $i, j \in [N] \times [r]$.
Repeat Step (ii) until the array of objective values $\{F(\lambda_{i}; k_{j})\}_{i, j}$ stops changing between successive sweeps across the 2D grid $\Lambda \times K$:
\item[(ii)] For $i \in [N], j \in [r]$ do the following:
\begin{compactitem}
\item[(a)] Set $(\lambda, k)  = (\lambda_{i}, k_{j})$ and use the DFO algorithm with (at most) four different neighborhood initializations $\hat{\B\beta}(\lambda_{a}; k_{b}),$
$(a,b) \in {\mathcal N}(i,j)$  where,
${\mathcal N}(i,j)$ are the neighbors of $({i}, {j})$.
For every $(a,b)$ in the neighborhood ${\mathcal N}(i,j)$, let $\hat{\B\beta}_{a,b}$ and~$F_{a,b}$ denote the corresponding estimate and objective value, respectively.
\item[(b)] Set $\hat{\B\beta} (\lambda_{i}; k_{j})$ equal to the estimate $\hat{\B\beta}_{a,b}$ with the smallest objective value: $F(\lambda_{i}; k_{j}) = \min \{ F_{a,b}:(a,b) \in {\mathcal N}(i,j)\}$.
\end{compactitem}
\end{compactitem}

We make a series of remarks pertaining to Algorithm~1:

\begin{compactitem}

\item If we denote one execution of Step-(ii) (formed by looping across all $i, j \in [N] \times [r]$) as a sweep, then successive sweeps may lead to
a strict improvement{\footnote{By construction, given $(i,j)\in [N] \times [r]$, the objective value $F(\lambda_{i}, k_{j})$ cannot increase between successive sweeps.}} in the objective values $\{F(\lambda_{i}, k_{j})\}_{i,j}$ for several $(i,j)$.

\item During the first sweep of Algorithm~1
many neighbors $\hat{\B\beta}(\lambda_{a}, r_{b})$ of $(i,j)$ are zero. After the first sweep, however,
all entries $(i,j)$ get populated.

\item The neighborhood initializations $\hat{\B\beta}(\lambda_{a}; k_{b})$ for $(a,b) \in {\mathcal N}(i,j)$ serve as excellent warm-starts
for~\eqref{best-subset-penalized} at $(\lambda_{i}, r_{j})$. This improves the overall runtime of the algorithm (as compared to independently computing the solutions on the 2D grid) and also
results in a solution with good objective values.

\end{compactitem}

\noindent {\textbf{A (randomized) local search heuristic.}}
We present a local-search heuristic, which, loosely speaking, is capable of navigating different parts of the model space
by perturbing the support of a DFO solution.
We draw inspiration from local search schemes commonly used in combinatorial optimization problems~\citep{aarts1997local,mladenovic1997variable}.
Our local search scheme works as follows: for every nonzero initialization $\hat{\B\beta}(\lambda_{a}, k_{b})$, we randomly swap roughly 50\% of the nonzero coefficients with an equal number of zero coefficients before passing the resulting estimate as an initialization to the DFO algorithm.
This stochastic search scheme is performed as a part of the 2D continuation scheme (described above) -- we register the estimate if it leads to an improvement in the objective value.



\section{Statistical Theory}
\label{sec:theory}

We study the performance of the proposed approach in the regression setting with deterministic design. In Sections \ref{sec.prelims}-\ref{sec.L1.bnds} we establish non-asymptotic oracle error bounds for the corresponding estimators.  In Section~\ref{sec.lower.bnds} we contrast the predictive performance of the new approach with that of best-subsets selection, by deriving novel lower-bounds on the prediction error of~$\widehat\bbeta_{\ell_0}$.  The comparison between the estimators is done for each fixed value of the model size tuning parameter~$k$. In Section~\ref{sec.choice.of.k} we analyze a BIC-type approach for selecting the optimal value of~$k$. Our results provide new insights on the benefits of additional regularization in best subset selection.

\subsection{Notation and preliminary results}
\label{sec.prelims}

We assume that the observed data follows the model
\begin{equation}
\label{main.model.theory}
\bY=\bff^*+\bepsilon.
\end{equation}
The components in the equation above are vectors in $\mathbb{R}^n$, vector~$\bff^*$ is an unknown deterministic mean, and the elements of~$\bepsilon$ are independent $N(0,\sigma^2)$ with $\sigma>0$.
A special case of~(\ref{main.model.theory}) is the linear model $\bff^*=\bX\bbeta^*$.  As before, we assume that the columns of~$\bX$ have unit $\ell_2$-norm.

We use the following notation for the regularized best-subsets solutions to Problem~\eqref{best-subset-penalized}:
\begin{equation}
\label{opt.problem.theory}
\widehat\bbeta_q = \argmin_{\bbeta} \|\bY-\bX\bbeta\|^2+\lambda\|\bbeta\|_q   \quad \text{s.t.}\quad \|\bbeta\|_0\le k, \qquad\;\text{for}\; q=1,2.
\end{equation}
The dependence of~$\widehat\bbeta_q$ on~$k$ and~$\lambda$ is understood implicitly. From here on, we drop the subscript in the notation $\|\cdot\|_2$, used for the Euclidean norm. To simplify the presentation, we refer to $\|\bff^*-\widehat\bbeta_q\|^2$ as the prediction error for~$\widehat\bbeta_q$, multiplying the usual prediction error by~$n$.
 Given an integer $s\in [p]$, we define $B_0(s)=\{\bu\in\mathbb{R}^p:\,\|\bu\|_0\le s\}$ and let~$\gamma_s$ denote the minimal $s$-sparse eigenvalue of~$\bX$:
\begin{equation*}
\gamma_s=\min_{\bu\ne\mathbf{0},\bu\in B_0(s)}\frac{\|\bX\bu\|}{\|\bu\|}.
\end{equation*}
Given a vector $\bu\in\mathbb{R}^p$, we write $u_1^\sharp,...,u_p^\sharp$ for a non-increasing rearrangement of $|u_1|,...,|u_p|$.  We say that a constant is \textit{universal} if it does not depend on other parameters, such as~$k$, $p$ or~$\lambda$. We write~$\gtrsim$ and~$\lesssim$ to indicate that inequalities~$\ge$ and~$\le$, respectively, hold up to positive universal multiplicative factors, and use~$\asymp$ when the two inequalities hold simultaneously. We use the notation $a\vee b= \max\{a,b\}$, $a\wedge b= \min\{a,b\}$, and treat algebraic expressions~of~the~form~$0\cdot\infty$~or~$0/0$~as~zero.

As is typical in high-dimensional regression settings, we establish the error bounds by conducting deterministic arguments on suitably chosen random events:
\begin{eqnarray*}
\mathcal{E}_s&=&\big\{\bepsilon^\top\bX\bu\le[4+\sqrt{2}]\sigma\sqrt{s\log(2ep/s)}\|\bu\|, \, \forall\bu\in B_0(s)  \big\}\\
\mathcal{F}&=&\big\{\bepsilon^\top\bX\bu\le[4+\sqrt{2}]\sigma\max\big(\sum\nolimits_{j=1}^p u_j^\sharp\sqrt{\log(2p/j)}\,,\,\sqrt{\log(1/\delta_0)}\|\bX\bu\|\big), \, \forall\bu\in \mathbb{R}^p\big\}\\
\mathcal{G}_s&=&\big\{\bepsilon^\top\bX\bu\le\sigma\sqrt{5s\log(ep/s)+\log(1/\delta_0)}\|\bX\bu\|, \, \forall\bu\in B_0(s)  \big\}\\
\mathcal{H}&=&\big\{\|\bX^\top\bepsilon\|_{\infty}\le\sigma\sqrt{2\log(2p)}+\sigma\sqrt{2\log(1/\delta_0)} \big\}.
\end{eqnarray*}
When $s/p$ and~$\delta_0$ are small, all four events hold with high probability.

\smallskip

\begin{thm} Suppose that $s\in[p]$ and $\delta_0\in(0,1]$.  Then,
\label{probs.thm}
\begin{equation*}
\mathbb{P}(\mathcal{E}_s)\ge 1-s/(4ep), \quad  \mathbb{P}(\mathcal{F})\ge 1-\delta_0/2, \quad \mathbb{P}(\mathcal{G}_s)\ge 1-\delta_0\quad\text{and}\quad \mathbb{P}(\mathcal{H})\ge 1-\delta_0.
\end{equation*}
\end{thm}
Some of the above probability bounds have appeared in the literature.  In particular, the bound for~$\mathcal{F}$, which is an important component of our analysis, was recently established in~\cite{bellec2018slope}.

\subsection{Results for the $\ell_2$-regularized best-subsets estimator}
\label{sec.L2.bnds}

We follow the common convention in the literature \citep[][for example]{dalalyan2017prediction} by referring to prediction error rates that involve terms of order~$\lambda^2$ as \textit{fast} and referring to prediction error rates that involve terms of order~$\lambda$ as \textit{slow}.  The slow rates are especially relevant to our study, because they tend to outperform the fast rates in the high-noise regimes.  The following result focuses on~$\widehat\bbeta_2$ and provides both the slow and the fast rate prediction error bounds.  We note that an important attractive feature of the last two error bounds in Theorem~\ref{slow.rate.thm} is the independence of the uncertainty parameter~$\delta_0$ from the tuning parameters~$\lambda$ and~$k$.  This feature allows us to control the expected prediction error, as we demonstrate in Corollary~\ref{Expect.slow.rate} below.

\smallskip

\begin{thm}
\label{slow.rate.thm}
(A) Slow rate. If $\lambda\ge[8+2\sqrt{2}]\sigma \sqrt{2k\log (ep/k)}$, then on the event~$\mathcal{E}_{2k}$,
\begin{equation*}
\|\bff^*-\bX\widehat\bbeta_2\|^2\le \inf_{\bbeta\in B_0(k)}\Big[\|\bff^*-\bX\bbeta\|^2 + 2\lambda\|\bbeta\|\Big];
\end{equation*}
and on the event~$\mathcal{F}$,
\begin{equation*}
\|\bff^*-\bX\widehat\bbeta_2\|^2\lesssim \inf_{\bbeta\in B_0(k)}\Big[\|\bff^*-\bX\bbeta\|^2 + \lambda\|\bbeta\|\Big]+\sigma^2\log(1/\delta_0).
\end{equation*}
(B) Fast rate. On the event~$\mathcal{G}_{2k}$,
\begin{equation*}
\|\bff^*-\bX\widehat\bbeta_2\|^2\lesssim \inf_{\bbeta\in B_0(k)}\|\bff^*-\bX\bbeta\|^2+\sigma^2k\log (ep/k) + \gamma_{2k}^{-2}\lambda^2+\sigma^2\log(1/\delta_0)
\end{equation*}
for every $\lambda\ge0$.
\end{thm}
The above result establishes oracle inequalities for the prediction error under potential model misspecification.  The added generality allows us to avoid restrictions on the model size parameter~$k$. This is relevant to the discussion in Section~\ref{sec.lower.bnds} on the relationship between decreasing~$k$ and the predictive performance of best-subsets. {We note that our oracle inequalities are restricted to~$B_0(k)$ for each fixed value of the model size tuning parameter~$k$. In Section~3.5 we present a data-driven approach for selecting~$k$ and establish oracle inequalities in a more general form.}

To illustrate the rates of convergence in Theorem~\ref{slow.rate.thm} more clearly, we consider the linear case, $\bff^*=\bX\bbeta^*$, and set~$\delta_0$ equal to some specific small values.

\smallskip

\begin{cor}
\label{lin.mod.l2}
Let $\bff^*=\bX\bbeta^*$ for some $\bbeta^*\in B_0(k)$. If $\lambda\ge[8+2\sqrt{2}]\sigma \sqrt{2k\log (ep/k)}$, then
\begin{equation*}
\|\bX\widehat\bbeta_2-\bX\bbeta^*\|^2\le 2\lambda\|\bbeta^*\|
\end{equation*}
with probability at least $1-k/(2ep)$, and
\begin{equation*}
\|\bX\widehat\bbeta_2-\bX\bbeta^*\|^2\lesssim \lambda\|\bbeta^*\|+\sigma^2\log(p)
\end{equation*}
with probability at least $1-1/p$.  Furthermore, with probability at least $1-(k/p)^k$,
\begin{equation*}
\|\bX\widehat\bbeta_2-\bX\bbeta^*\|^2\lesssim \sigma^2k\log (ep/k) + \gamma_{2k}^{-2}\lambda^2
\quad\,\text{and}\,\quad \|\widehat\bbeta_2-\bbeta^*\| \lesssim \gamma_{2k}^{-1}\sigma\sqrt{k\log (ep/k)} + \gamma_{2k}^{-2}\lambda
\end{equation*}
for every $\lambda\ge0$.
\end{cor}

We make the following observations regarding the established error bounds for~$\widehat\bbeta_2$.

\smallskip

\begin{rem}
Letting $k=k^*$, we note that the fast prediction error rate, $\sigma^2k^*\log (ep/k^*)$, matches the minimax rate over $\bbeta^*\in B_0(k^*)$ \citep{rigollet2011exponential, lounici1, raskutti2011minimax}.
\end{rem}

\smallskip

\begin{rem}
When $\lambda=0$, the fast rate part of Corollary~\ref{lin.mod.l2} yields the prediction and estimation error bounds for the best-subsets estimator, $\widehat\bbeta_{\ell_0}$.
\end{rem}

\smallskip

\begin{rem}
The slow rate for~$\widehat\bbeta_2$ is $\sigma\sqrt{k^*\log(ep/k^*)}\|\bbeta^*\|$, which improves on the prediction error bound for~$\widehat\bbeta_{\ell_0}$ when $\|\bbeta^*\|/\sigma\lesssim \sqrt{k^*\log(ep/k^*)}$ with a sufficiently small universal constant.
\end{rem}

\smallskip

{
\begin{rem}
The lower-bound on~$\lambda$ needed for the slow rate results contains the unknown parameter~$\sigma$, i.e., the standard deviation of the noise in the model. The noise variance can be estimated by employing a preliminary regression estimator \citep[see, for example, the discussion in][]{belloni2013least} that is unrestricted in terms of the model size.
In practice, parameter~$\lambda$ can be tuned based on a separate validation set (or by cross-validation), leading to the best $k$-sparse model with respect to the validation error.
\end{rem}
}

{
The error rates presented above can also apply to approximate solutions, obtained after an early termination of the MIO solver.  Upon termination, the solver provides the upper and lower bounds on the value of the objective in~(\ref{opt.problem.theory}). We denote these bounds by $UB$ and $LB$, respectively, and write $\tau=(UB-LB)/UB$ for the corresponding optimality gap. The next result focuses on an approximate $\ell_2$-regularized best-subsets solution~$\widetilde{\bbeta}_2$ in the linear setting of Corollary~\ref{lin.mod.l2}.

\smallskip
\begin{cor}
\label{lin.mod.opt.gap}
Let $\bff^*=\bX\bbeta^*$ for some $\bbeta^*\in B_0(k)$ and suppose that $\tau\le 1-c$ for some positive universal constant~$c$. Then, with probability at least $1-1/p$,
\begin{equation*}
\|\bX\widetilde{\bbeta}_2-\bX\bbeta^*\|^2\lesssim \lambda\|\bbeta^*\|+\sigma^2\big[\log(p)+\tau n\big] \quad\text{for}\;
\lambda\ge[8+2\sqrt{2}]\sigma \sqrt{2k\log (ep/k)}.
\end{equation*}
In addition, with probability at least $1-(k/p)^k$,
\begin{equation*}
\|\bX\widetilde{\bbeta}_2-\bX\bbeta^*\|^2\lesssim \sigma^2\big[k\log (ep/k)+\tau n\big] + \gamma_{2k}^{-2}\lambda^2 \quad \text{for every}\; \lambda\ge0.
\end{equation*}
\end{cor}
We note that~$\widetilde{\bbeta}_2$ achieves the second slow error rate in Corollary~\ref{lin.mod.l2} when $\tau\lesssim\log(p)/n$, and it achieves the corresponding fast error rate when $\tau\lesssim k\log(ep/k)/n$. Furthermore, we show in the proof of Corollary~\ref{lin.mod.opt.gap} that the multiplicative increase in the slow rate error bound relative to the case $\tau=0$ is at most $1+\frac{\tau}{1-\tau}\big\{1\vee\frac{n}{58\log(p)}\big\}$; the corresponding multiplicative increase in the fast rate error bound is at most $1+\frac{\tau}{1-\tau}\big\{1\vee\frac{n}{43k\log(ep/[2k])}\big\}$. These expressions illustrate the trade-off between the optimality gap and the quality of the prediction error bounds.
}

The next result bounds the expected prediction error of~$\widehat\bbeta_2$.

\smallskip

\begin{cor}
\label{Expect.slow.rate}
If $\lambda\ge[8+2\sqrt{2}]\sigma \sqrt{2k\log (ep/k)}$, then
\begin{equation*}
\mathbb{E}\|\bff^*-\bX\widehat\bbeta_2\|^2\lesssim \inf_{\bbeta\in B_0(k)}\Big[\|\bff^*-\bX\bbeta\|^2 + \lambda\|\bbeta\|\Big]+\sigma^2.
\end{equation*}
Furthermore, for every $\lambda\ge0$,
\begin{equation*}
\mathbb{E}\|\bff^*-\bX\widehat\bbeta_2\|^2\lesssim \inf_{\bbeta\in B_0(k)}\|\bff^*-\bX\bbeta\|^2 +\sigma^2k\log (ep/k) + \gamma_{2k}^{-2}\lambda^2.
\end{equation*}
\end{cor}
Comparing the slow rate bounds in Theorem~\ref{slow.rate.thm} and Corollary~\ref{Expect.slow.rate}, we note that the additional~$\sigma^2$ term in the corollary matches the expected prediction error rate for the oracle least-squares estimator, achieved in the setting where $\|\bbeta^*\|_0$ is bounded above by a universal constant.

\subsection{Results for the $\ell_1$-regularized best-subsets estimator}
\label{sec.L1.bnds}

There exists extensive literature \citep[for example,][]{bickel1,koltchinskii2011nuclear,bartlett2012,sun2012scaled,belloni2014pivotal,dalalyan2017prediction} on the prediction error bounds for the Lasso, which is an $\ell_1$-regularized least-squares estimator. The following theorem focuses on the $\ell_1$-regularized estimator with an additional~$\ell_0$ constraint.  It establishes both the slow and the fast rate prediction error bounds for~$\widehat\bbeta_1$. { Like the estimator~$\widehat\bbeta_1$, the presented oracle inequalities are restricted to~$B_0(k)$ for each fixed value of the model size tuning parameter~$k$. In this respect, they are not as strong as the bounds in the literature that are stated without such a restriction.  In Section~\ref{sec.choice.of.k} we analyze a data-driven approach for selecting the optimal value of~$k$ and establish oracle inequalities in a more general form.}

\smallskip

\begin{thm}
\label{L1.rate.thm}
(A) Slow rate. If $\lambda= 2\sigma \sqrt{2\log (2p)}+2\sigma \sqrt{2\log (1/\delta_0)}$, then on the event~$\mathcal{H}$,
\begin{equation*}
\|\bff^*-\bX\widehat\bbeta_1\|^2\le \inf_{\bbeta\in B_0(k)}\Big[\|\bff^*-\bX\bbeta\|^2 + 2\lambda\|\bbeta\|_1\Big].
\end{equation*}
If $\lambda\ge[8+2\sqrt{2}]\sigma \sqrt{\log (2p)}$, then on the event~$\mathcal{F}$,
\begin{equation*}
\|\bff^*-\bX\widehat\bbeta_1\|^2\lesssim \inf_{\bbeta\in B_0(k)}\Big[\|\bff^*-\bX\bbeta\|^2 + \lambda\|\bbeta\|_1\Big]+\sigma^2\log(1/\delta_0).
\end{equation*}
(B) Fast rate. On the event~$\mathcal{G}_{2k}$,
\begin{equation*}
\|\bff^*-\bX\widehat\bbeta_1\|^2\lesssim \inf_{\bbeta\in B_0(k)}\|\bff^*-\bX\bbeta\|^2+\sigma^2k\log (ep/k) + \gamma_{2k}^{-2}\lambda^2k+\sigma^2\log(1/\delta_0)
\end{equation*}
for every $\lambda\ge0$.
\end{thm}

Focusing on the linear case, $\bff^*=\bX\bbeta^*$, we make the following observations.

\smallskip

\begin{rem}
 The slow rate prediction error bound for~$\widehat\bbeta_1$ is $\sigma\sqrt{\log(ep)}\|\bbeta^*\|_1$, which is better than the $\sigma^2k^*\log (ep/k^*)$ bound for best-subsets when $\|\bbeta^*\|_1/\sigma\lesssim k^* \log (ep/k^*) / \sqrt{\log (ep)}$ with a sufficiently small universal constant.
\end{rem}

\smallskip

\begin{rem}
As is the case with~$\widehat\bbeta_2$, the fast prediction error rate for~$\widehat\bbeta_1$ matches the minimax rate over $\ell_0$-balls.
The slow rate for~$\widehat\bbeta_1$ matches the corresponding rate for the Lasso and the minimax lower bound over $\ell_1$-balls derived in \cite{raskutti2011minimax}.  This rate is slightly worse than the corresponding minimax rate established in \cite{rigollet2011exponential}. However, we note that the latter rate can be derived for~$\widehat\bbeta_1$ with an appropriate tuning of the parameter~$k$, using the arguments in the proof of Corollary 4.1 in \cite{rigollet2011exponential}, which bounds the prediction error of a modified BIC estimator.
\end{rem}

\smallskip

{
\begin{rem}
Similarly to the~$\ell_2$ case (Corollary~\ref{lin.mod.opt.gap}), the established error rates can also apply to solutions obtained after an early termination of the MIO solver. More specifically, if the optimality gap, $\tau$, is bounded away from one, then the approximate solution achieves the second slow error rate in Theorem~\ref{L1.rate.thm} when $\tau\lesssim\log(1/\delta_0)/n$, and it achieves the corresponding fast error rate when $\tau\lesssim k\log(ep/k)/n$.
\end{rem}
}

\smallskip

\begin{rem}
Similarly to Corollary~\ref{lin.mod.l2}, the fast rate part of Theorem~\ref{L1.rate.thm} implies an estimation error bound: $\|\widehat\bbeta_1-\bbeta^*\| \lesssim \gamma_{2k}^{-1}\sigma\sqrt{k\log (ep/k)} + \gamma_{2k}^{-2}\lambda\sqrt{k}$.
\end{rem}

\smallskip

{
\begin{rem}
The first  slow rate bound in Theorem~\ref{L1.rate.thm} can be potentially improved \citep[see, for example, the discussion in][]{belloni2014pivotal} by  replacing the approximation $\sqrt{2\log (2p)}+\sqrt{2\log (1/\delta_0)}$, used in the definition of~$\lambda$, directly with the $(1-\delta_0)$ quantile of $\|\bX^\top\bepsilon/\sigma\|_{\infty}$. As before, $\sigma$ can be estimated by employing a preliminary regression estimator, unrestricted in terms of the model size.
In practice, $\lambda$ can be tuned based on a separate validation set or by cross-validation.
\end{rem}
}

The next result uses Theorems~\ref{probs.thm} and~\ref{L1.rate.thm} to bound the expected prediction error for~$\widehat\bbeta_1$.

\smallskip

\begin{cor}
\label{L1.Expect.slow.rate}
If $\lambda\ge[8+2\sqrt{2}]\sigma \sqrt{\log (2p)}$, then
\begin{equation*}
\mathbb{E}\|\bff^*-\bX\widehat\bbeta_1\|^2\lesssim \inf_{\bbeta\in B_0(k)}\Big[\|\bff^*-\bX\bbeta\|^2 + \lambda\|\bbeta\|_1\Big]+\sigma^2.
\end{equation*}
Furthermore, for every $\lambda\ge0$,
\begin{equation*}
\mathbb{E}\|\bff^*-\bX\widehat\bbeta_1\|^2\lesssim \inf_{\bbeta\in B_0(k)}\|\bff^*-\bX\bbeta\|^2+\sigma^2k\log (ep/k) + \gamma_{2k}^{-2}\lambda^2k.
\end{equation*}
\end{cor}

We now compare the slow rate prediction error bounds for the two proposed estimators:~$\widehat\bbeta_1$ and~$\widehat\bbeta_2$.  In the case where all the non-zero coefficients of~$\bbeta^*$ are of the same order of magnitude, the prediction error rate for~$\widehat\bbeta_2$ is superior to the one for~$\widehat\bbeta_1$, because the former replaces the $\log(ep)$ term with~$\log(ep/k^*)$.  Alternatively, the slow rate for~$\widehat\bbeta_1$ is better when the ratio $\|\bbeta^*\|_1/\|\bbeta^*\|$ is sufficiently small.  The following result formalizes the last observation in the asymptotic setting.

\smallskip

\begin{cor}
\label{cor.comp.L1.L2}
Denote the slow prediction error rates for~$\widehat\bbeta_1$ and~$\widehat\bbeta_2$ by $SR_1$ and $SR_2$, respectively. Suppose that $k=k^*$, $\bff^*=\bX\bbeta^*$ and
\begin{equation*}
{\|\bbeta^*\|_1}/\big(\sqrt{k^*}\|\bbeta^*\|\big)=o\big(\sqrt{\log(p/k^*)/\log(p)}\big)
\end{equation*}
as~$p\rightarrow\infty$.  Then, $SR_1/SR_2\rightarrow0$.
\end{cor}

In the next section we complement the slow rate prediction error bounds for~$\widehat\bbeta_2$ and~$\widehat\bbeta_1$ with a corresponding lower-bound for~$\widehat\bbeta_{\ell_0}$.

\subsection{Lower bounds for the best-subsets estimator}
\label{sec.lower.bnds}

Focusing on the linear setting and comparing the slow rate prediction error bound for~$\widehat\bbeta_2$ in Corollary~\ref{lin.mod.l2} with the one provided for~$\widehat\bbeta_{\ell_0}$ by the fast rate part of the same result, we note that the former bound is superior when $\|\bbeta^*\|/\sigma\lesssim\sqrt{k\log(ep/k)}$ with a sufficiently small constant.  The following novel result demonstrates that in this regime of low $\|\bbeta^*\|/\sigma$ the above comparison is meaningful, because the error bound for~$\widehat\bbeta_{\ell_0}$ is tight.

\smallskip

\begin{thm}
\label{best.sub.lower.bnd}
Suppose that $k\in[p]$ and $\|\bbeta^*\|/\sigma\lesssim\gamma_k\sqrt{k\log(ep/k)}$ with a sufficiently small universal constant.  Then, there exists a positive universal constant~$c$, such that
\begin{equation*}
\|\bX\hat\bbeta_{\ell_0}-\bX\bbeta^*\|^2\gtrsim\sigma^2\gamma^2_k k{\log (ep/k)}
\end{equation*}
with (high) probability of at least $1-2({ep}/k)^{-c\gamma_{k}^2 k}-({ep}/k)^{-k}$.
\end{thm}

Suppose that $\gamma_k$ is bounded away from zero by a positive universal constant.  Note that this holds under the sparse eigenvalue condition, which is standard in the literature (see the discussion in Section~8 of~\cite{bellec2018slope}, for example).  In particular, this condition holds with high probability for a wide class of random matrices~$\bX$ with i.i.d.\;rows, provided $k\log(ep/k)\lesssim n$ with an appropriate universal constant \citep{lecue2017sparse}.  Under this setting, we make the following key observations.

\smallskip

\begin{rem}
Combining the upper-bound for~$\widehat\bbeta_{\ell_0}$ from Corollary~\ref{lin.mod.l2} with the lower-bound from Theorem~\ref{best.sub.lower.bnd} yields $\|\bX\hat\bbeta_{\ell_0}-\bX\bbeta^*\|^2\asymp\sigma^2k{\log (ep/k)}$.  Comparing this prediction error to the slow rate prediction error bound for~$\widehat\bbeta_2$, we conclude that
\begin{equation}
\label{PE.ratio}
\|\bX\bbeta^*-\bX\widehat\bbeta_{\ell_0}\|^2/\|\bX\bbeta^*-\bX\widehat\bbeta_2\|^2\gtrsim \big(\sigma/\|\bbeta^*\|\big)\sqrt{k\log (ep/k)}
\end{equation}
with high probability.
\end{rem}

\smallskip

\begin{rem}
In the regime of interest, where $\|\bbeta^*\|/\sigma\lesssim\sqrt{k\log(ep/k)}$, the ratio of prediction errors in~(\ref{PE.ratio}) can be made arbitrarily large by decreasing~$\|\bbeta^*\|/\sigma$ or increasing~$k$.  Similarly, Theorem~\ref{L1.rate.thm} implies that the prediction error for~$\widehat\bbeta_1$ is smaller than the one for~$\widehat\bbeta_{\ell_0}$ in the regime of low $\|\bbeta^*\|_1/\sigma$.  These observations are supported empirically, as illustrated by the left column in Figure~\ref{fig-one}, where the predictive performance of~$\widehat\bbeta_{\ell_0}$ steadily deteriorates relative to that of~$\widehat\bbeta_{2}$ and~$\widehat\bbeta_{1}$  as~$k$ increases.
\end{rem}

The lower-bound in Theorem~\ref{best.sub.lower.bnd}, together with the companion upper-bound implied by Corollary~\ref{lin.mod.l2}, suggests that in the setting where~$\|\bbeta^*\|/\sigma$ is low, the prediction error for~$\widehat\bbeta_{\ell_0}$ could be reduced by decreasing~$k$ below~$k^*$.  Thus, decreasing the model size parameter~$k$ may have a regularizing effect on the best-subsets estimator.  However, if we tune~$k$ in order to improve the predictive performance, then we lose the attractive feature of subset selection that allows the user to select the model size based on external considerations.  In contrast, estimator~$\widehat\bbeta_2$ is regularized via the tuning parameter~$\lambda$, for each given model size~$k$.  Moreover, the next example illustrates that, even with optimal data-dependent choice of~$k$, best subset selection does not achieve the $\sigma \sqrt{k^*\log (p/k^*)}\|\bbeta^*\|$ prediction error rate available for~$\widehat\bbeta_2$.

\textbf{Example.} Suppose that all pairwise correlations among the predictors are equal to a fixed universal constant $\rho\in(0,1)$.  Recall the notation $k^*=\|\bbeta^*\|_0$, let $k^*>0$ and assume that each nonzero element of~$\bbeta^*$ is equal to $b\sigma\sqrt{\log(ep)}/{k^*}$ for some positive~$b$.

\smallskip

\begin{prop}
\label{prop1}
Let~$\delta\in(0,1]$ be a fixed universal constant. Under the setting of the Example, there exist positive universal constants~$b_0$ and $a$, such that if $b\in[\delta b_0,b_0]$, then
\begin{equation*}
\min_{k \in \{0, 1, \ldots, p\} }\|\bX\bbeta^*-\bX\widehat\bbeta_{\ell_0}\|^2\gtrsim \sigma \sqrt{k^*\log(ep)}\|\bbeta^*\|
\end{equation*}
with probability at least $1-2(ep)^{-a}$. Moreover, the result holds uniformly over~$\bbeta^*$.
\end{prop}
We note that, under the setting of the Example and up to universal multiplicative constants, the above lower-bound matches the minimax rate on the intersection of $\ell_0$ and $\ell_1$ balls \citep[Section 5.2]{rigollet2011exponential}.  Comparing this lower-bound with the $\sigma \sqrt{k^*\log (ep/k^*)}\|\bbeta^*\|$ upper-bound in the slow rate part of Corollary~\ref{lin.mod.l2},
we conclude that in general the best-subsets estimator is not able to achieve the slow rate of $\ell_2$-regularized best-subsets estimator.
We emphasize that the lower-bound in Proposition~\ref{prop1} holds with high probability, and is uniform over~$k$ and~$\bbeta^*$.  In particular, even if best-subsets were able to choose an optimal~$k$ for each given sample, the prediction error rate for the resulting ``oracle'' estimator would still be worse than the one for~$\widehat\bbeta_2$.

The next result shows that for larger~$k$ the difference between the prediction errors for~$\widehat\bbeta_{\ell_0}$ and~$\widehat\bbeta_2$ is substantially greater than the one suggested by the uniform lower-bound in Proposition~\ref{prop1}.

\smallskip

\begin{prop}
\label{prop2}
Suppose that $k\in[p]$. Under the setting of the Example, there exist positive universal constants~$b_0$, $k_0$ and $a$, such that if either $b\le b_0$ or~$\max\{k^*,k\}\ge k_0$, then
\begin{equation*}
\|\bX\hat\bbeta_{\ell_0}-\bX\bbeta^*\|^2\gtrsim\sigma^2 k{\log (ep/k)}
\end{equation*}
with probability at least $1-3(ep/k)^{-a k}$.
\end{prop}

We now compare the prediction errors for~$\widehat\bbeta_{\ell_0}$ and~$\widehat\bbeta_2$ in the concrete case where $k=k^*$.  Proposition~\ref{prop2} and the slow rate part of Corollary~\ref{lin.mod.l2} imply that
\begin{equation*}
{\|\bX\bbeta^*-\bX\widehat\bbeta_{\ell_0}\|^2}/{\|\bX\bbeta^*-\bX\widehat\bbeta_2\|^2}\gtrsim  k^*[\log(ep/k^*)/\log(ep)]^{1/2}
\end{equation*}
with high probability.  In particular, if we let $k^*=O(p^{1-c})$ for some positive~$c$, then the lower-bound in the above display grows linearly in~$k^*$.

\subsection{Data-driven choice of~$k$}
\label{sec.choice.of.k}

In this section we study a BIC-type approach for selecting the model size~$k$. We define
\begin{eqnarray*}
\widehat\bbeta_2^{\text{B}} &=& \argmin_{\bbeta} \|\bY-\bX\bbeta\|^2+\lambda_{\bbeta}\|\bbeta\|  +\mu_{\bbeta}\|\bbeta\|_0\\
\widehat\bbeta_1^{\text{B}} &=& \argmin_{\bbeta} \|\bY-\bX\bbeta\|^2+\lambda\|\bbeta\|_1  +\mu_{\bbeta}\|\bbeta\|_0,
\end{eqnarray*}
where $\lambda_{\bbeta}=a\sqrt{\|\bbeta\|_0\log (ep/\|\bbeta\|_0)}$ and $\mu_{\bbeta}=b \log ({ep}/{\|\bbeta\|_0})$ for some nonnegative~$a$ and~$b$.
The above optimization problems are equivalent to first solving the corresponding constrained problems~(\ref{opt.problem.theory}), for each~$k$, and then identifying the optimal model size~$k$ via BIC-type penalization. The value of~$\lambda$ in the corresponding constrained formulation for~$\widehat\bbeta_2^{\text{B}}$ is  $a\sqrt{k\log (ep/k)}$.

The following result establishes general oracle inequalities for~$\widehat\bbeta_2^{\rm{B}}$ and~$\widehat\bbeta_1^{\rm{B}}$.  To simplify the presentation, we focus on the expected prediction error.

\smallskip

\begin{thm}
\label{pen2.thm}
There exist universal constants~$a_0$, $b_0$ and $c_0$, such that if $a\ge a_0\sigma$ or $b\ge b_0\sigma^2$, then
\begin{equation*}
\mathbb{E}\|\bff^*-\bX\widehat\bbeta_2^{\rm{B}}\|^2\lesssim \inf_{\bbeta\in \mathbb{R}^p}\Big[\|\bff^*-\bX\bbeta\|^2 + \lambda_{\bbeta}\|\bbeta\|  +\mu_{\bbeta}\|\bbeta\|_0\Big]+\sigma^2;
\end{equation*}
and if $\lambda\ge c_0\sigma\sqrt{\log(ep)}$ or $b\ge b_0\sigma^2$, then
\begin{equation*}
\mathbb{E}\|\bff^*-\bX\widehat\bbeta_1^{\rm{B}}\|^2\lesssim \inf_{\bbeta\in \mathbb{R}^p}\Big[\|\bff^*-\bX\bbeta\|^2 + \lambda\|\bbeta\|_1  +\mu_{\bbeta}\|\bbeta\|_0\Big]+\sigma^2.
\end{equation*}
\end{thm}
{ The next result, which focuses on the linear case for concreteness, shows that the new estimators achieve the error rates in Corollaries~\ref{Expect.slow.rate} and~\ref{L1.Expect.slow.rate} while producing model sizes of the same order as the true model size~$k^*$.

\smallskip

\begin{cor}
\label{data.driven.cor}
Let $\bff^*=\bX\bbeta^*$ and consider the universal constants that appear in the statement of Theorem~\ref{pen2.thm}. If $a_0\sigma\le a\lesssim\sigma$ and $b \asymp \big(\lambda_{\bbeta^*}\|\bbeta^*\|+\sigma^2\big)/\big\{[k^*\vee1]\log (ep/[k^*\vee1])\big\}$, then
\begin{equation*}
\mathbb{E}\|\bff^*-\bX\widehat\bbeta_2^{\rm{B}}\|^2   \lesssim \sigma\sqrt{k^*\log (ep/k^*)}\|\bbeta^*\|+\sigma^2 \quad \text{and} \quad
\mathbb{E}\|\widehat\bbeta_2^{\rm{B}}\|_0\lesssim k^*\vee1.
\end{equation*}
If $a \lesssim {\sigma^2\sqrt{k^*\log (ep/k^*)}}/{\|\bbeta^*\|}$ and $b_0\sigma^2\le b\lesssim\sigma^2$, then
\begin{equation*}
\mathbb{E}\|\bff^*-\bX\widehat\bbeta_2^{\rm{B}}\|^2   \lesssim \sigma^2k^*\log (ep/k^*)+\sigma^2 \quad \text{and} \quad
\mathbb{E}\|\widehat\bbeta_2^{\rm{B}}\|_0\lesssim k^*\vee1.
\end{equation*}
If $c_0\sigma\sqrt{\log(ep)}\le\lambda\lesssim\sigma\sqrt{\log(ep)}$ and $b \asymp \big(\lambda\|\bbeta^*\|_1+\sigma^2\big)/\big\{[k^*\vee1]\log (ep/[k^*\vee1])\big\}$, then
\begin{equation}
\label{date.dr.slow1}
\mathbb{E}\|\bff^*-\bX\widehat\bbeta_1^{\rm{B}}\|^2   \lesssim\sigma\sqrt{\log(ep)}\|\bbeta^*\|_1+\sigma^2
\quad \text{and} \quad
\mathbb{E}\|\widehat\bbeta_1^{\rm{B}}\|_0\lesssim k^*\vee1.
\end{equation}
If $\lambda \lesssim {\sigma^2 k^*\log (ep/k^*)}/{\|\bbeta^*\|_1}$ and $b_0\sigma^2\le b\lesssim\sigma^2$, then
\begin{equation}
\label{date.dr.fast1}
\mathbb{E}\|\bff^*-\bX\widehat\bbeta_1^{\rm{B}}\|^2   \lesssim\sigma^2k^*\log (ep/k^*)+\sigma^2
\quad \text{and} \quad
\mathbb{E}\|\widehat\bbeta_1^{\rm{B}}\|_0\lesssim k^*\vee1.
\end{equation}
\end{cor}
}

It is useful to compare~$\widehat\bbeta_1^{\rm{B}}$ to the related Lasso estimator. We first note that the slow rate in~(\ref{date.dr.slow1}) also holds for the Lasso, as a consequence of the second bound in Theorem~\ref{pen2.thm} when~$b=0$.  The fast rate in~(\ref{date.dr.fast1}) holds for the Lasso as well~\citep[Corollary~4.4]{bellec2018slope}, however, under a ``strong restricted eigen value condition''.  In contrast, all the error bounds in this section hold without imposing any assumptions on the design beyond the usual normalization of the columns of~$\bX$.  This can be viewed as a non-trivial advantage of $\ell_0$-based approaches over Lasso-type methods: \cite{zhang2017optimal} gives examples of design matrixes for which the Lasso\footnote{The lower-bound holds for a wide range of coordinate-separable M-estimators, including popular nonconvex regularizers such as SCAD and MCP.} prediction error is lower-bounded by a constant multiple of $\sqrt{n}$, which is generally much larger than the fast rate error bound in~(\ref{date.dr.fast1}). Similarly, the sparsity bounds for the Lasso estimator \citep{belloni2013least} require sparse eigen value conditions, while the corresponding bounds the proposed approach hold without any additional assumptions on the design.

\section{Related work and connections to existing estimators}\label{sec:broader}
The literature on penalized estimation in high-dimensional regression is extensive. Here we discuss a subset of this work that is closely related to the topic of our paper.

When $q=2$, estimator~\eqref{best-subset-penalized} is related\footnote{A convex relaxation of~\eqref{mio-form-1} with $q=2$, obtained by relaxing $z_{j} \in \{0, 1\}$ to $z_{j} \in [0,1]$, leads to a slight modification of the elastic net optimization problem, where the squared-$\ell_{2}$-penalty is replaced by the $\ell_{2}$-penalty.} to the elastic net estimator~\cite{ZH2005}.  Similarly, when $q=1$, a relaxation of~\eqref{mio-form-1} leads to the Lasso problem. However, as we demonstrate in Section~\ref{sec:expts}, the operating characteristics of estimator~\eqref{best-subset-penalized} are quite different from these relaxations.

Estimator~\eqref{best-subset-penalized} bears similarities with the nonconvex approaches in~\cite{FF93,huang2016mnet,zou2009adaptive,liu2007variable,fan2013asymptotic}, however, the particular form of~\eqref{best-subset-penalized} is not considered in these works.
Despite apparent similarities, our work is different in terms of motivation, context and computational methods. More specifically, our primary motivation is to \textit{regularize} the overfitting behavior of best subsets selection and obtain sparse models with good predictive power. From a computational standpoint, our MIO framework delivers a \emph{global} solution for the corresponding optimization problem.

\cite{zou2009adaptive,huang2016mnet} propose improvements over the elastic net by replacing the $\ell_{1}$-penalty with more aggressive penalties (for example, adaptive Lasso and MCP).
They consider the
penalized formulation, different from the cardinality constrained version~\eqref{best-subset-penalized}. While these works focus on improved estimation accuracy in low-noise regimes, the resulting estimators may also perform well in the high-noise settings.
\cite{fan2013asymptotic} impose both a concave penalty and the $\ell_1$-penalty on~$\B\beta$, demonstrating theoretically that their estimator combines the predictive strength of the $\ell_1$ regularization with the variable selection strength of the nonconvex regularization.
\cite{liu2007variable} impose a convex combination of the $\ell_{0}$ and the $\ell_{1}$ penalties on~$\B\beta$, and study statistical properties of their estimator in the low-dimensional setting.
There are differences in the computational approaches as well:
\cite{liu2007variable} propose using a piecewise linear approximation to the $\ell_0$-penalty for computational purposes; their numerical experiments are mostly limited to the case $p \leq 15$.
\cite{huang2016mnet} and \cite{fan2013asymptotic} rely on local approximations to
nonconvex optimization problems, which may potentially lead to sub-optimal local solutions.

{
Our approach has interesting connections with Bayesian procedures that use sparsity-inducing prior distributions for the regression coefficients -- for example, the spike-and-slab priors~\cite{mitchell1988bayesian,polson2019bayesian,soussen2011bernoulli}. In the Bernoulli-Gaussian mixture model~\cite{soussen2011bernoulli}, each coefficient follows a mixture distribution involving a point mass at zero and a zero mean Gaussian distribution:  $\beta_{j}|\theta, \sigma_{\beta} \sim (1-\theta) \delta_0 + \theta N(0, \sigma_{\beta}^2)$. One may represent~$\beta_j$  as a product of two independent random variables: $\beta_{j}=\gamma_{j}\alpha_{j}$, where
$\gamma_{j}|\theta \sim \text{Bernoulli}(\theta)$,
$\alpha_{j}|\sigma_{\beta} \sim N(0, \sigma^2_{\beta})$. The corresponding MAP estimator
then minimizes
\begin{equation*}
\| \M{y} - \M{X}\B\beta\|^2 + \lambda_{1} \| \B\alpha \|^2 + \lambda_{2} \| \B{\gamma}\|_0
\end{equation*}
with respect to variables $(\B\beta, \B\gamma, \B\alpha)$, for a suitable choice of parameters $\lambda_{1}, \lambda_{2}$. The above problem is an $\ell_0$-penalized version of Problem~\eqref{subset-lq} with $q=2$, in which the squared $\ell_{2}$-penalty replaces the $\ell_{2}$-penalty.
Such problems are known to pose computational challenges in large-scale settings. \cite{polson2019bayesian} study a special case of this problem with $\lambda_{1} \approx 0$, which corresponds to the high-SNR regime, and consider a number of approximate  algorithms (for example, proximal gradient~\cite{bertsimas2015best} and single-best-replacement~\cite{soussen2011bernoulli}) for the Lagrangian version of Problem~\eqref{subset-1}.
Another possibility is to use the Bernoulli-Laplace prior for the regression coefficients~\cite{amini2012analog,polson2019bayesian} -- the corresponding MAP formulation leads to an $\ell_{0}$-penalized form of Problem~\eqref{subset-lq} with $q=1$.
Our proposed algorithms may potentially be used to obtain (near-optimal or optimal) solutions for both of these problems.

Another popular approach is to employ continuous spike-and-slab priors, such as a mixture of two Laplace distributions. When $q=1$, the penalized modification of estimator~\eqref{best-subset-penalized} corresponds to the limiting case in which the spike distribution is a point mass. Importantly, our estimator~\eqref{best-subset-penalized} is constrained rather than penalized, providing a direct control over the sparsity level.  When the mixture weight in the aforementioned Laplace mixture follows its own prior distribution, the resulting approach is the powerful spike-and-slab Lasso procedure of \cite{rockova2018spike}. Some other state-of-the-art Bayesian shrinkage methods include the horseshoe regression \citep{carvalho2010horseshoe} and the empirical Bayes method of \cite{martin2017empirical}. These methods are known to improve on the predictive performance of the global shrinkage approaches such as ridge regression \citep{bhadra2019prediction,martin2020empirical}. In particular, \cite{martin2020empirical} propose a Monte-Carlo scheme to approximate the predictive density, allowing for uncertainty quantification.
From an algorithmic standpoint, the main difference between our approach and the related Bayesian methods for computing MAP estimators is our use of mixed integer programming.  Furthermore, our theoretical analysis focuses on the low-SNR regime. To the best of our knowledge, the earlier works discussed above do not consider the low-SNR regime in their theoretical development.
}

The topic of this paper is closely related to the interesting recent work of~\cite{hastie2020best}, where the authors also observe that in the low-SNR regimes the Lasso leads to better predictive models than best subset selection, while the reverse is true in the high-SNR regimes. As a compromise between the two approaches, \cite{hastie2020best} propose a variant{\footnote{This is given by  a convex combination of the Lasso estimator and its polished version (obtained by performing a least squares fit on the Lasso support).}}  of relaxed Lasso~\cite{meinshausen2007relaxed}. Interestingly, the original form of the relaxed Lasso estimator can be interpreted as a feasible solution to Problem~\eqref{best-subset-penalized}, with $q=1$, for a suitable choice of tuning parameters~$k$ and~$\lambda$.
The key advantages of our approach are as follows.  Unlike relaxed Lasso, estimator~\eqref{best-subset-penalized} is given by a transparent optimization formulation with an explicit control on the support size.  We conduct an extensive theoretical analysis of the predictive properties of estimator~\eqref{best-subset-penalized}, including its superior performance relative to best-subsets in high-noise regimes.  To our knowledge, similar results are not available for the relaxed Lasso estimator.

After an earlier version of this paper became publicly available, some interesting follow-up work has been conducted with the focus on the computational aspects of the regularized best-subset estimators \citep[for example,][]{hazimeh2018fast,atamturk2019rank,hazimeh2020sparse}.

\input{results1.tex}

\section*{Acknowledgements}
We thank the anonymous referees for their constructive comments that helped us improve the paper.

\renewcommand{\baselinestretch}{1.25}

\bibliographystyle{plainnat_my}

{{\small{\bibliography{Peterbib,rahul_dbm3}}}}

\newpage

\renewcommand{\baselinestretch}{1.5}

\begin{appendix}

\input{Supplement_July2021.tex}

\end{appendix}

\end{document}

%% file: results1.tex
\section{Experiments}\label{sec:expts}
We explore the properties of our estimator empirically on synthetic datasets with varying values of~$n$, $p$, SNR and correlations among the predictors, as well as on several real datasets. {An implementation of the algorithms we propose in this paper is available on github\footnote{Link to repository: \url{https://github.com/antoine-dedieu/subset_selection_with_shrinkage}}}.



\subsection{Synthetic Datasets}

We generate the rows of the model matrix $\M{X}$ as~$n$ independent realizations from a $p$-dimensional multivariate Gaussian distribution with mean zero and covariance matrix $\B{\Sigma} = (\sigma_{jk})$. We standardize the columns of $\mathbf{X}$ to have zero mean and unit $\ell_2$-norm, and generate $\mathbf{y} = \mathbf{X}\B{\beta}^* +\B{\epsilon}$  with ${\epsilon}_{i} \stackrel{\text{iid}}{\sim} N(0,\sigma^2)$ and $\B{\beta}^* \in \mathbb{R}^p$.   Recall that we define
$\text{SNR} = \|\mathbf{X} \B{\beta}^{*}\|_2^2 / \| \B\epsilon\|_{2}^2$ and let $k^* = \|\B{\beta}^{*}\|_0$ denote the true number of nonzeros.
We consider the following examples:


\noindent \textbf{Example 1.}  $\sigma_{jk} =\rho^{ |j-k| }$ (with the convention $0^0=1$),
$\beta^*_j = 1$ for $k^*=7$ equispaced values in~$[p]$ and $\beta^*_j = 0$ otherwise.

\noindent \textbf{Example 2.} $\sigma_{jk} = \rho+(1-\rho)I\{j=k\}$, $\beta^*_j = 1$ for $j \leq k^*=7$ and $\beta^*_j = 0$ otherwise.

In the above examples all the nonzero coefficients in $\B\beta^*$ have the same magnitude.  We focus on this setting to get a clear understanding of how our proposed estimator regulates the overfitting behavior of best-subsets and compares with estimators such as ridge regression and the Lasso, as the SNR is varied. In our simulations, we also vary the values of~$\rho,n$ and~$p$.

We conduct a comparison across the following methods:

\begin{description}
\item[(L1+L0)] Estimator~\eqref{best-subset-penalized} with $q=1$. The 2D grid of tuning parameters has~$\lambda$ taking values in a geometrically spaced
sequence $\left\{\lambda_{i}\right\}_1^{100}$, with $\lambda_1 = \|\mathbf{X}^\top\mathbf{y} \|_{\infty}$ and  $\lambda_{100} \sim 10^{-4} \lambda_1$, while~$k$ takes values in $\{0,\ldots, 15\}$.
\item[(L2+L0)] Estimator~\eqref{best-subset-penalized} with $q=2$. The 2D grid was similar to the above, with $\lambda_1=\|\mathbf{X}^\top\mathbf{y} \|_{2}$, which ensures a zero solution.
\item[(L0)] Best-subsets estimator~\eqref{subset-1} with $k \in \{0, \ldots, 15 \}$.
\item[(L1)] The Lasso estimator given by Problem~\eqref{subset-lq} with $q=1$ on a grid of 100 values of~$\lambda$.
\item[(L1P)] Polished version of the Lasso estimator, computed as the least-squares estimator on the support of every L1 solution.
\item[(L2)] Ridge regression estimator given by Problem~\eqref{subset-lq} with $q=2$ on a grid of 100 values of~$\lambda$.
\item[(L1+L2)] Elastic net estimator \citep{ZH2005}. For each value of parameter $\lambda$, we consider a sequence of~20 values $\alpha \in  [0.05, 0.95]$ for weighting the~$\ell_1$ and~$\ell^2_2$ penalties.
\end{description}

The estimators in~\eqref{best-subset-penalized} are computed via 3 rounds of Algorithm~1 (Neighborhood Continuation) with stochastic local search, as described in Section~\ref{nbhd-conti-1}.
Let $\{\hat{\B\beta}(\lambda, k)\}$ denote the corresponding 2-dimensional family of solutions.
The discrete first order algorithm (DFO) is run until reaching the convergence threshold of $\tau=10^{-3}$ or a maximum of 1000 iterations, whichever is earlier. 
Once the family $\{\hat{\B\beta}(\lambda, k)\}$ is obtained, the best pair $(\hat{\lambda}, \hat{k})$ is chosen on a held-out validation set as discussed below.  For this choice of $(\hat{\lambda}, \hat{k})$, we solve the MIO formulation~\eqref{mio-form-1} with a time-limit of 30 minutes\footnote{We use a Python interface to the Gurobi solver for our experiments.} -- the resultant solutions are referred to
as L1+L0 or L2+L0.
We obtain the L0  solution in a similar fashion, using $\hat{\B\beta}(\lambda_{N}, k)$ from Problem~\eqref{best-subset-penalized} with $q=1$
to warm-start the DFO.
Methods L1, L1P, L2 and L1+L2 are computed using Python's~{\texttt{scikit-learn}} suite of algorithms.


\begin{figure}[bp!]
\renewcommand{\baselinestretch}{1.25}
	\centering
	\begin{tabular}{l c c c}
			\multicolumn{4}{c} { \sf \small{Example~1: Small settings: $n=50, p=100$} }\\
		&\sf {\small{$\rho=0.5, \text{SNR}=1$}} &  \sf {\small{$\rho=0.2, \text{SNR}=2$}} & \sf {\small{$\rho=0.5, \text{SNR}=3$}}\\
		\rotatebox{90}{\sf {\small{~~~~~~~~~Prediction Error}}}&
		\includegraphics[width=0.3\textwidth,height=0.18\textheight,  trim =1.8cm 1.cm 2cm 2cm, clip = true ]{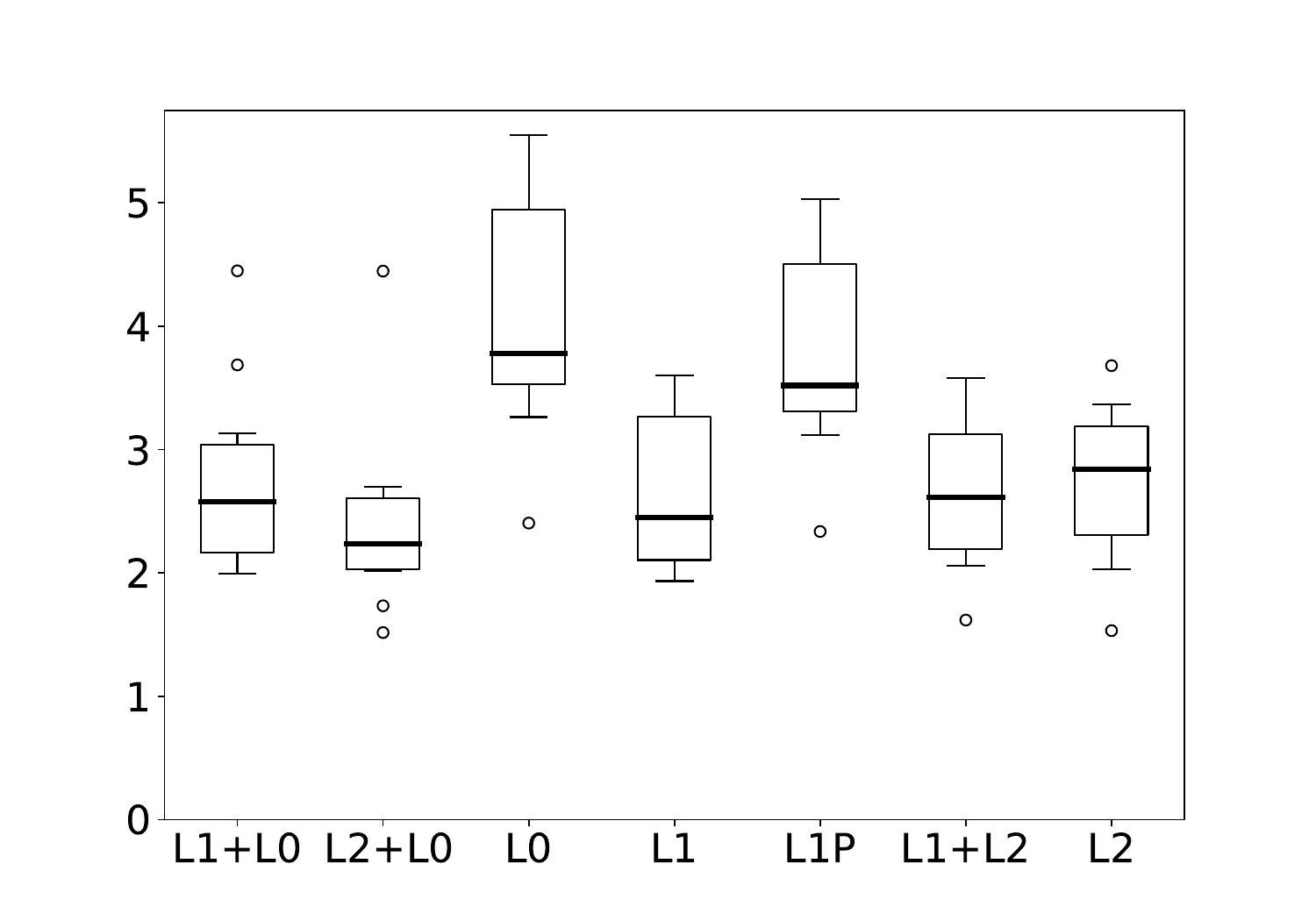}&
		\includegraphics[width=0.3\textwidth,height=0.18\textheight,  trim =1.2cm 1cm 2cm 2cm, clip = true ]{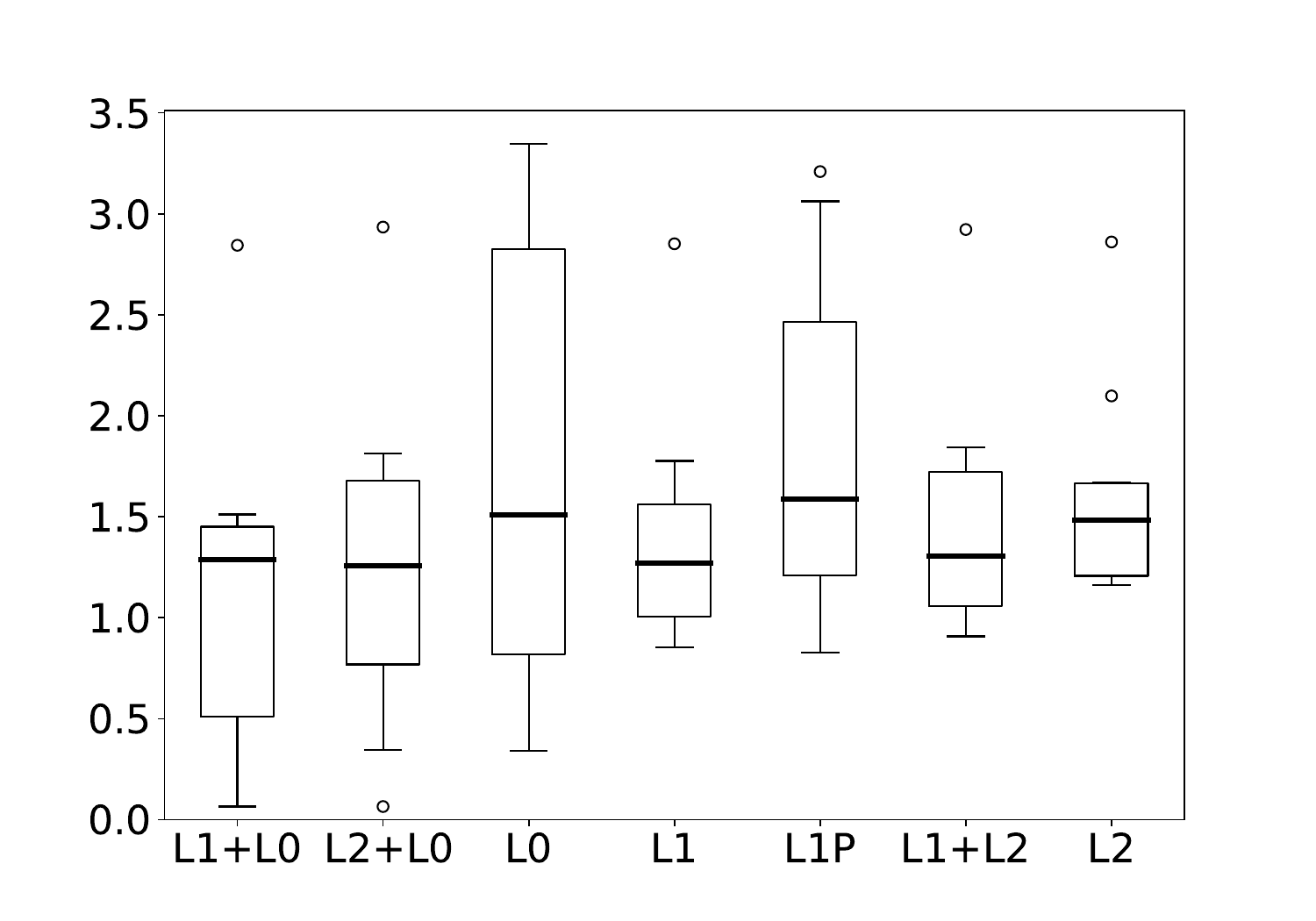}&
		\includegraphics[width=0.3\textwidth,height=0.18\textheight,  trim =1.2cm 1cm 2cm 2cm, clip = true ]{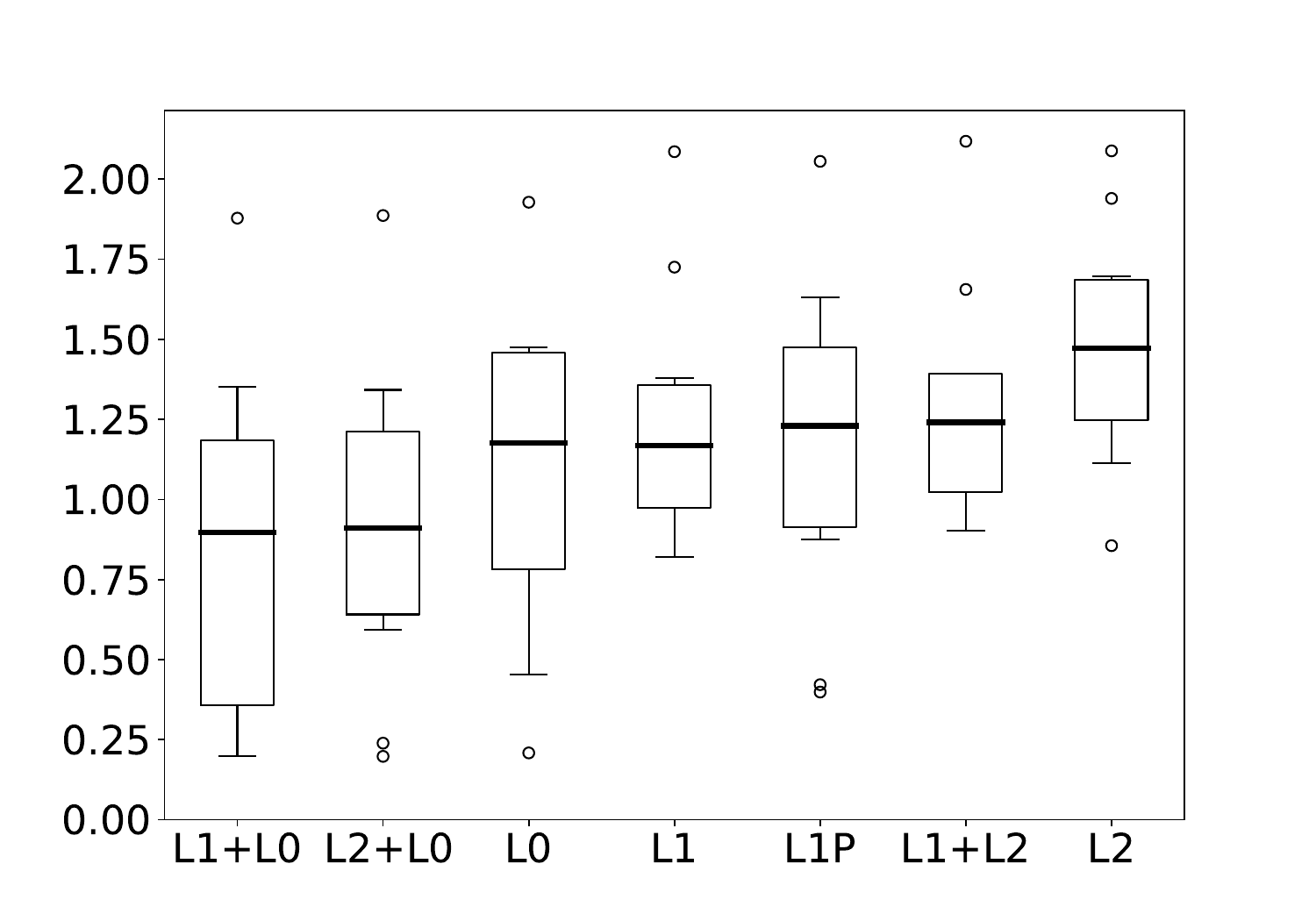}\\
		
		\rotatebox{90}{\sf {\small{~~~~~~~\# nonzeros}}}&
		\includegraphics[width=0.3\textwidth,height=0.18\textheight,  trim =1.8cm 1cm 2cm 1cm, clip = true ]{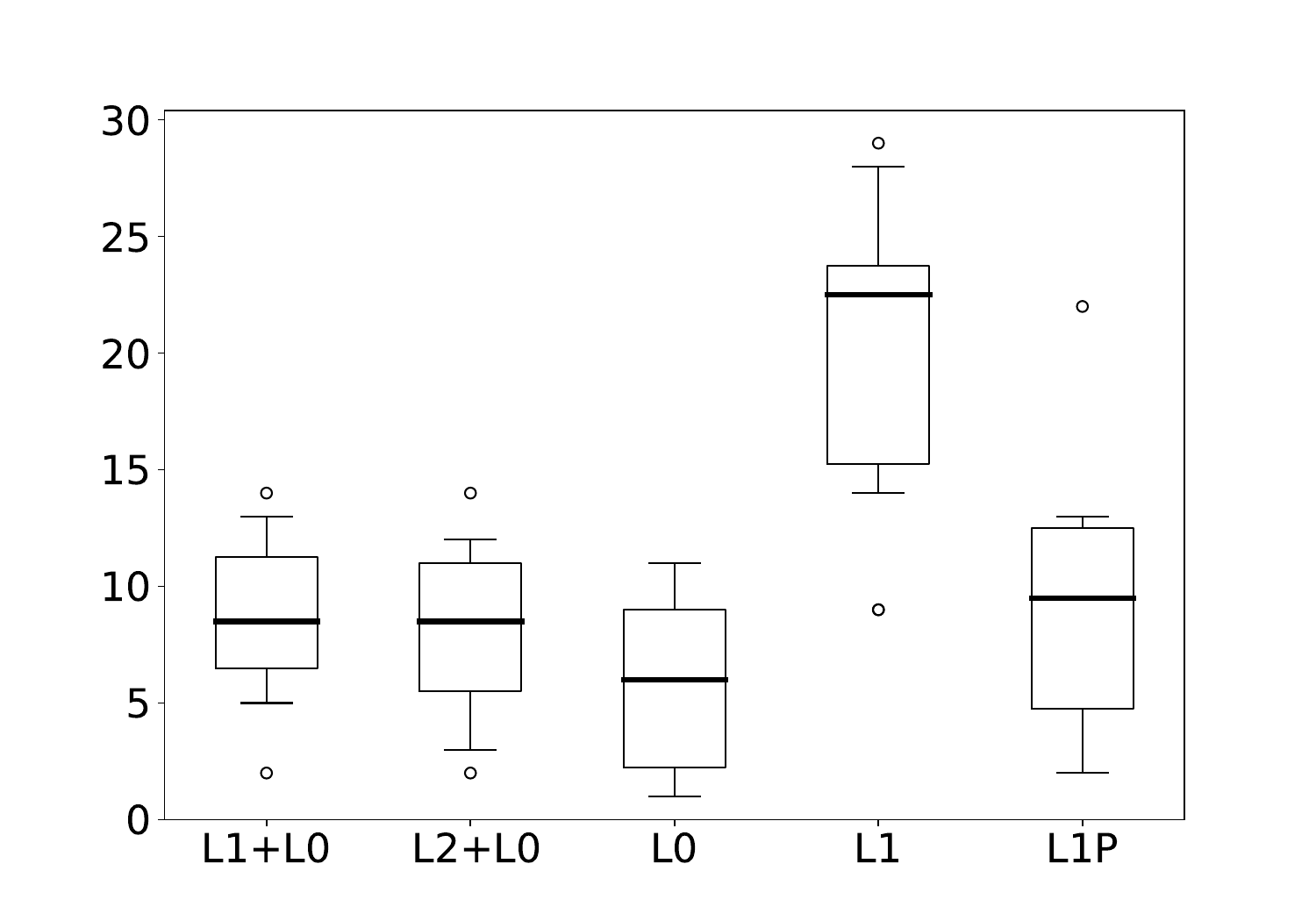}&
		\includegraphics[width=0.3\textwidth,height=0.18\textheight,  trim =1.2cm 1cm 2cm 1cm, clip = true ]{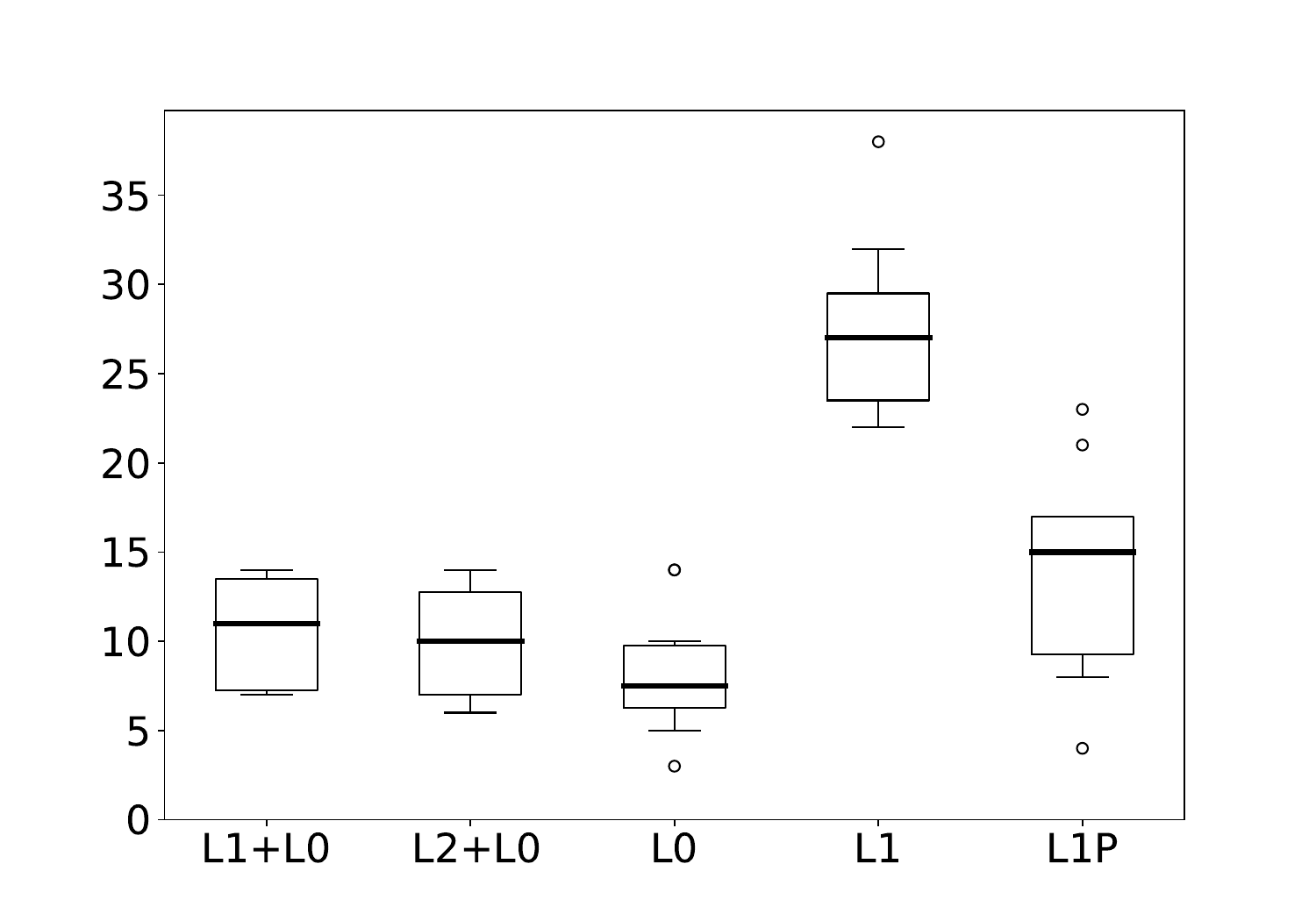}&
		\includegraphics[width=0.3\textwidth,height=0.18\textheight,  trim =1.2cm 1cm 2cm 1cm, clip = true ]{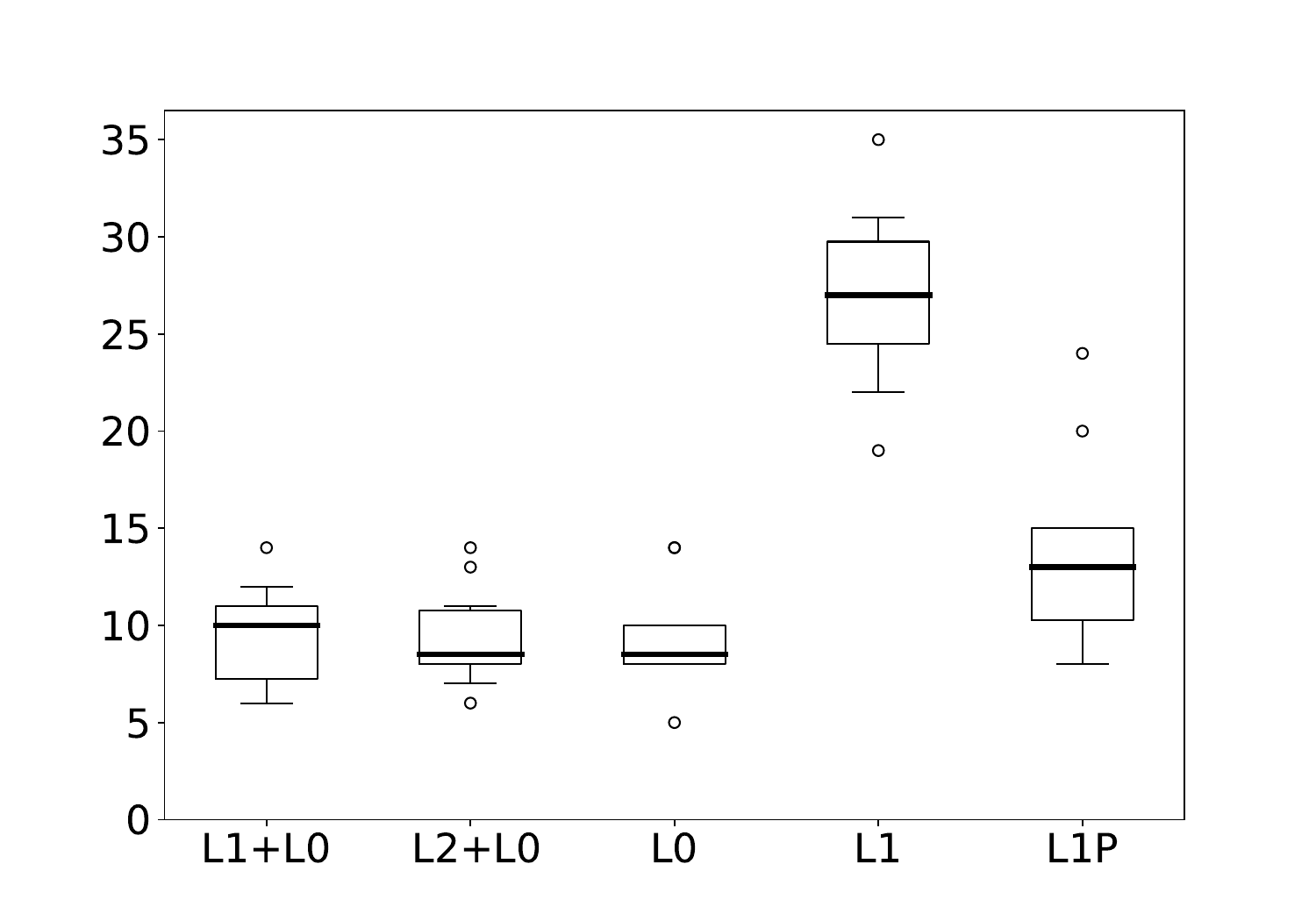}\\
		 \vspace{-18pt}\\

		\multicolumn{4}{c} { \sf {Example~1: Large settings: $n=100, p=1000$} }\\
		&\sf {\small{$\rho=0.2, \text{SNR}=1$}} &  \sf {\small{$\rho=0.2, \text{SNR}=2$}} & \sf {\small{$\rho=0.8, \text{SNR}=3$}}\\
		\rotatebox{90}{\sf {\small{~~~~~~~~~Prediction Error}}} &
		\includegraphics[width=0.3\textwidth,height=0.18\textheight,  trim =1.8cm 1cm 2cm 2cm, clip = true ]{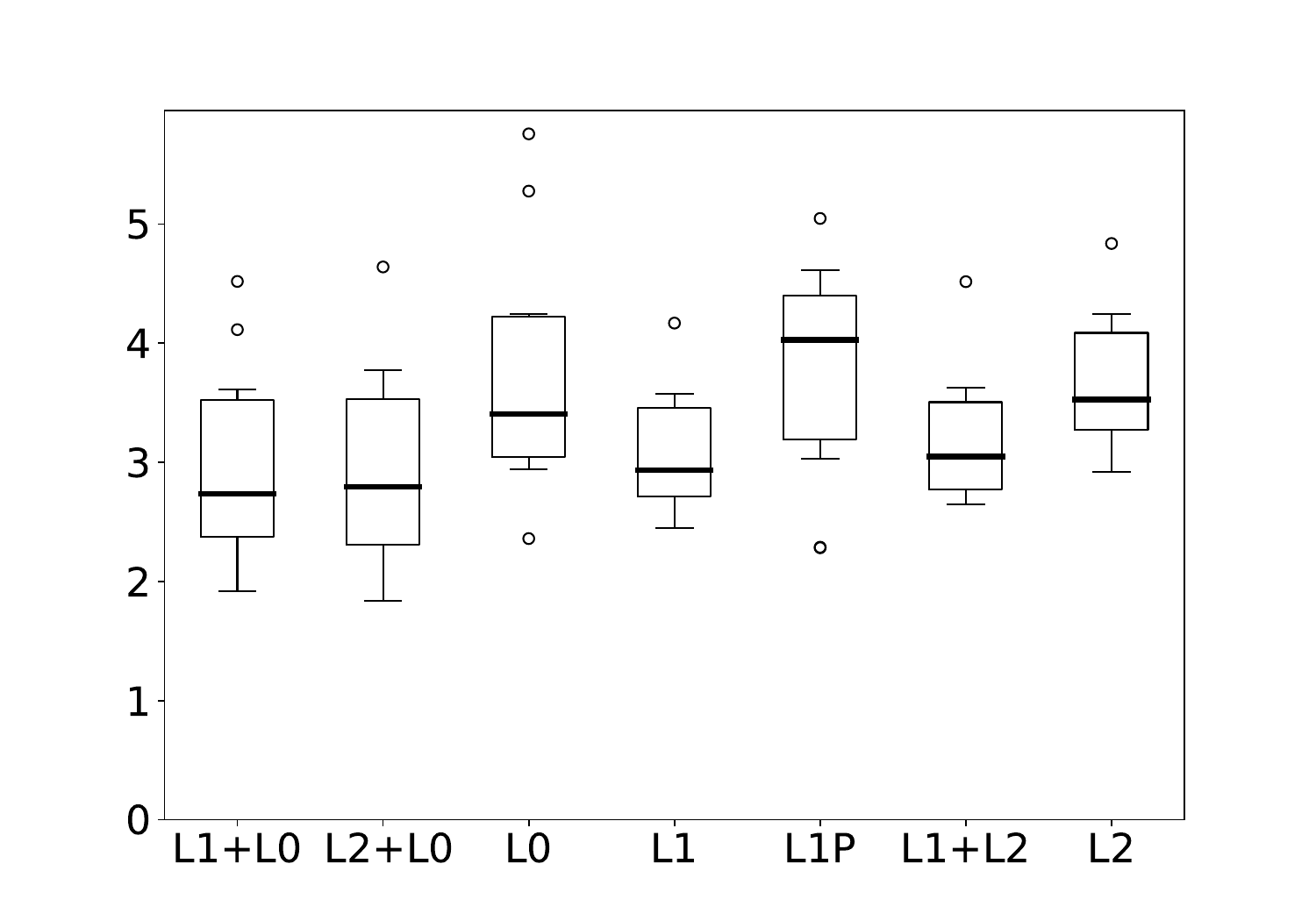}&
		\includegraphics[width=0.3\textwidth,height=0.18\textheight,  trim =1.2cm 1cm 2cm 2cm, clip = true ]{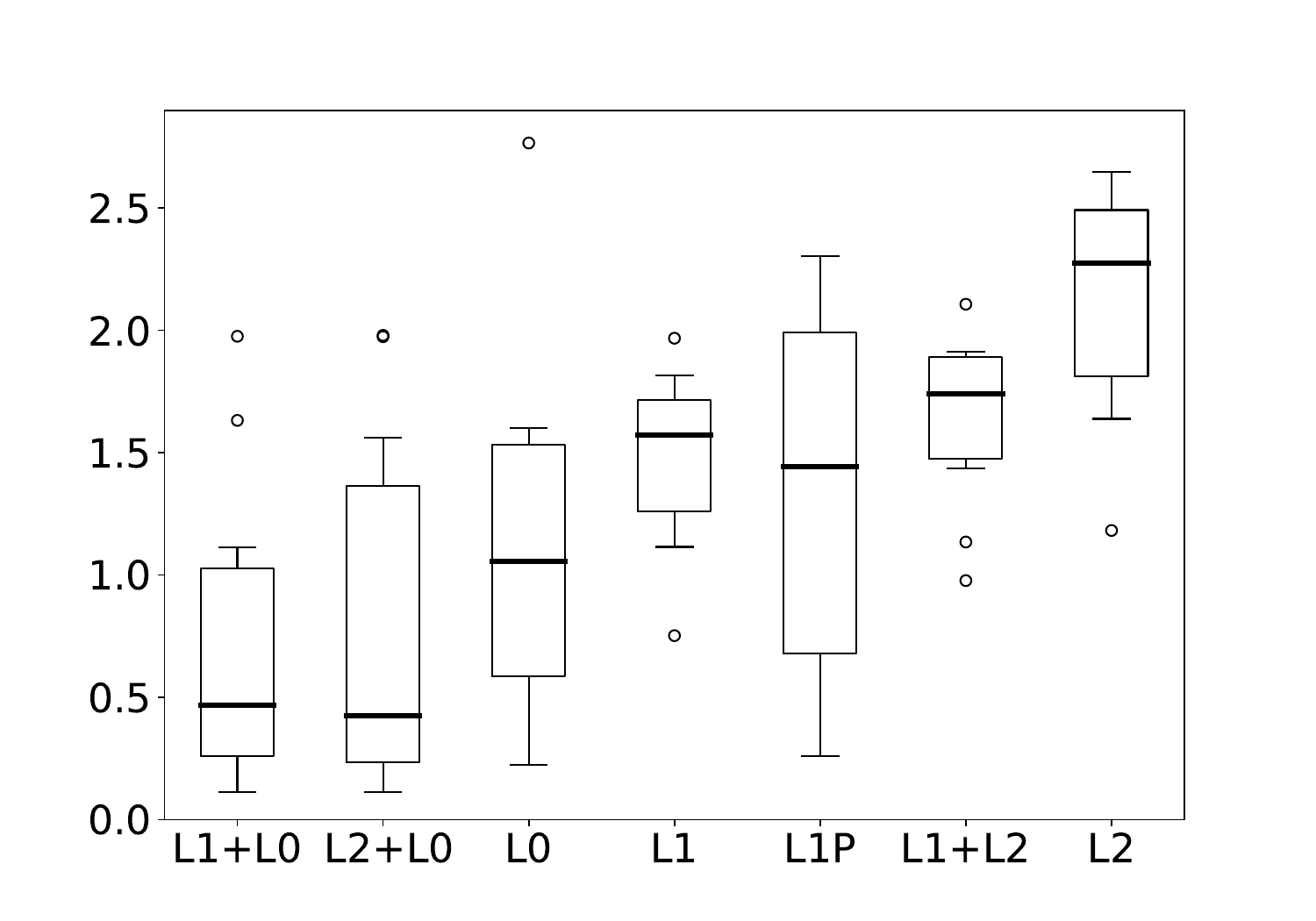}&
		\includegraphics[width=0.3\textwidth,height=0.18\textheight,  trim =1.2cm 1cm 2cm 2cm, clip = true ]{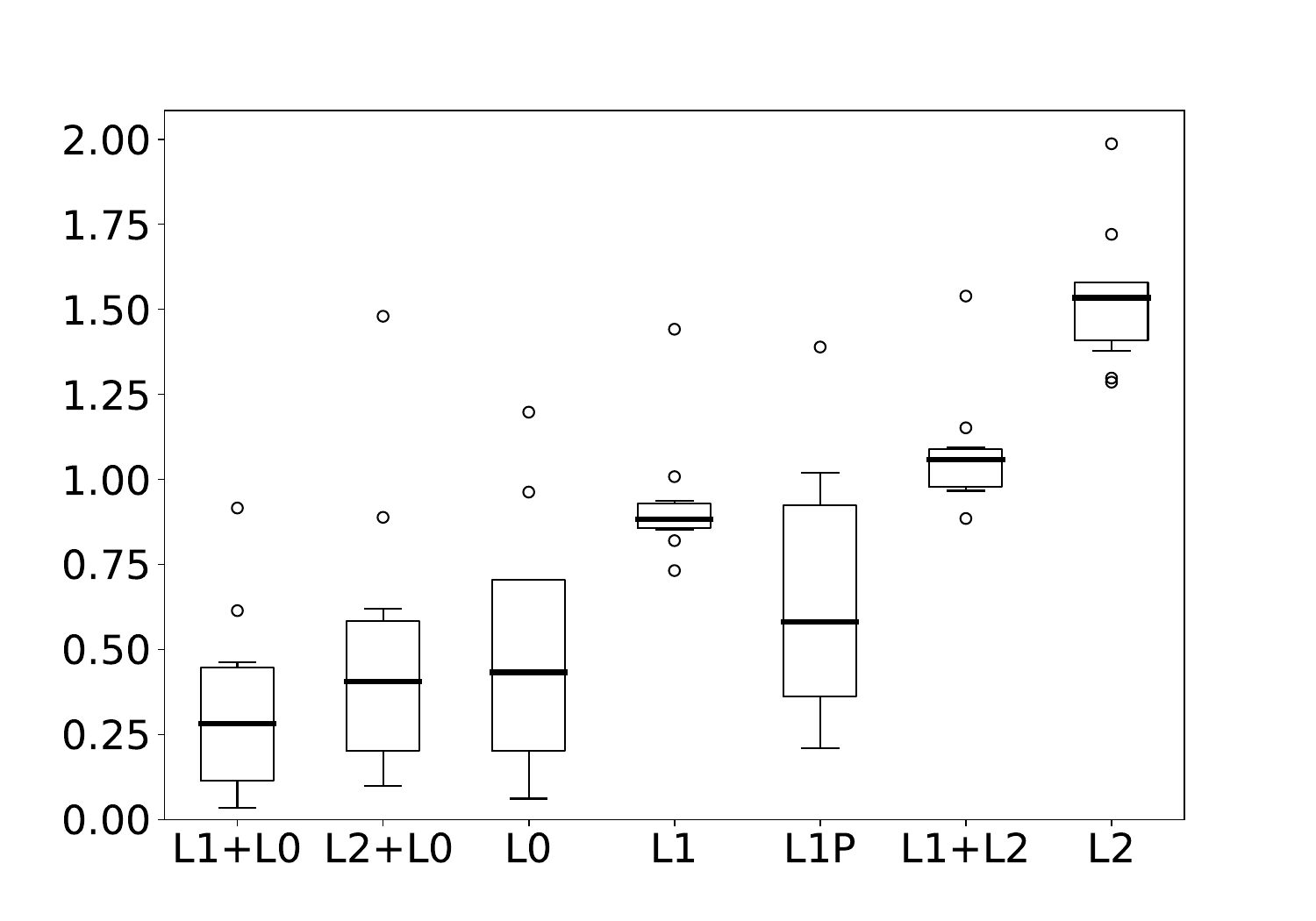}\\
		
		\rotatebox{90}{\sf {\small{~~~~~~~\# nonzeros}}}&
		\includegraphics[width=0.3\textwidth,height=0.18\textheight,  trim =1.8cm 1cm 2cm 1cm, clip = true ]{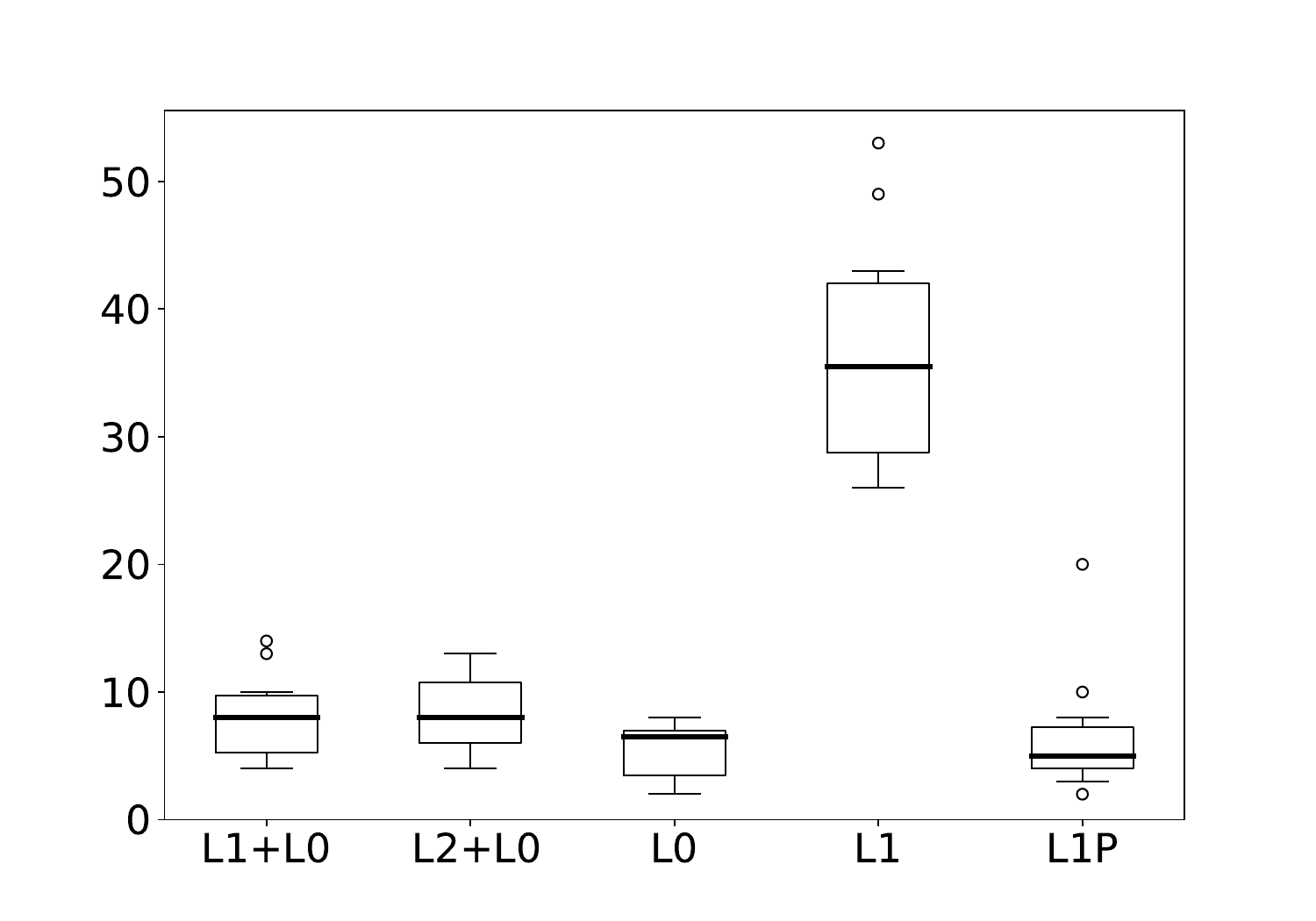}&
		\includegraphics[width=0.3\textwidth,height=0.18\textheight,  trim =1.2cm 1cm 2cm 1cm, clip = true ]{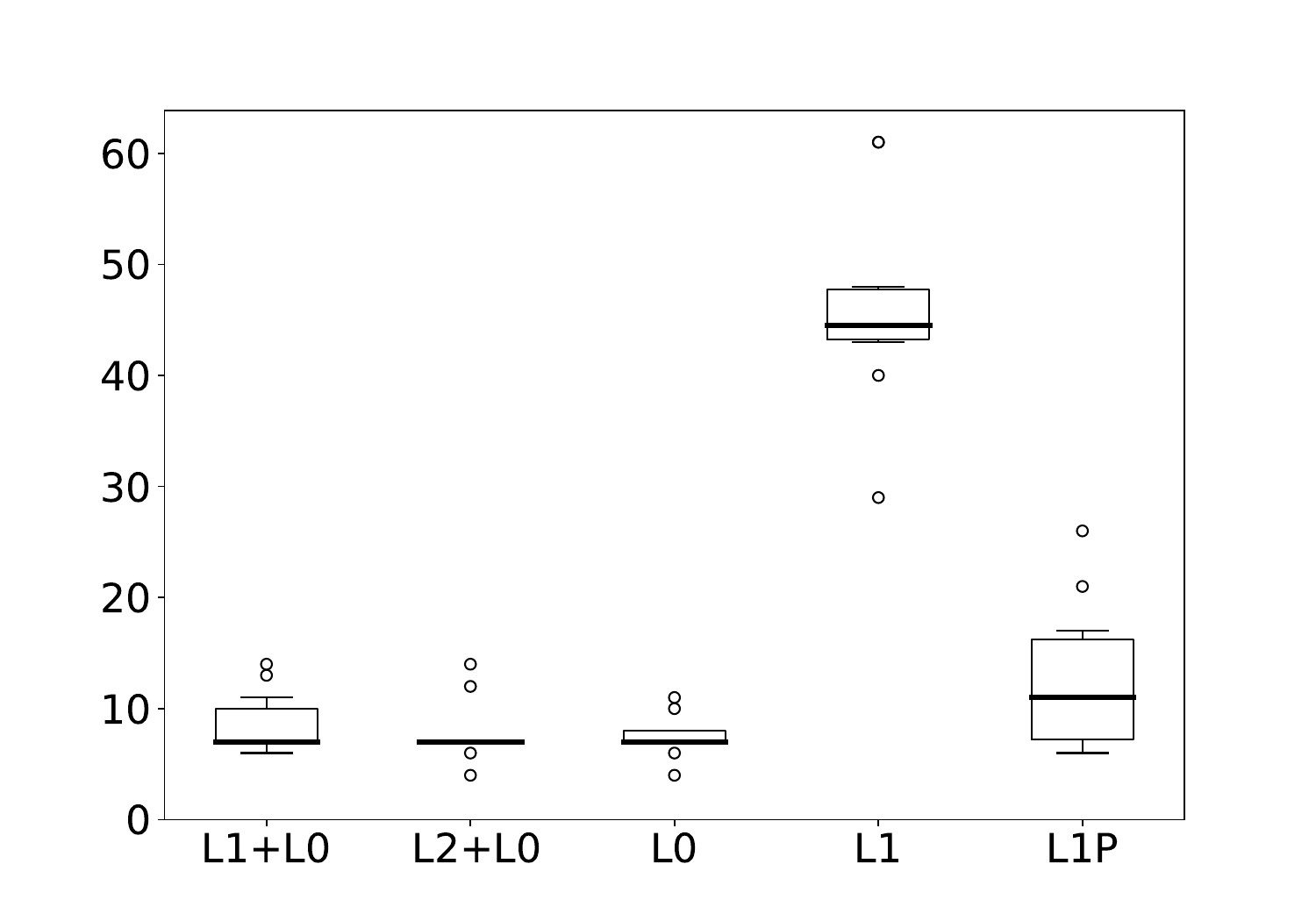}&
		\includegraphics[width=0.3\textwidth,height=0.18\textheight,  trim =1.2cm 1cm 2cm 1cm, clip = true ]{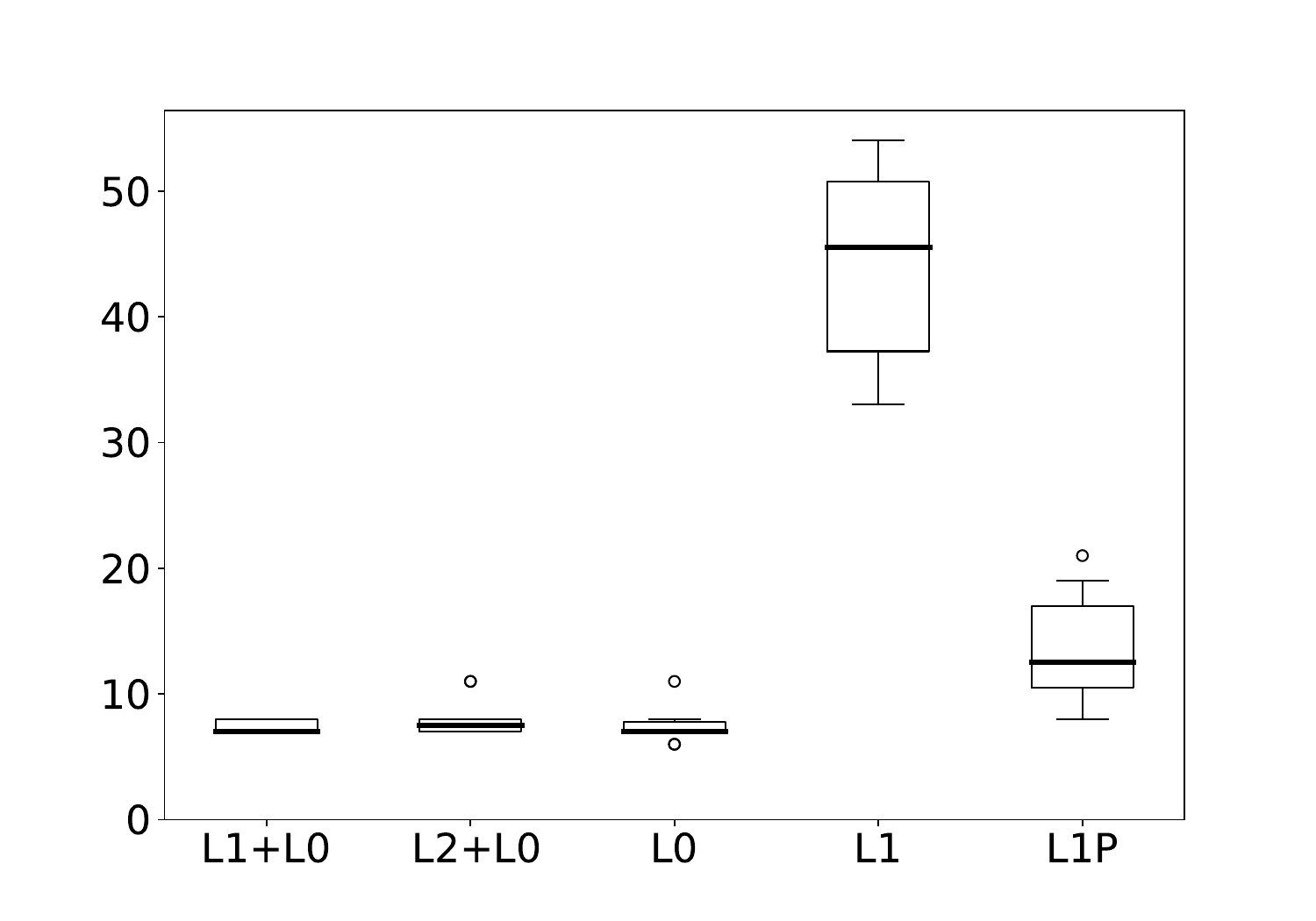}
			\end{tabular}
	\caption{{\small{Example 1 simulations for different values of $n$, $p$, $\rho$, and SNR.
Prediction error refers to the best predictive models obtained after tuning on a separate validation set. \# nonzeros refers to the corresponding
number of nonzero coefficients.
For low SNR values, L0 led to poor predictive models and was outperformed by~L1 and~L2.
Overall, the best predictive models were produced by L1+L0/L2+L0 -- in some instances they were comparable to the best  L1/L2 models, but much sparser.
}}}
	\label{fig: example1}
\end{figure}

\begin{figure}[bp!]
\renewcommand{\baselinestretch}{1.25}
	\centering
	\begin{tabular}{l c c c}
		\multicolumn{4}{c} { \sf \small{Example~2: Small settings: $n=50, p=100$} }\\
		&\sf {\small{$\rho=0.2, \text{SNR}=1$}} &  \sf {\small{$\rho=0.2, \text{SNR}=2$}} & \sf {\small{$\rho=0.2, \text{SNR}=3$}}\\
		\rotatebox{90}{\sf {\small{~~~~~~Prediction Error}}}&
		\includegraphics[width=0.3\textwidth,height=0.18\textheight,  trim =1.8cm 1cm 2cm 2cm, clip = true ]{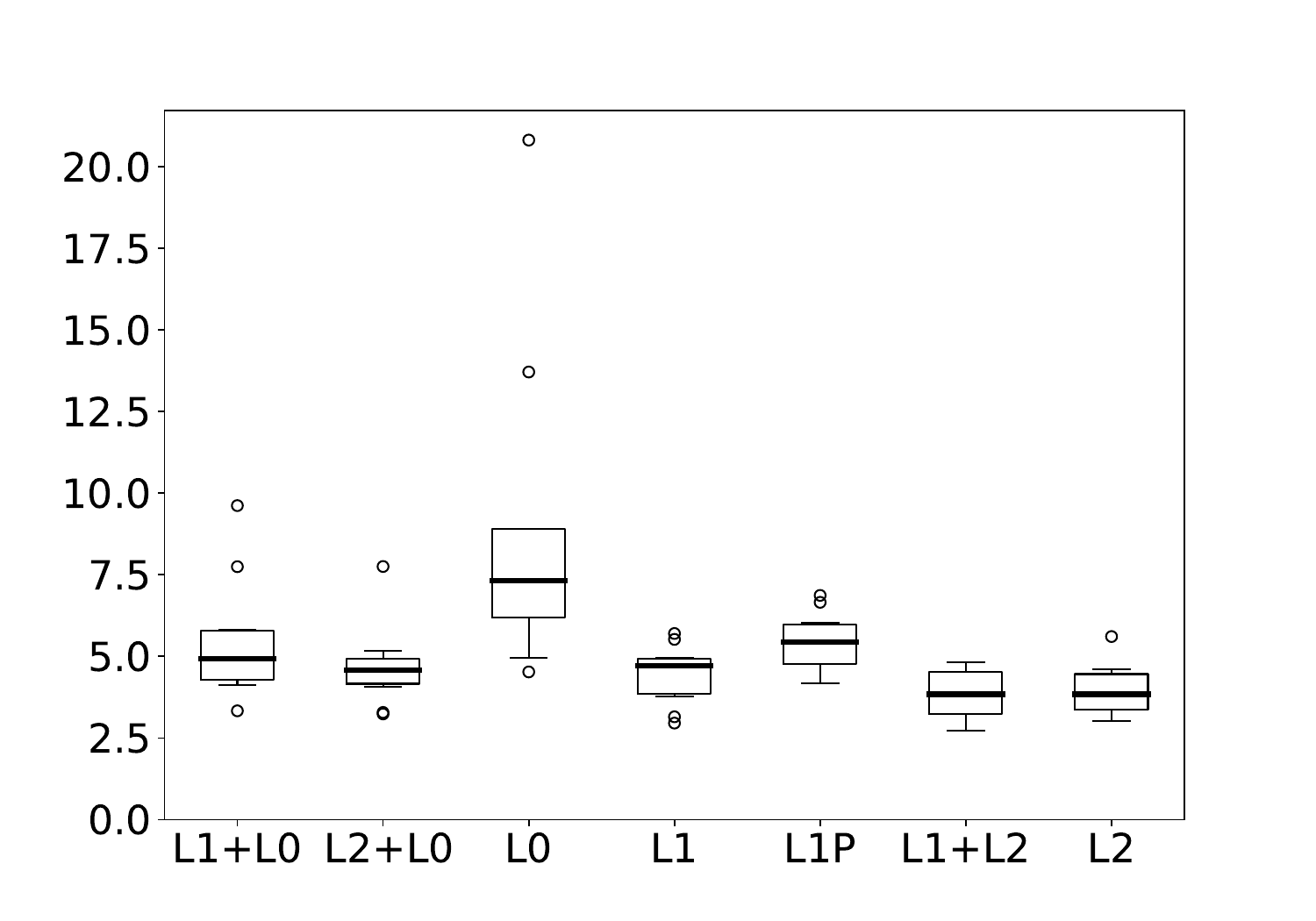}&
		\includegraphics[width=0.3\textwidth,height=0.18\textheight,  trim =1.2cm 1cm 2cm 2cm, clip = true ]{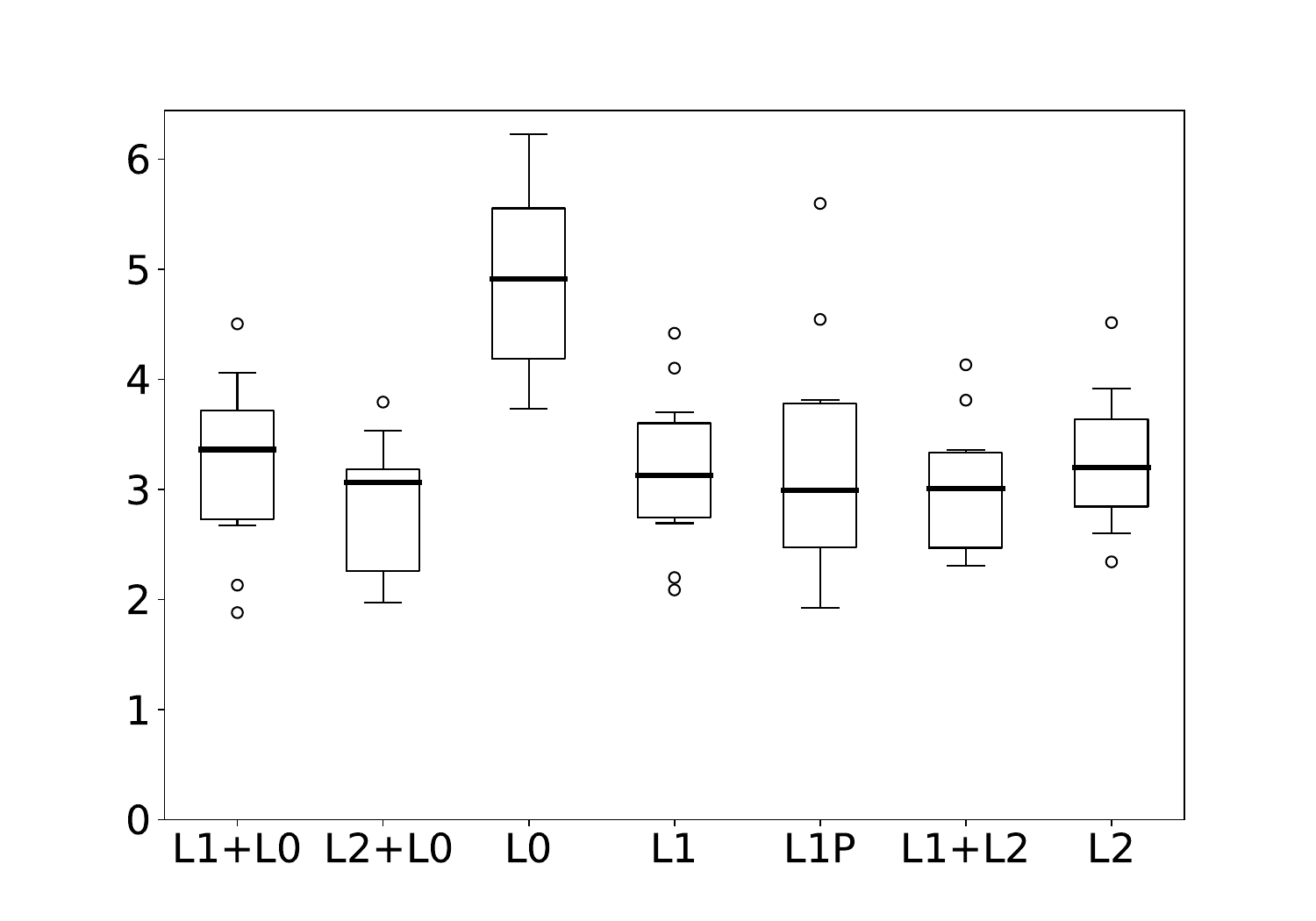}&
		\includegraphics[width=0.3\textwidth,height=0.18\textheight,  trim =1.2cm 1cm 2cm 2cm, clip = true ]{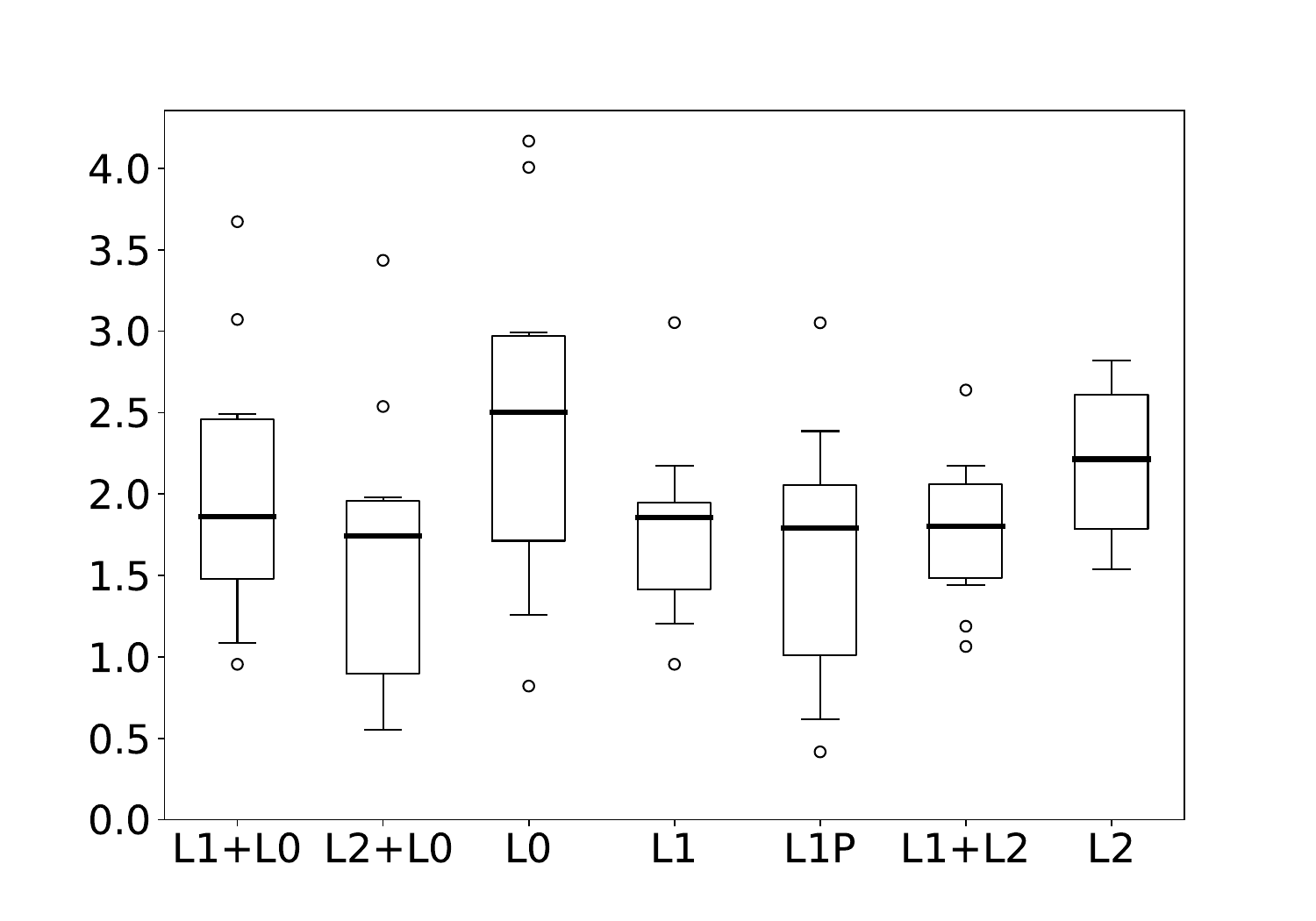}\\
		
		\rotatebox{90}{\sf {\small{~~~~~~~\# nonzeros}}}&
		\includegraphics[width=0.3\textwidth,height=0.18\textheight,  trim =1.8cm 1cm 2cm 2cm, clip = true ]{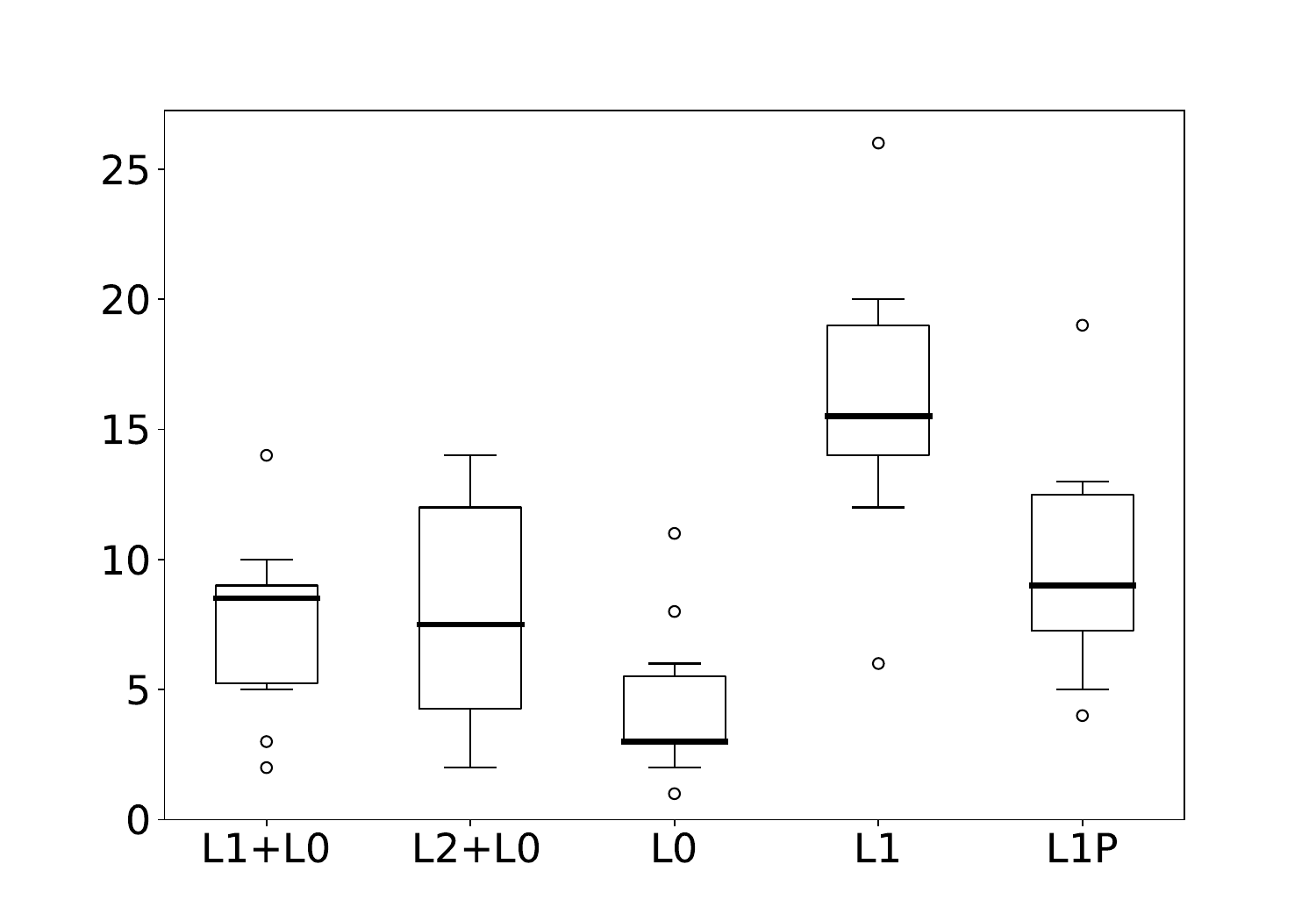}&
		\includegraphics[width=0.3\textwidth,height=0.18\textheight,  trim =1.2cm 1cm 2cm 2cm, clip = true ]{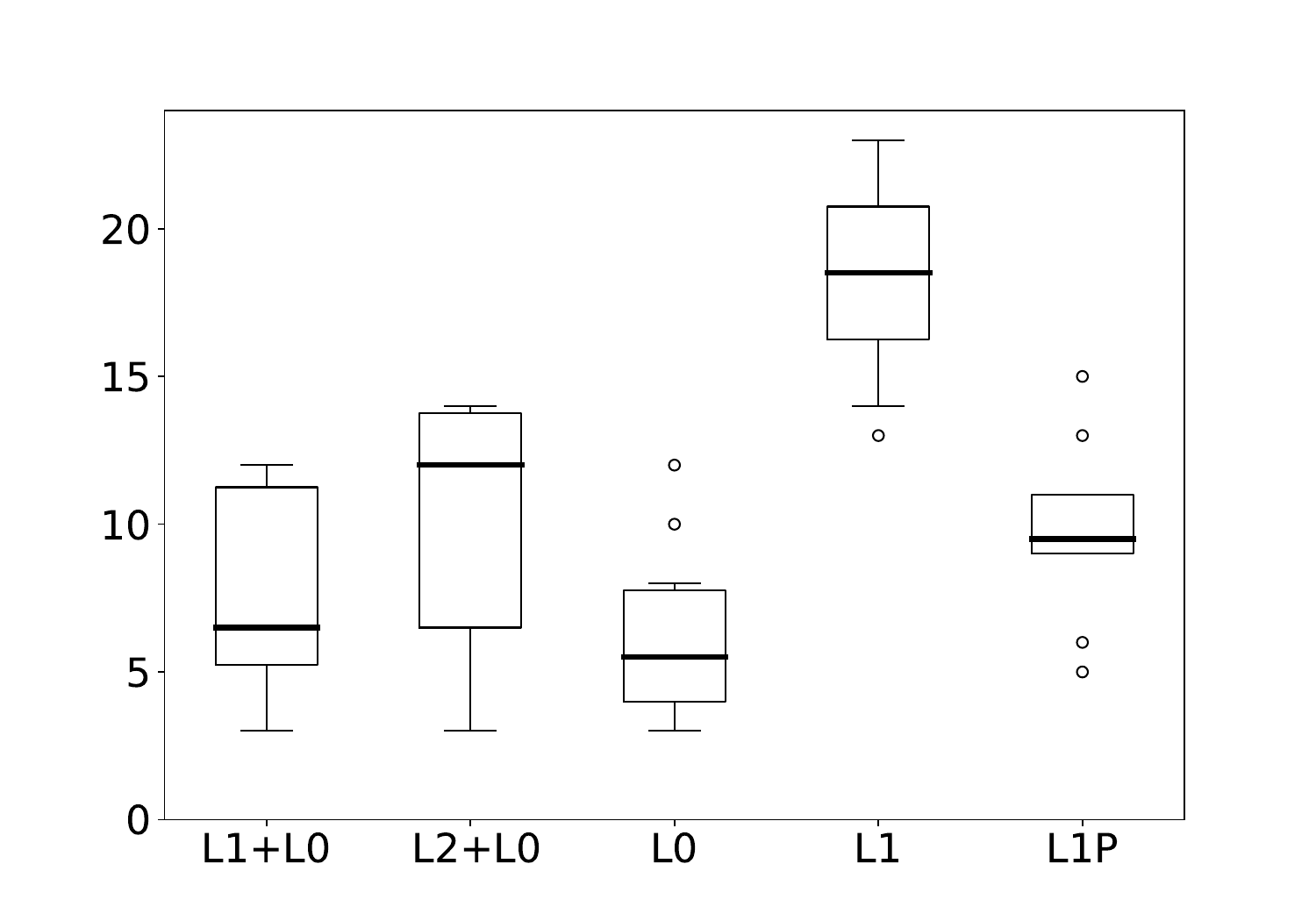}&
		\includegraphics[width=0.3\textwidth,height=0.18\textheight,  trim =1.2cm 1cm 2cm 2cm, clip = true ]{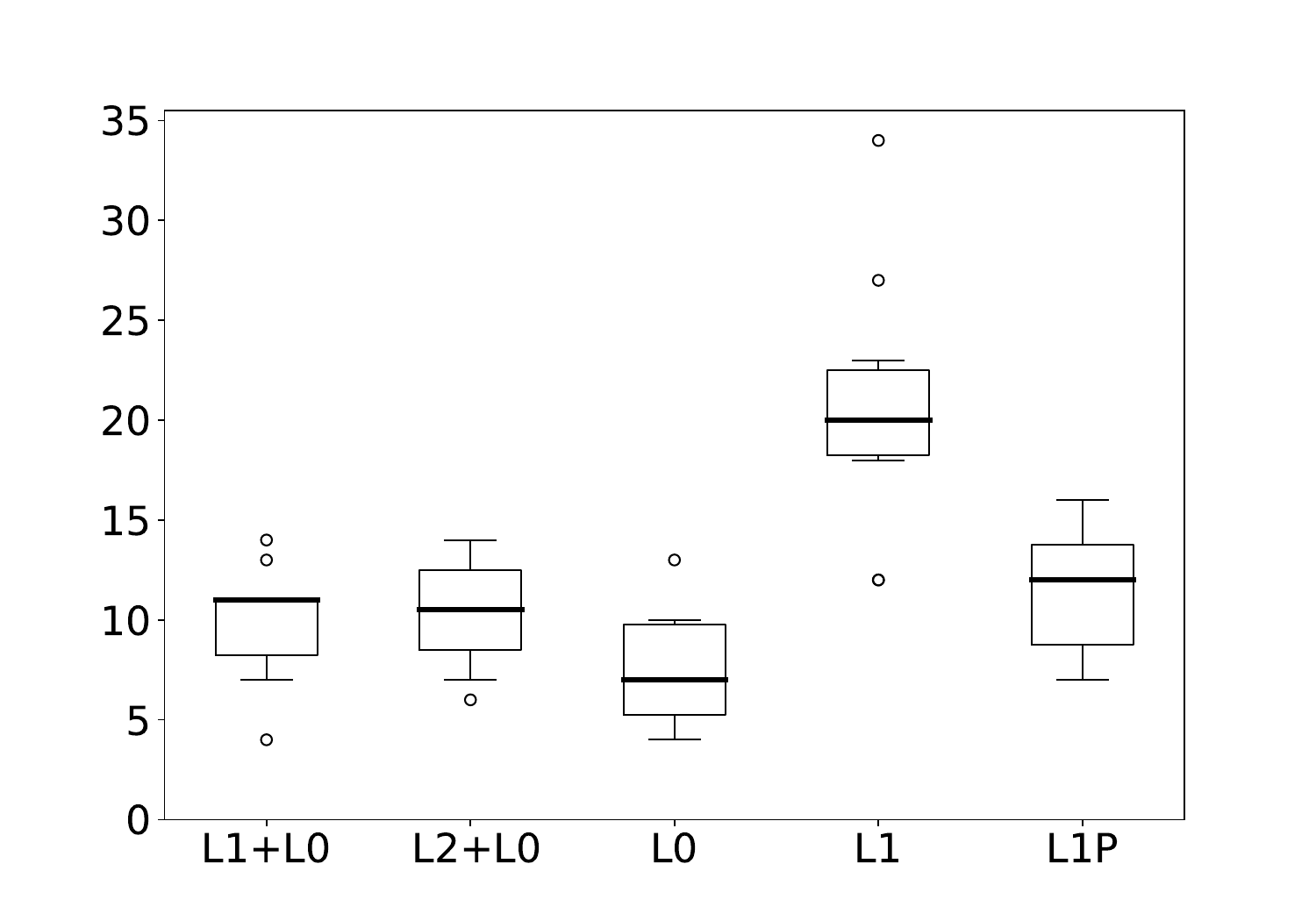}\\
        \vspace{-18pt}\\
		
		\multicolumn{4}{c} { \sf \small{Example~2: Large settings: $n=100, p=1000$} }\\
		&\sf {\small{$\rho=0.0, \text{SNR}=1$}} &  \sf {\small{$\rho=0.2, \text{SNR}=1$}} & \sf {\small{$\rho=0.2, \text{SNR}=3$}}\\
		\rotatebox{90}{\sf {\small{~~~~~~Prediction Error}}}&
		\includegraphics[width=0.3\textwidth,height=0.18\textheight,  trim =1.8cm 1cm 2cm 2cm, clip = true ]{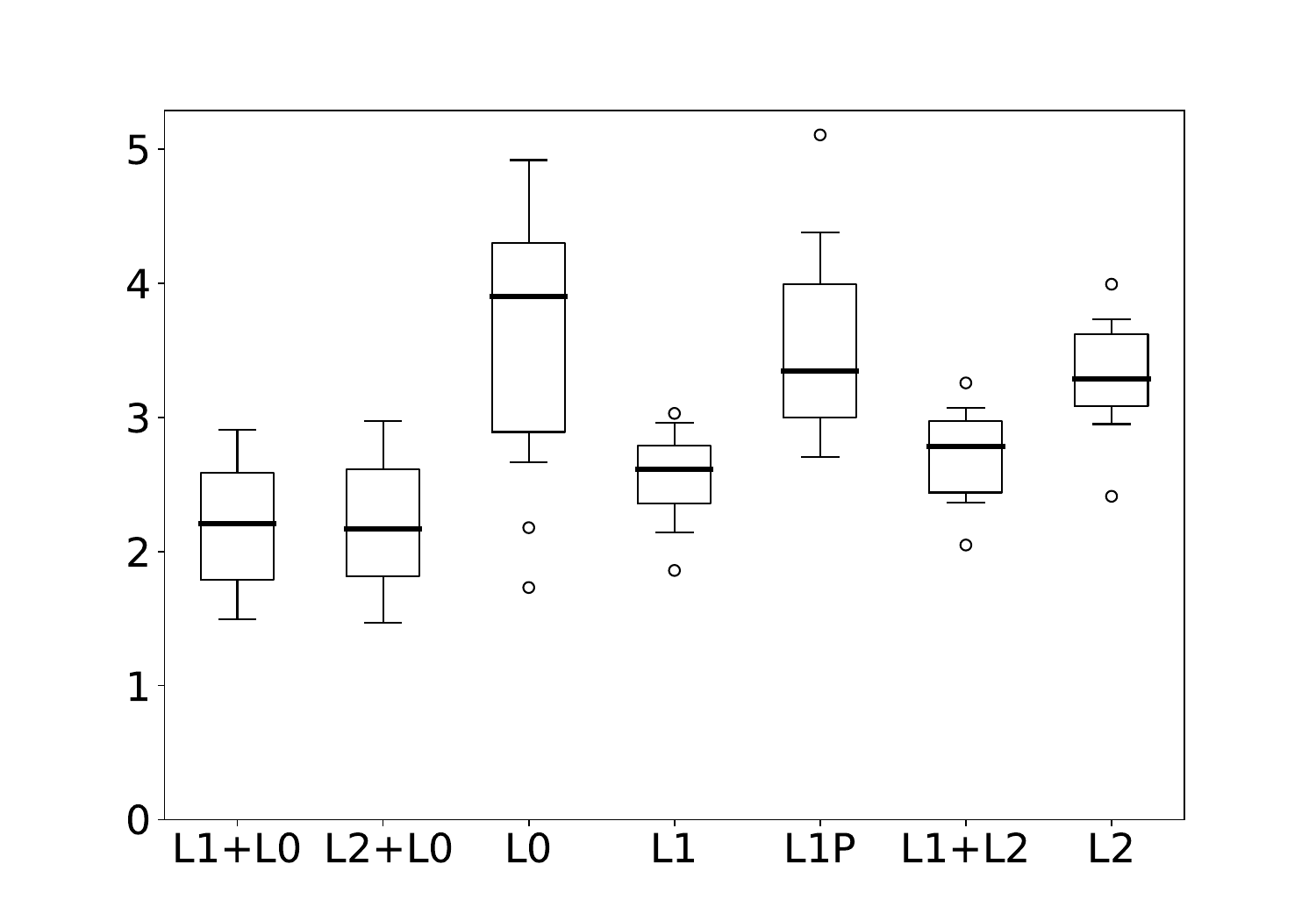}&
		\includegraphics[width=0.3\textwidth,height=0.18\textheight,  trim =1.2cm 1cm 2cm 2cm, clip = true ]{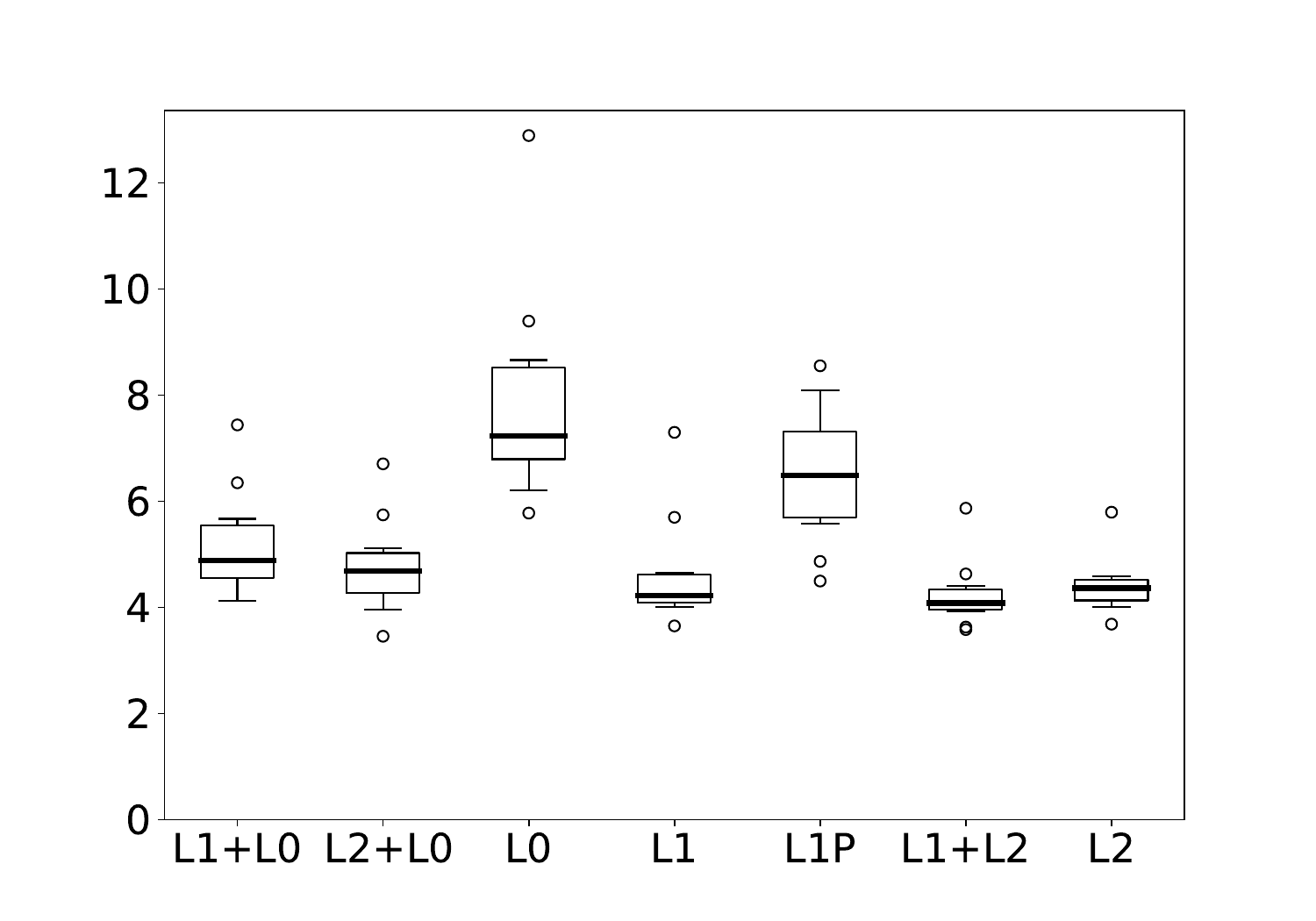}&
		\includegraphics[width=0.3\textwidth,height=0.18\textheight,  trim =1.2cm 1cm 2cm 2cm, clip = true ]{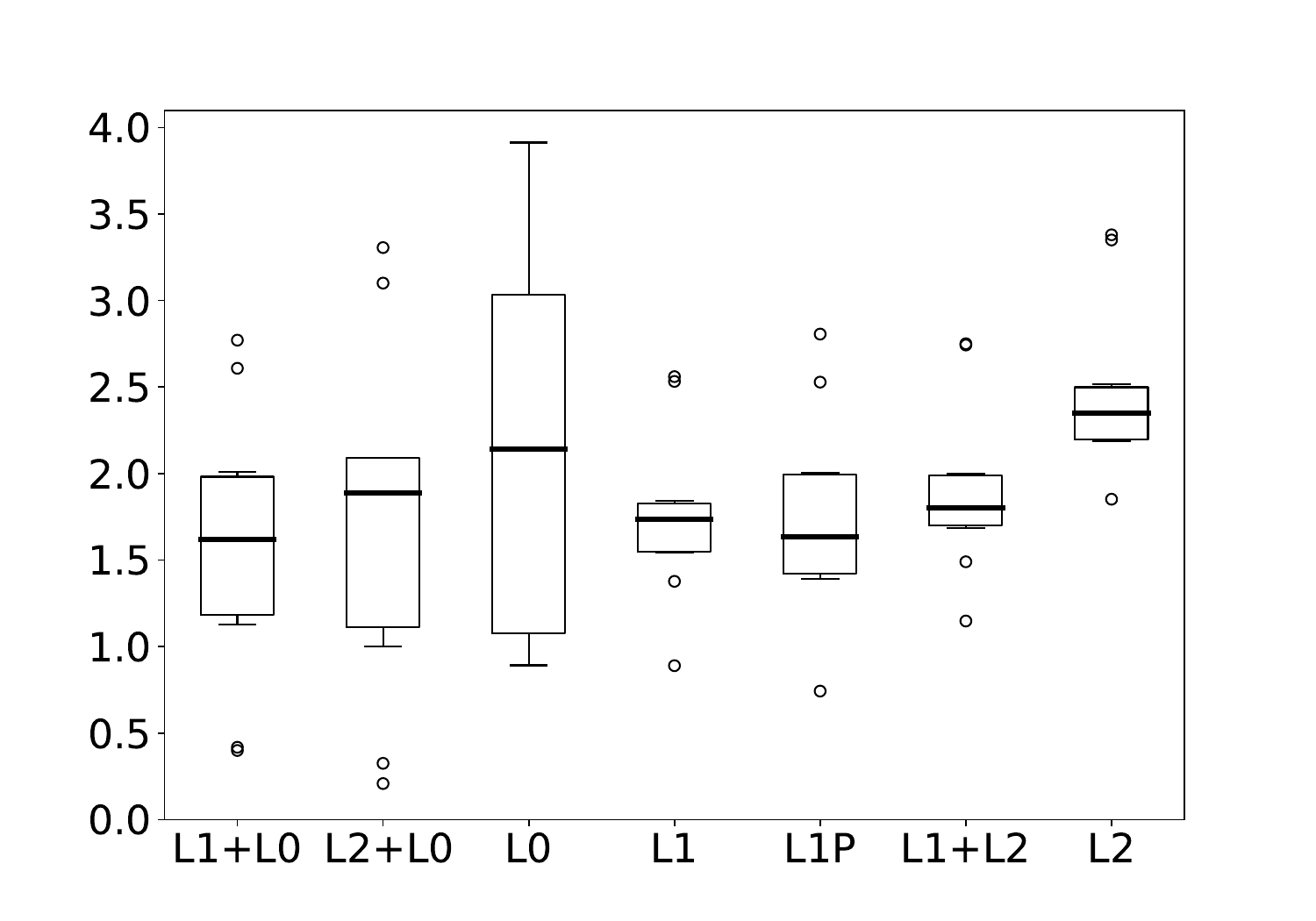}\\
		
		\rotatebox{90}{\sf {\small{~~~~~~~\# nonzeros}}}&
		\includegraphics[width=0.3\textwidth,height=0.18\textheight,  trim =1.8cm 1cm 2cm 2cm, clip = true ]{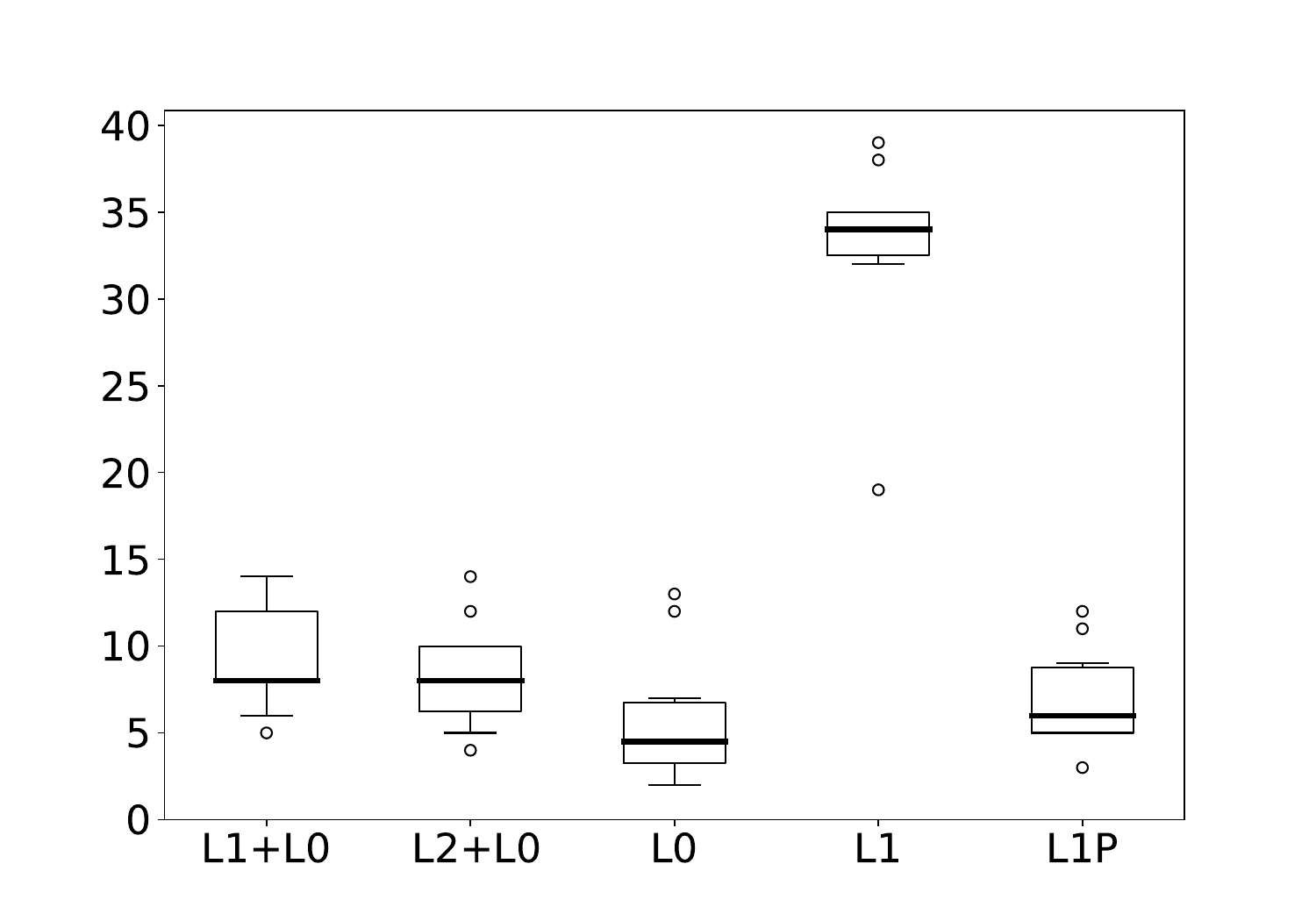}&
		\includegraphics[width=0.3\textwidth,height=0.18\textheight,  trim =1.2cm 1cm 2cm 2cm, clip = true ]{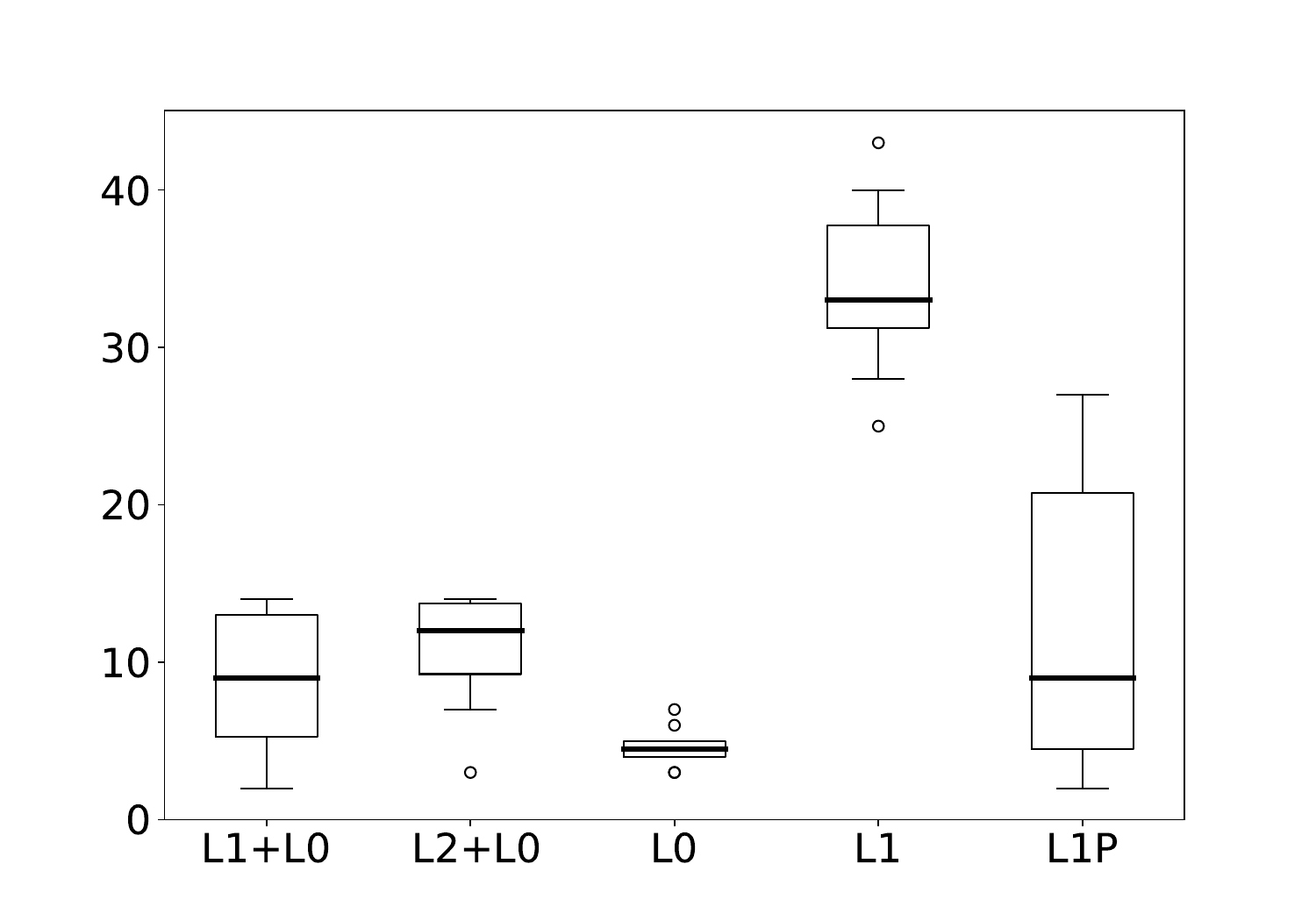}&
		\includegraphics[width=0.3\textwidth,height=0.18\textheight,  trim =1.2cm 1cm 2cm 2cm, clip = true ]{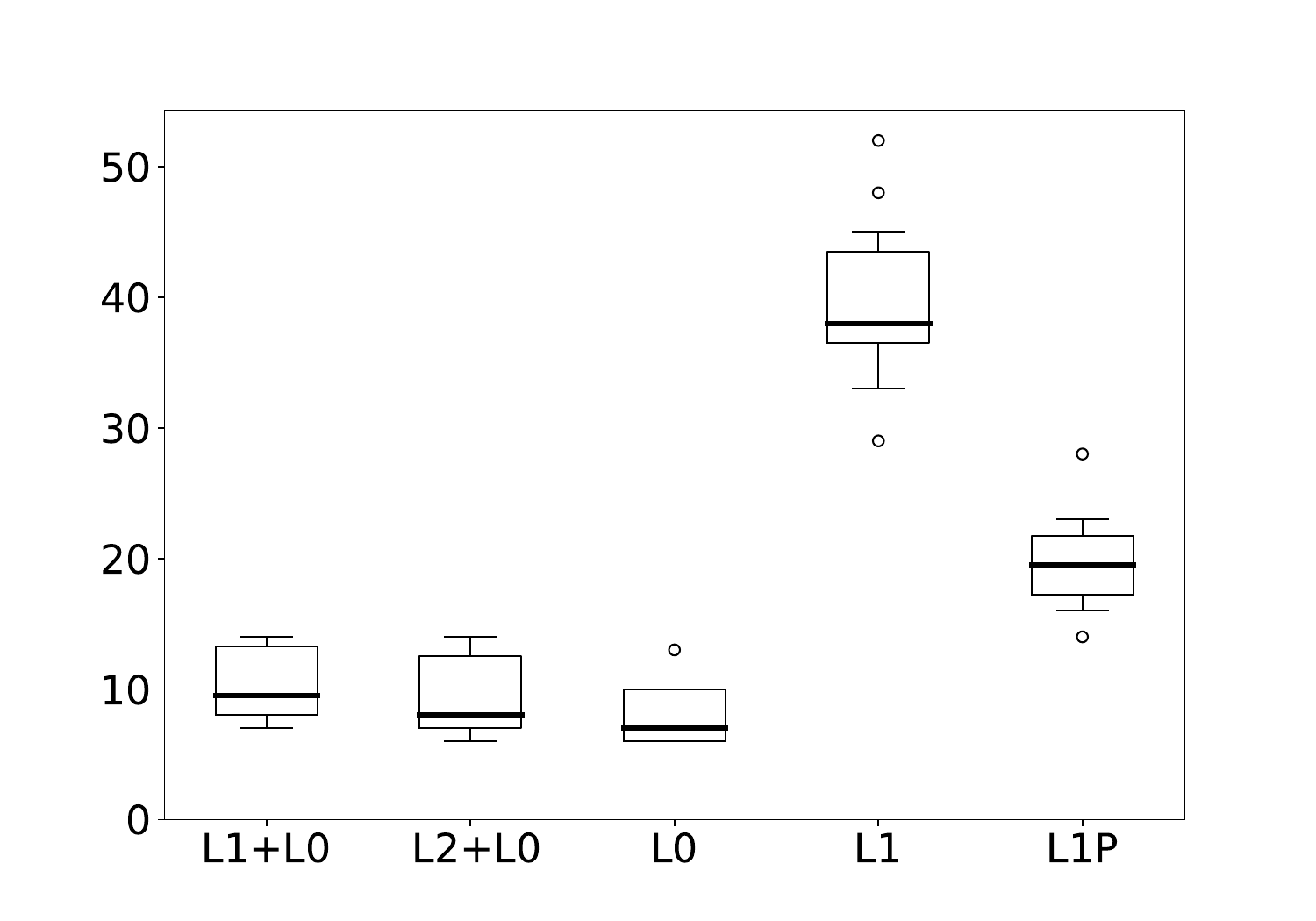}
		
	\end{tabular}
	\caption{
	{\small{Experimental results for Example~2. The results are qualitatively similar to Figure~\ref{fig: example1} -- however, this example is ``harder'' than Example~1 due to the increased correlation among the features -- a larger nominal value of SNR is required before L0 matches the performance of L1+L0/L2+L0. 	The L1+L0/L2+L0 methods performed the best in terms of obtaining a good predictive model that is also sparse -- the model sizes were larger than~$k^*$ but smaller than those available from the best L1 models.} }
	}
	\label{fig: example2}
\end{figure}

\begin{figure}[tp!]
\renewcommand{\baselinestretch}{1.25}
	\centering
		\begin{tabular}{c c c c}
		\multicolumn{2}{c}{Example~1} & \multicolumn{2}{c}{Example~2}  \\
		\sf {\scriptsize{Prediction error}} &\sf {\scriptsize{\# nonzeros }} &  \sf {\scriptsize{Prediction error}} &\sf {\scriptsize{\# nonzeros}}\\
		\includegraphics[width=0.228\textwidth,height=0.2\textheight,  trim = 1.4cm 1cm 1.9cm 1.3cm, clip = true ]{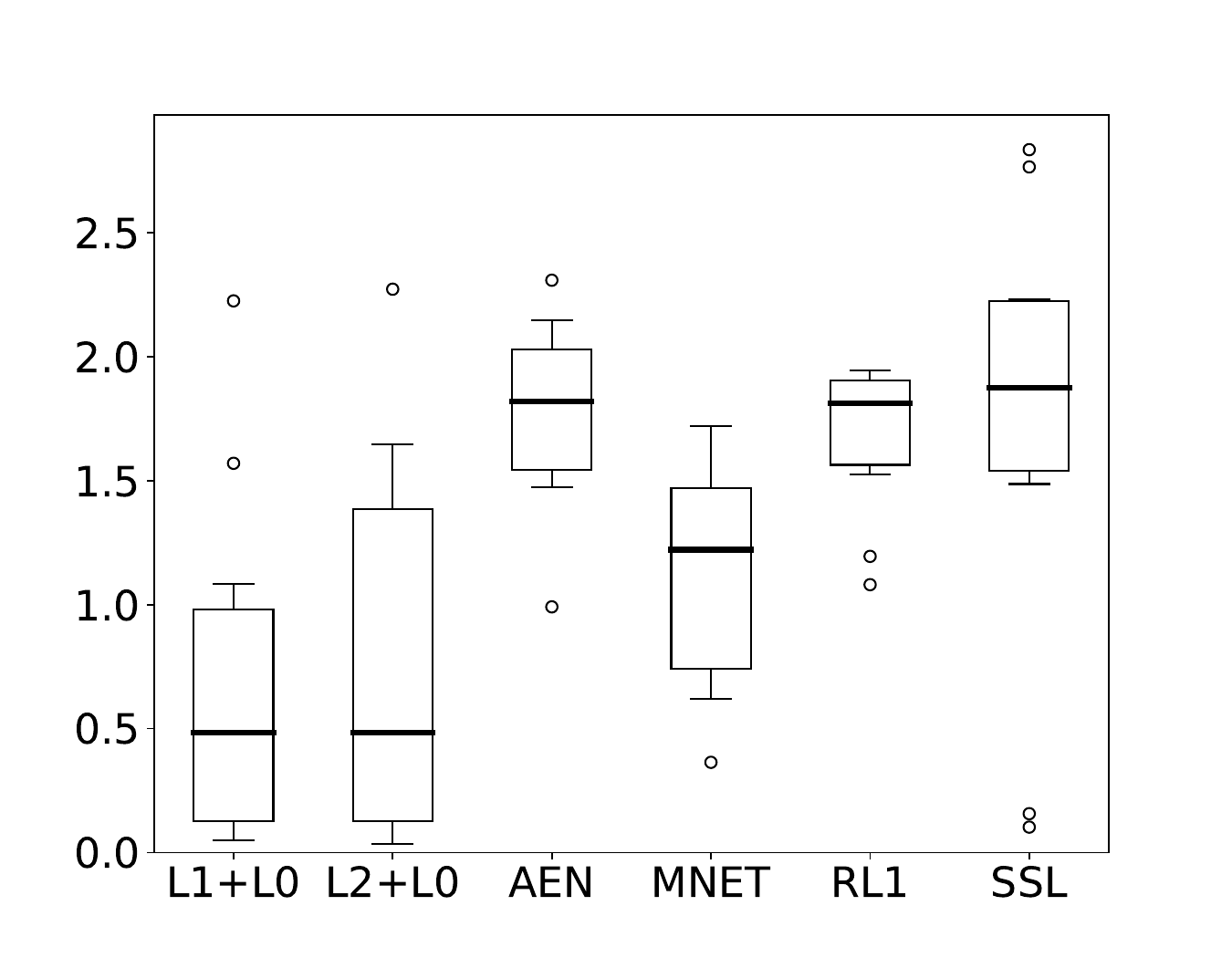}&
		\includegraphics[width=0.228\textwidth,height=0.2\textheight,  trim =1.4cm 1cm 1.9cm 1.3cm, clip = true ]{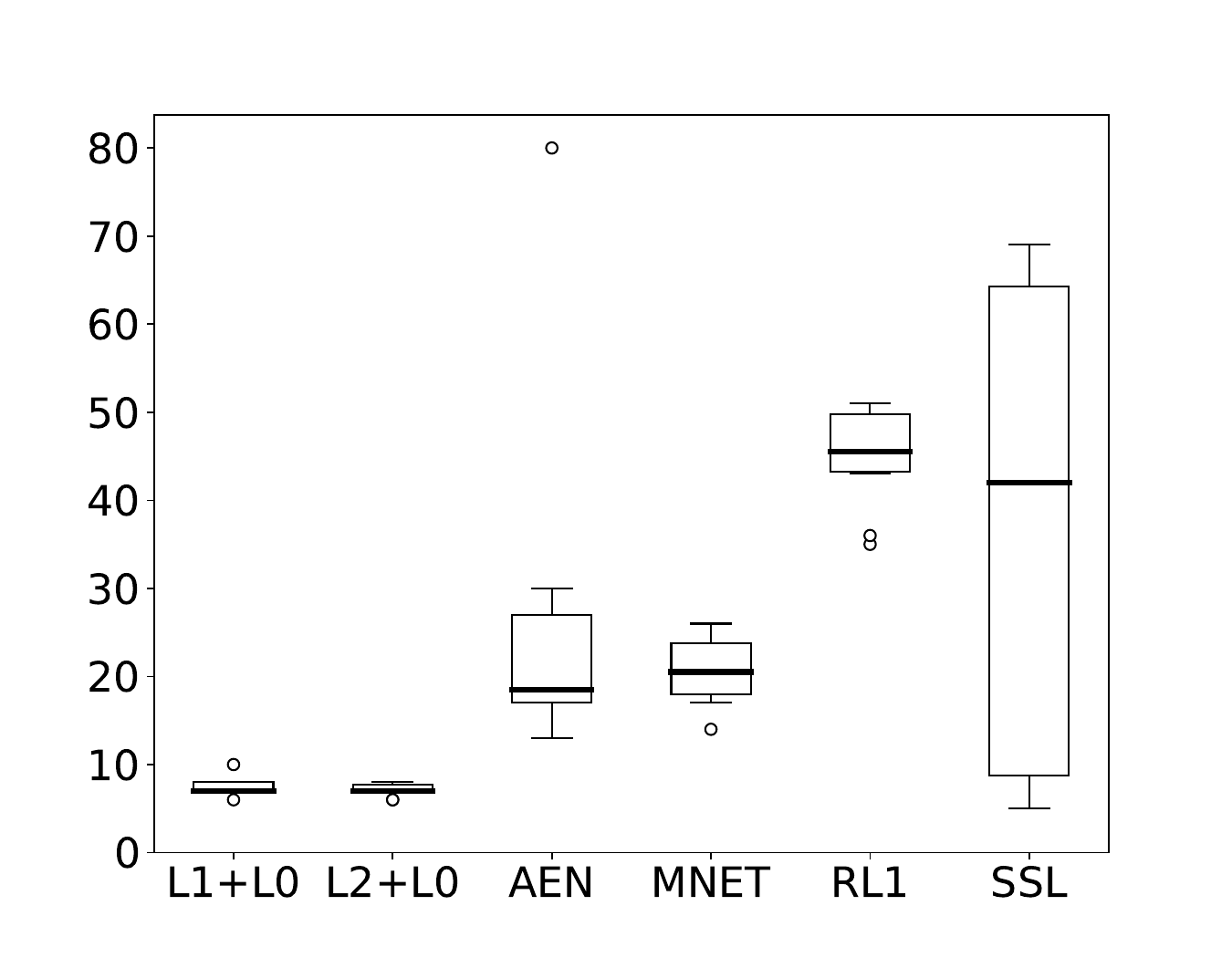}&
		\includegraphics[width=0.228\textwidth,height=0.2\textheight,  trim =1.4cm 1cm 1.9cm 1.3cm, clip = true ]{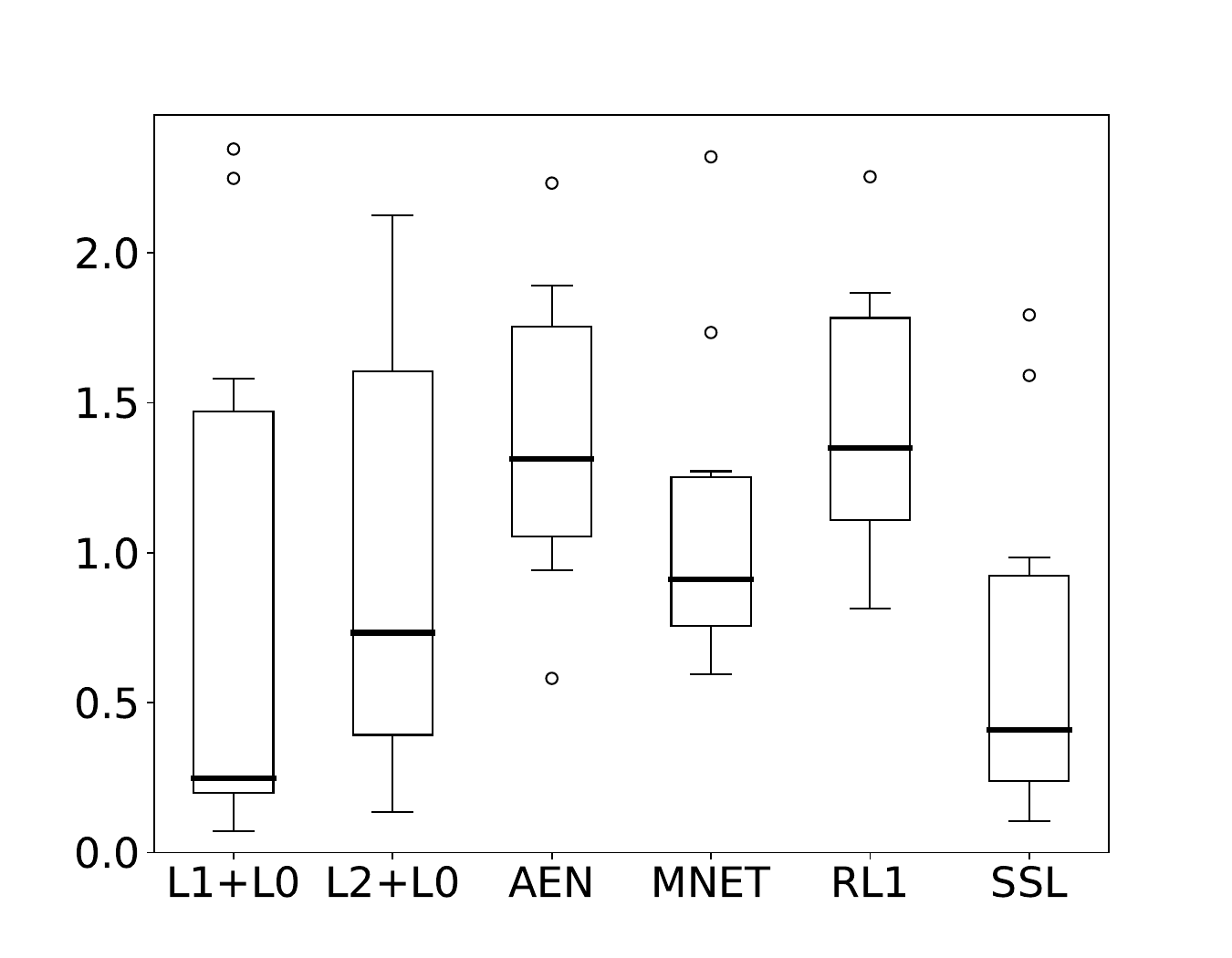}&
		\includegraphics[width=0.228\textwidth,height=0.2\textheight,  trim =1.4cm 1cm 1.9cm 1.3cm, clip = true ]{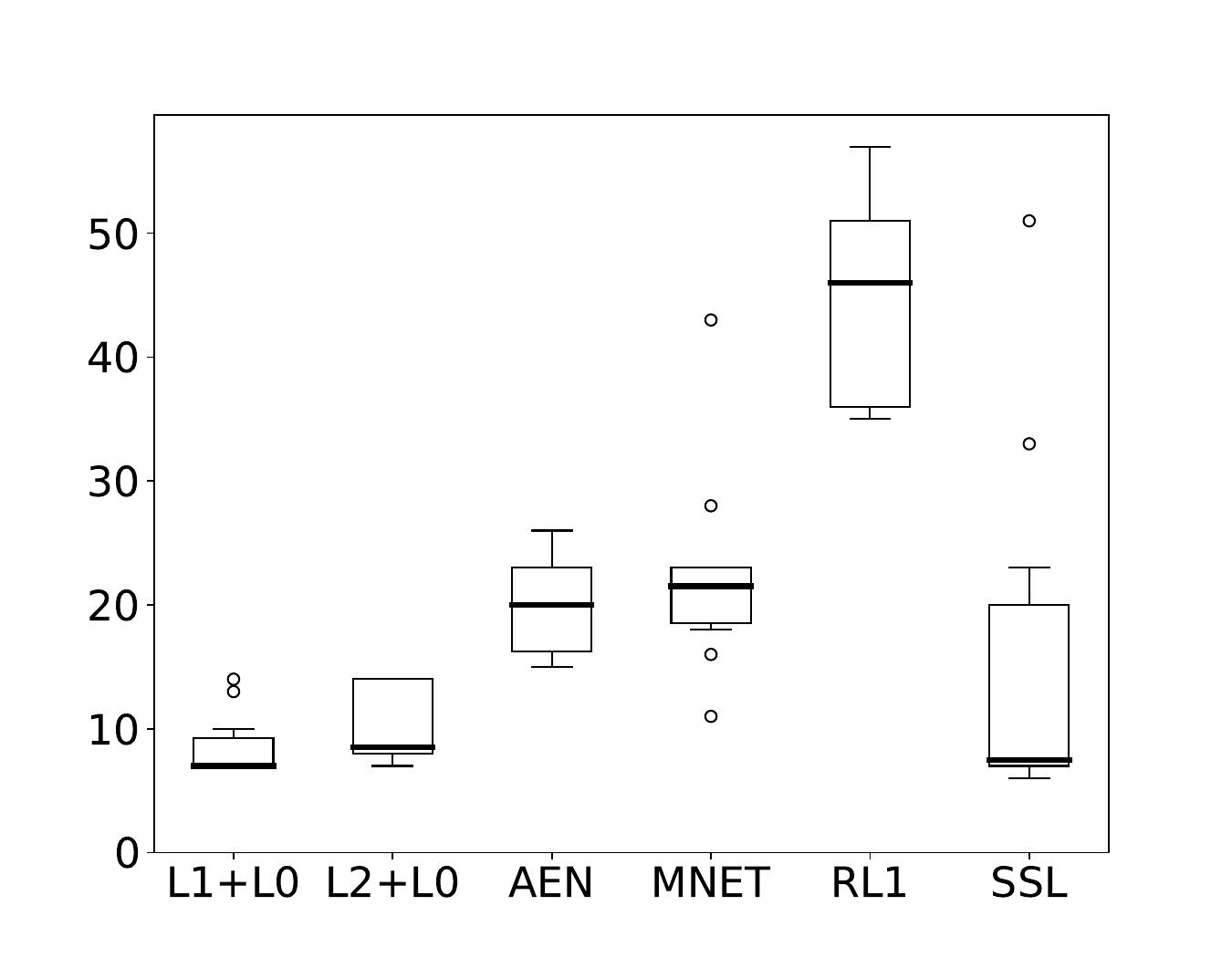}\\
	\end{tabular}
	\caption{{Experimental results for the proposed methods, L1+L0 and L2+L0, as well as adaptive elastic net (AEN), Mnet, relaxed Lasso (RL1), and spike-and-slab Lasso (SSL) methods (as described in the text).  Here, $\rho=0.2$, $\text{SNR}=2$ for Example 1 and $\rho=0.1$, $\text{SNR}=3$ for Example 2; $n=100, \ p=1000$ in both settings. 
	Overall, our proposed approach performed favorably in terms of both the model sparsity and the prediction accuracy.}}
	\label{fig: exampleA}
\end{figure}

\noindent {\textbf{Selecting the tuning parameters.}} For each of the above methods, we pick the estimator that minimizes the least squares criterion on a validation set simulated as
$\mathbf{y} = \mathbf{X}\B{\beta}^{*} +\B{\epsilon}$, with the fixed~$\M{X}$ and an independent realization of $\B\epsilon$, with the same SNR.
For each selected estimator we compute the prediction error, $\| \mathbf{X} \hat{\B{\beta}} - \mathbf{X} \B{\beta}^* \|_2^2/n$, and the associated number of nonzero regression coefficients.
Figures \ref{fig: example1}, \ref{fig: example2} and \ref{fig: exampleA} summarize the results  via box plots, in which the boxes extend from the lower to the upper quartile of the data with a line at the median, to aggregate the results over the ten independent simulations.  We do not display the sparsity levels of L1+L2 and L2, as these methods are considerably denser than L1, which, in turn, produces the densest solutions among the remaining methods in the examples we consider.

\noindent {\textbf{Summary of observations.}}
We summarize our general observations below:
\begin{itemize}
	\item When the noise level is high (SNR=1), L0 performs poorly in terms of prediction accuracy. To mitigate its overfitting behavior, L0 attempts to regularize by selecting very sparse models -- the best predictive model for L0 has fewer nonzeros than~$\B\beta^*$.  In this setting, methods L1 and L2 work better than L0 in terms of the prediction accuracy. However, the estimated models are rather dense. The polished version of the Lasso, L1P, selects a model that is sparser than the Lasso but suffers in prediction accuracy.

The two new methods, L1+L0 and L2+L0, display the best prediction accuracy overall. They \emph{fix} the overfitting behavior of L0 via the additional shrinkage. This observation agrees with the theoretical results and the discussion in Sections~\ref{sec.L2.bnds}-\ref{sec.lower.bnds}.  The best predictive models available from L1+L0/L2+L0 are similar in performance to the best predictive models available via L1 and L2, however, the new methods lead to estimators that are significantly sparser.    The L0 models are sparser than those for L1+L0 and L2+L0, however, L0 suffers in terms of the prediction accuracy.  In summary, the new L1+L0/L2+L0 methods significantly improve upon the predictive performance of L0 at the cost of marginally decreasing the model sparsity.


	\item As SNR increases, L1+L0 and L2+L0 become more similar to L0, in terms of both sparsity and the prediction accuracy.  Additional shrinkage marginally helps the prediction accuracy, and the model sparsity becomes comparable to that of L0, with the model size concentrating around $\| \B\beta^*\|_{0}$.  This observation is consistent with the results in the fast rate parts of Theorems~\ref{slow.rate.thm} and~\ref{L1.rate.thm}. L1 performs better than both L1+L2 and L2; it also benefits from polishing -- L1P gets closer to L0 in terms of the prediction accuracy but selects a denser model.
	
	

\end{itemize}

{
In the Supplementary Material, we discuss additional experiments corresponding to the challenging ultra-high dimensional setting \citep{verzelen2012minimax} with $k^*\log(p/k^*)>n/2$. These experiments provide further support for the observations listed above.
}

\noindent {\textbf{Comparisons with adaptive elastic net (AEN), Mnet, relaxed Lasso and spike-and-slab Lasso (SSL).}}
We present simulation results that compare our proposal with methods Mnet~\citep{huang2016mnet}, AEN~\citep{zou2009adaptive}, relaxed Lasso~\citep{hastie2020best}, 
{
and SSL~\citep{rockova2018spike} .
}
Mnet and AEN reduce the estimation error of elastic net, and encourage greater sparsity, by using a nonconvex penalty on~$\B\beta$ instead of the usual $\ell_{1}$-norm.  The proposed estimator with $q=2$ is a natural alternative to Mnet and AEN in the regimes where these methods are found to be useful -- however, our motivation for estimator~\eqref{best-subset-penalized} is different.
Empirically, we observe important differences in the statistical performance of Mnet, AEN and our approach.
These differences are likely a consequence of (a) the optimization algorithms{\footnote{\cite{huang2016mnet} use a coordinate descent method directly on the $\ell_{2}^2$+MCP penalized problem; \cite{zou2009adaptive} work with the $\ell_{2}^2$ + adaptive Lasso regularized least squares, which is a convex problem.}} and (b) the exact forms of the estimators, including the choice of the penalty function.

Figure \ref{fig: exampleA} compares the methods on the data generated as per Examples~1 and~2, with $n=100$ and $p=1000$.
For AEN, we used 
R package {\texttt{gcdnet}} with 	weights chosen based on Example~1 in~\cite{zou2009adaptive}. For Mnet, we used 
R package {\texttt{ncvreg}}, with the MCP penalty and ridge regularization.   For the relaxed Lasso, we implemented the code in~\cite{hastie2020best}; 
{
and for SSL, we used R package {\texttt{SSLASSO}}.
}
For 
AEN, Mnet, and relaxed Lasso, 
we used the same number of tuning parameters as for our proposed methods\footnote{For Mnet, we used 15 values for the tuning parameter that combines the ridge and MCP penalties, and 100 values for the MCP penalty weight. We made a similar choice for AEN. For the relaxed Lasso, we used 15 values for the weight in the convex combination, and 100 tuning parameter values for the Lasso.}. As before, the tuning parameters were selected based on a held-out validation set. {For SSL, we used the default settings of R package \texttt{SSLASSO} (with the exception of the variance parameter, set to be unknown).}  In summary, estimator~\eqref{best-subset-penalized} produced models with significantly fewer nonzeros and overall better predictive performance.

{
In the Supplementary Material, we compare estimator~\eqref{best-subset-penalized} to two additional state-of-the-art Bayesian shrinkage methods -- the horseshoe regression \citep{carvalho2010horseshoe} and the empirical Bayes method of \cite{martin2017empirical}, which were outperformed in our experiments by the spike-and-slab Lasso approach considered in Figure~\ref{fig: exampleA}. }

\begin{figure}[h!]
\renewcommand{\baselinestretch}{1.25}
	\centering
	\begin{tabular}{c c c c}
		\multicolumn{2}{c}{\sf {\scriptsize{Triazine: $n=93, p=560$}}} & \multicolumn{2}{c}{ \sf {\scriptsize{Riboflavin: $n=35, p=4,088$}} }  \\
		\sf {\scriptsize{Prediction error}} &\sf {\scriptsize{\# nonzeros }} &  \sf {\scriptsize{Prediction error}} &\sf {\scriptsize{\# nonzeros}}\\
		\includegraphics[width=0.22\textwidth,height=0.18\textheight,  trim = 0.5cm 1cm 2.5cm 2cm, clip = true ]{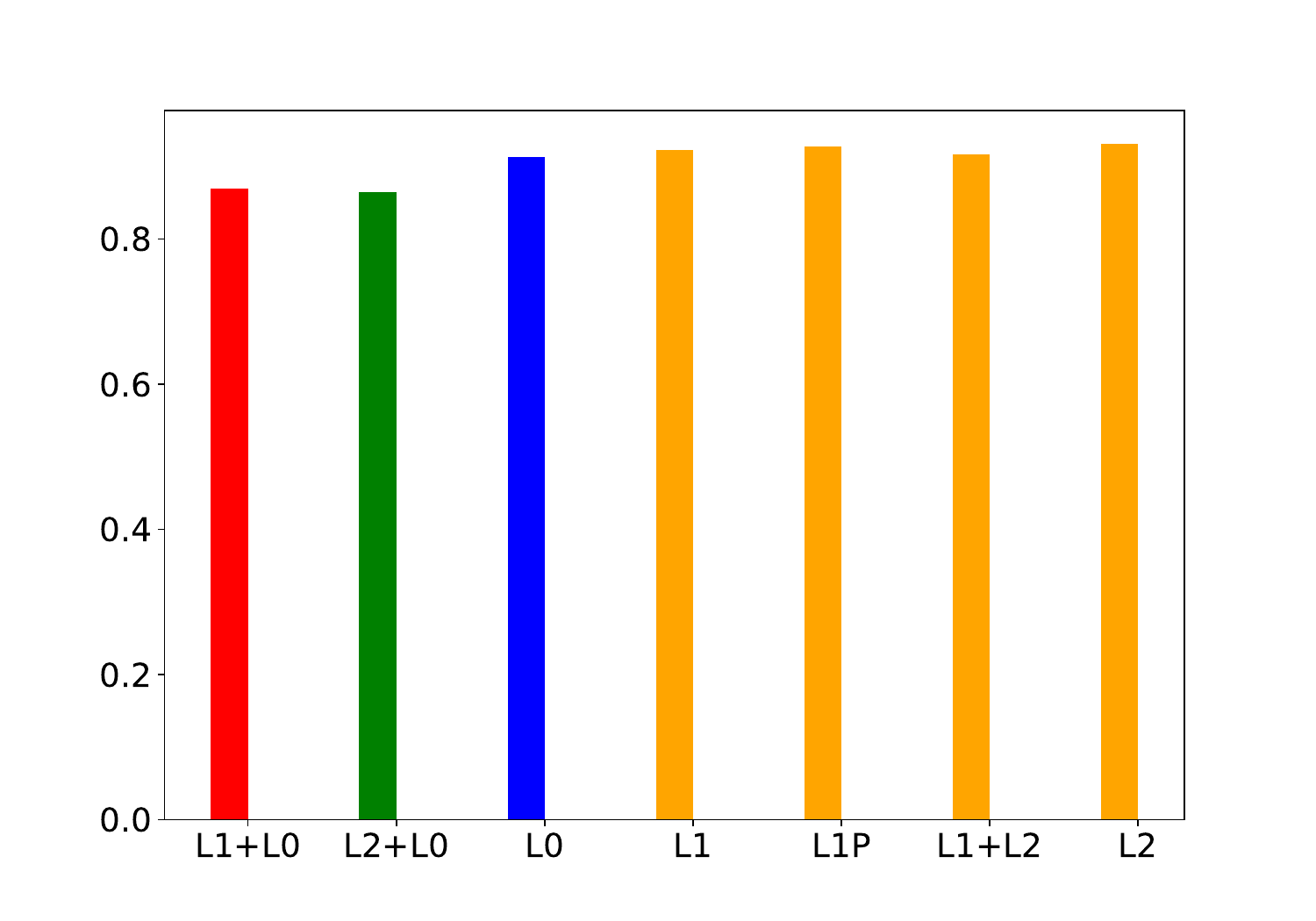}&	
		\includegraphics[width=0.22\textwidth,height=0.18\textheight,  trim =0.5cm 1cm 2.5cm 2cm, clip = true ]{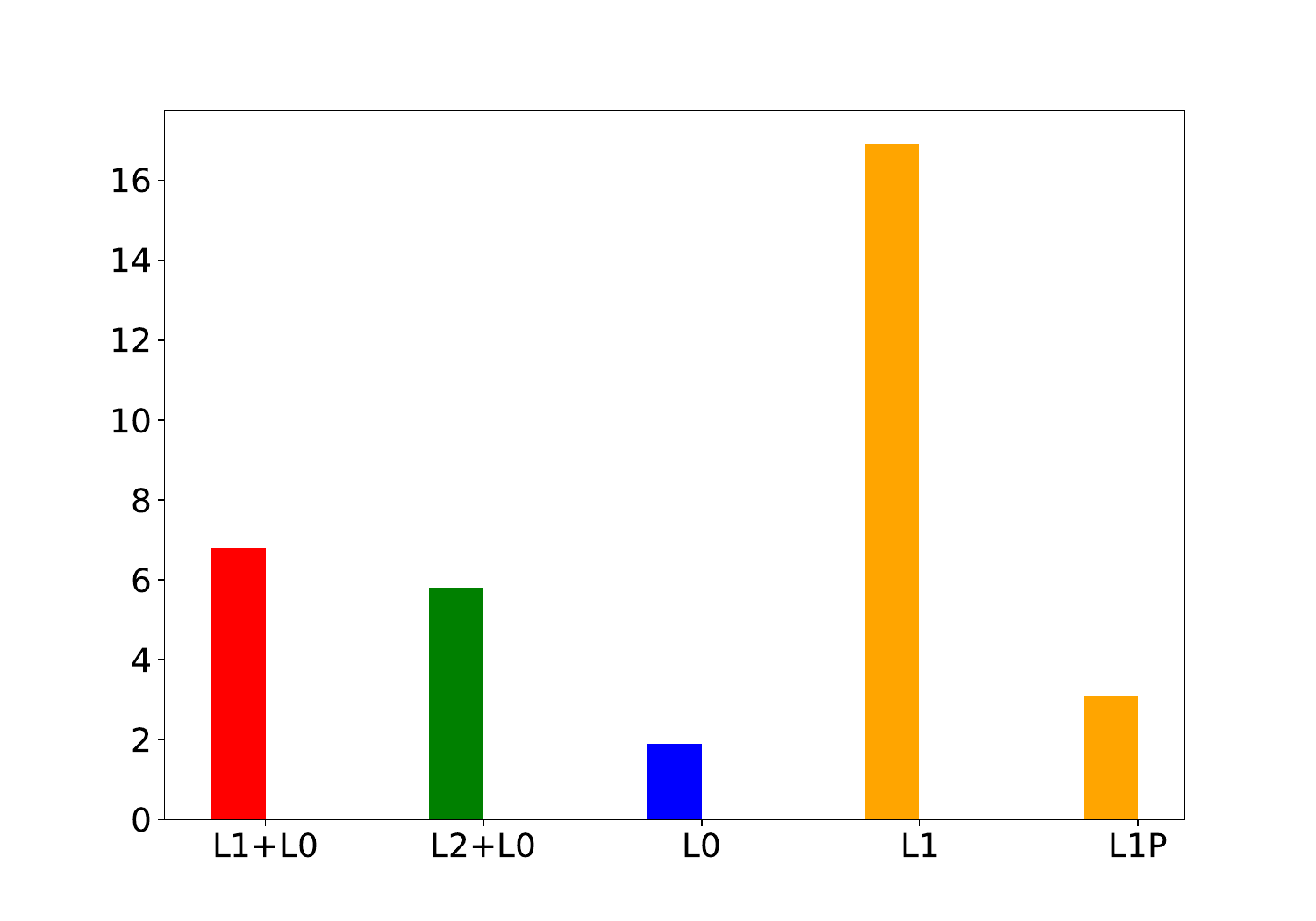}&
		\includegraphics[width=0.22\textwidth,height=0.18\textheight,  trim =.5cm 1cm 2.5cm 2cm, clip = true ]{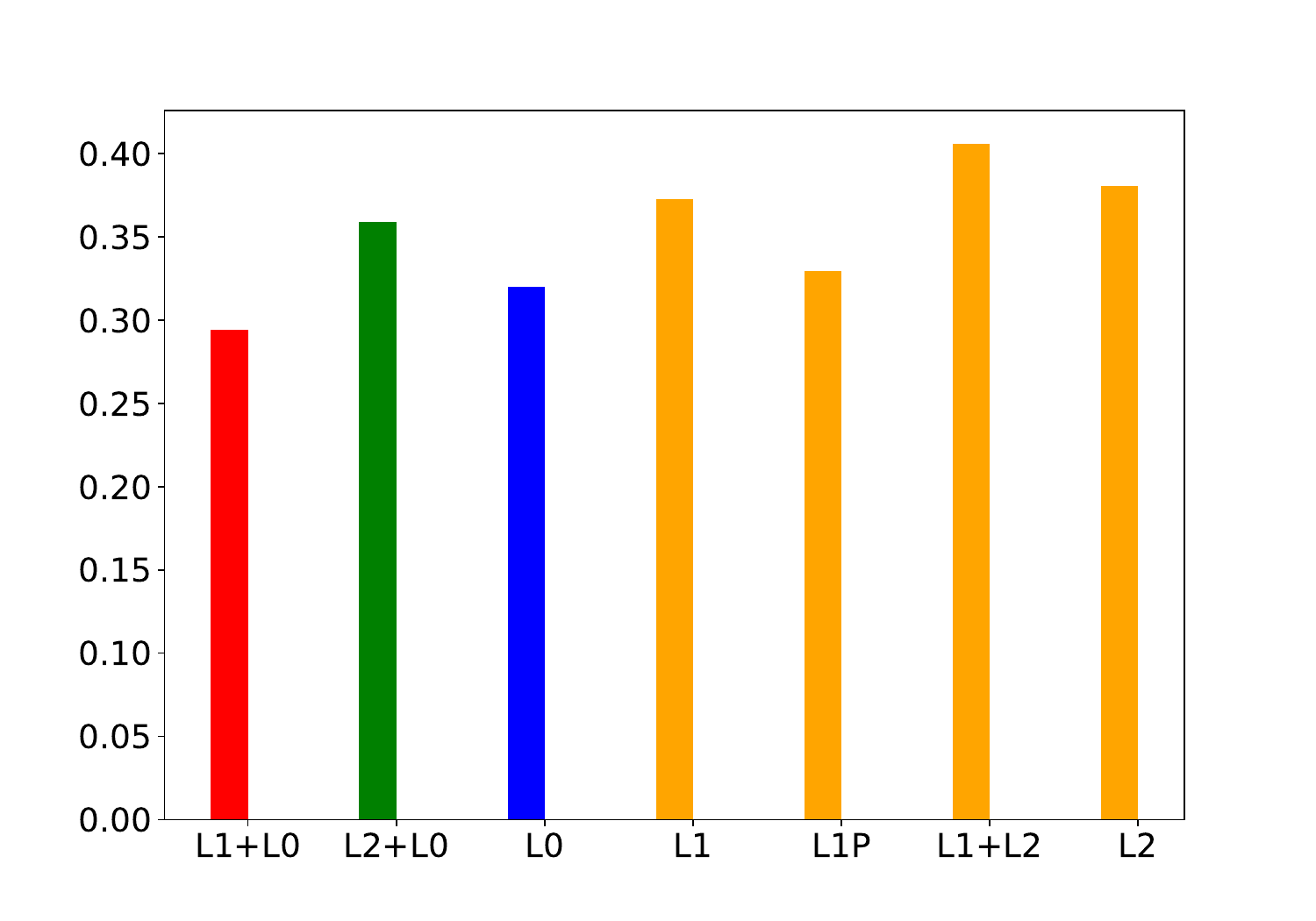}&
		\includegraphics[width=0.22\textwidth,height=0.18\textheight,  trim =.5cm 1cm 2.5cm 2cm, clip = true ]{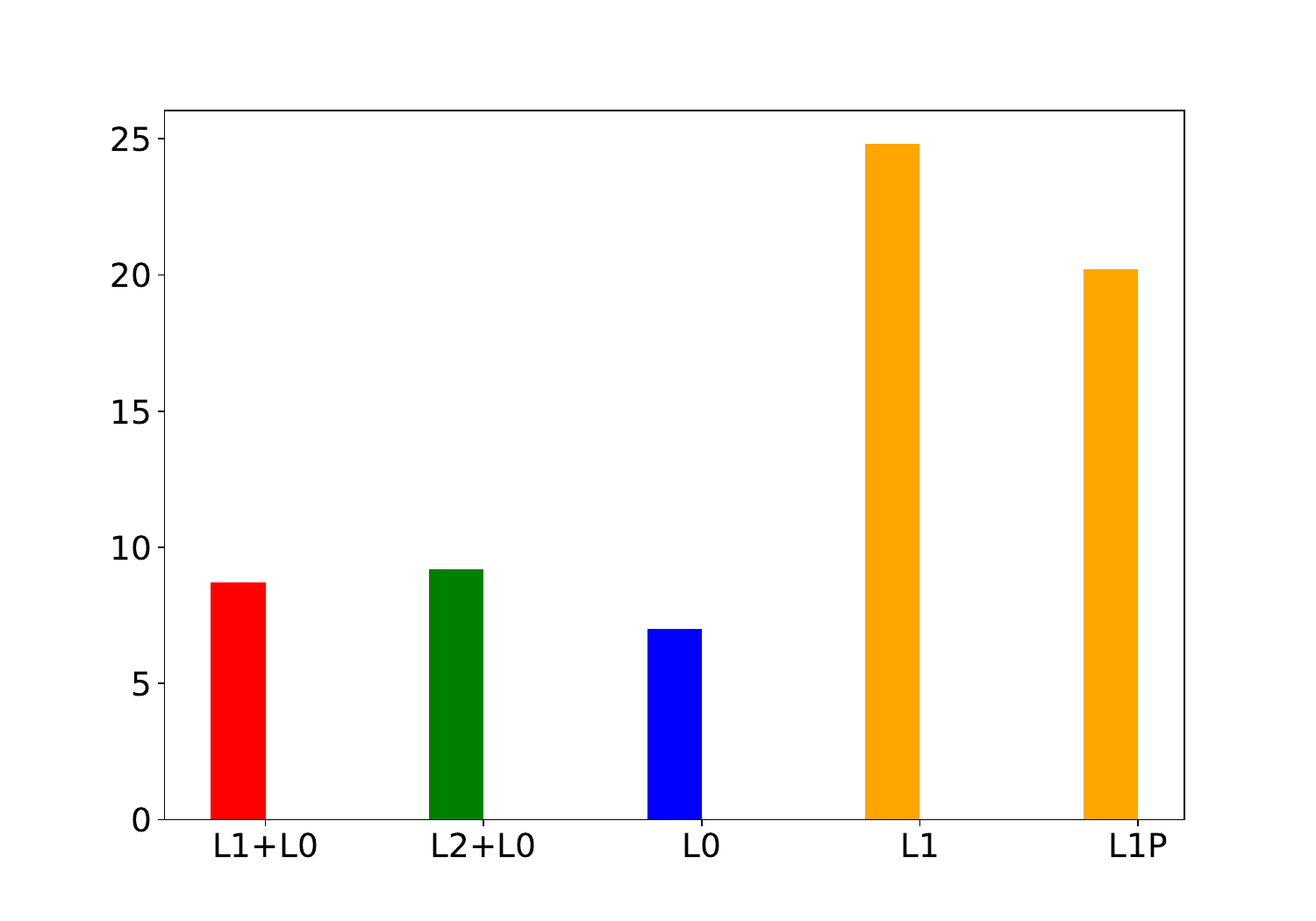}\\
				\multicolumn{2}{c}{\sf {\scriptsize{Leukemia: $n=35, p=2,000$}}} & \multicolumn{2}{c}{\sf {\scriptsize{Rat: $n=60, p=18,975$}} }  \\
		\sf {\scriptsize{Prediction error}} &\sf {\scriptsize{\# nonzeros }} &  \sf {\scriptsize{Prediction error}} &\sf {\scriptsize{\# nonzeros}}\\
		\includegraphics[width=0.22\textwidth,height=0.18\textheight,  trim = 0.5cm 1cm 2.5cm 2cm, clip = true ]{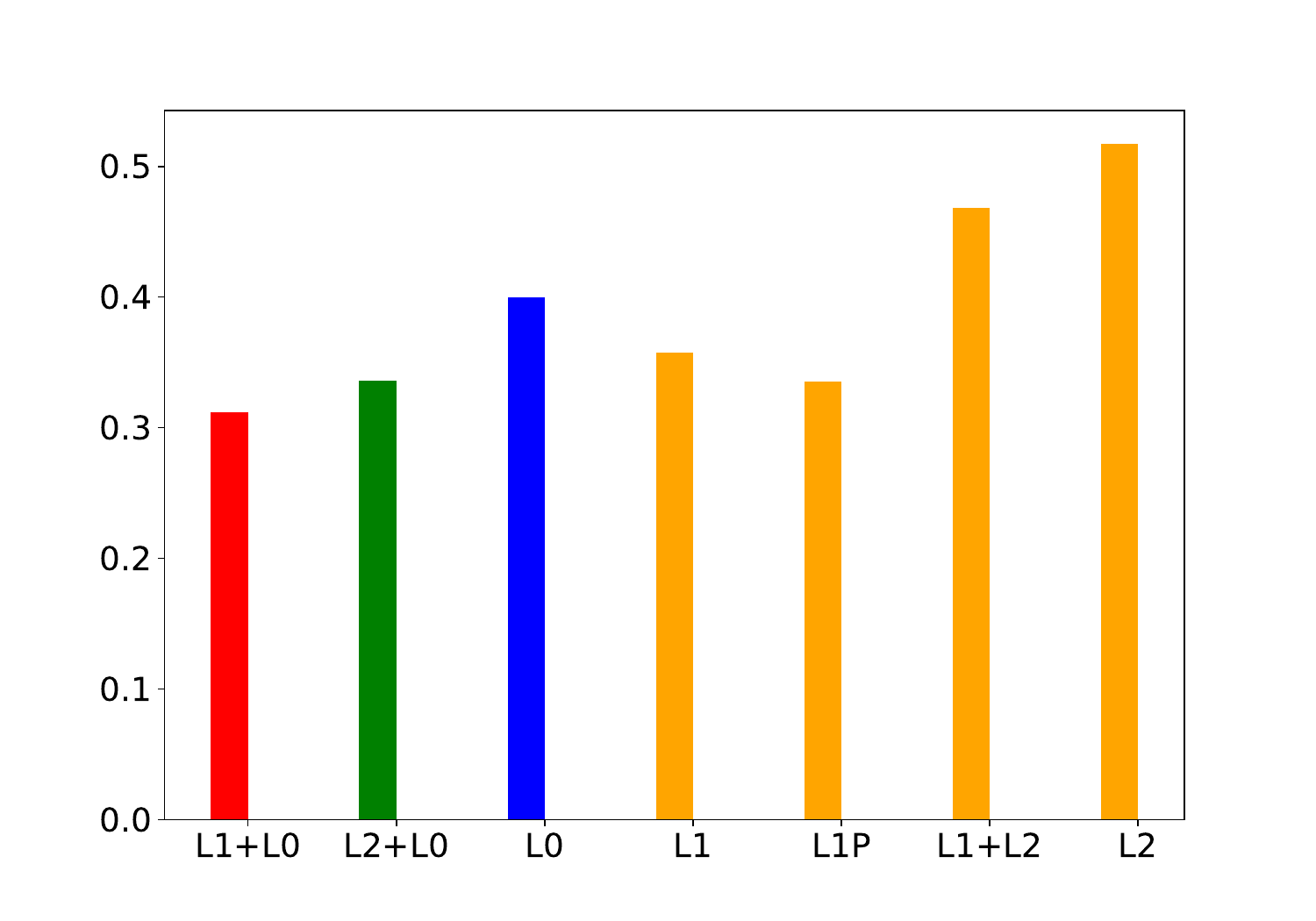}&
		\includegraphics[width=0.22\textwidth,height=0.18\textheight,  trim =0.5cm 1cm 2.5cm 2cm, clip = true ]{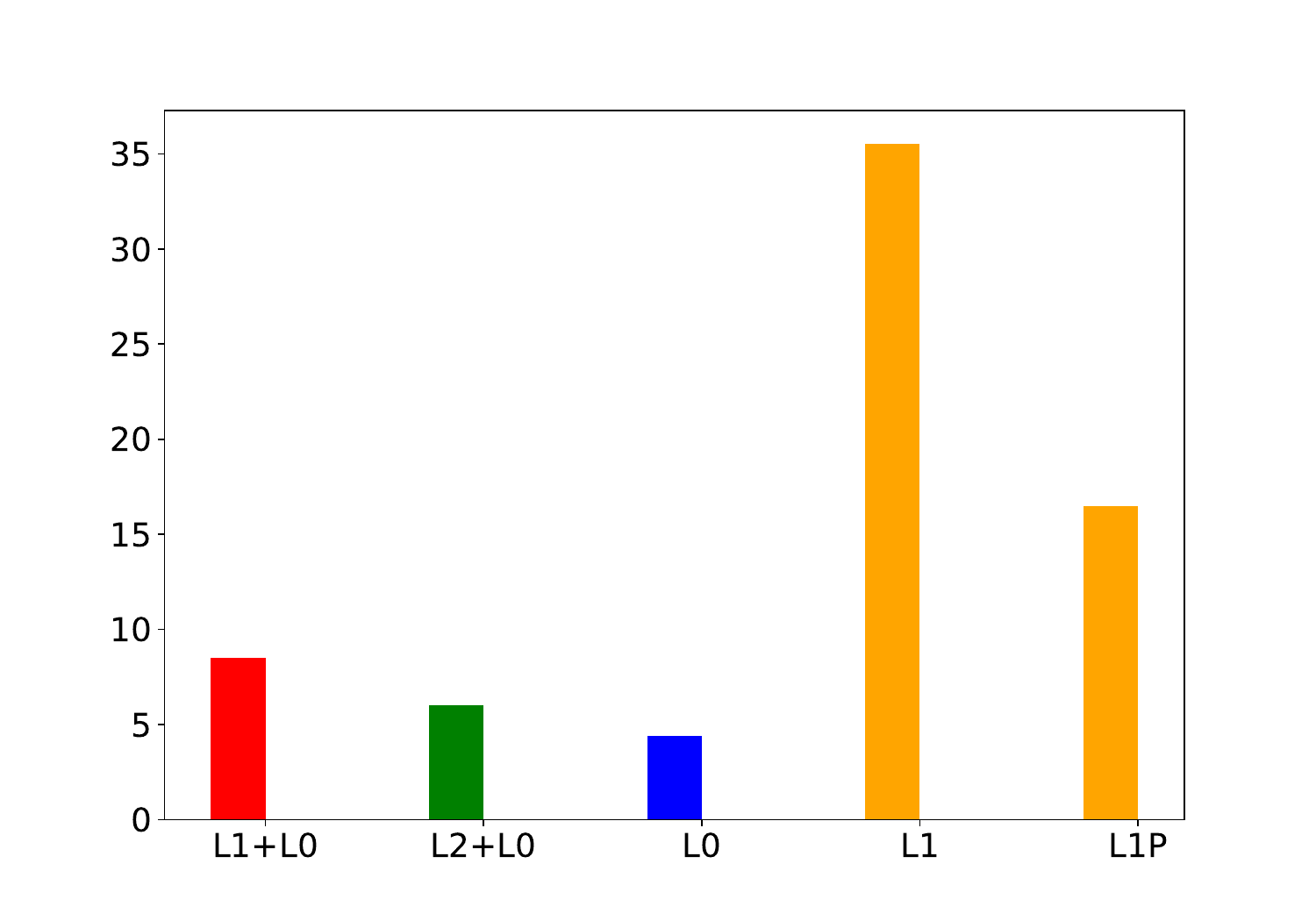}&
		\includegraphics[width=0.22\textwidth,height=0.18\textheight,  trim =.5cm 1cm 2.5cm 2cm, clip = true ]{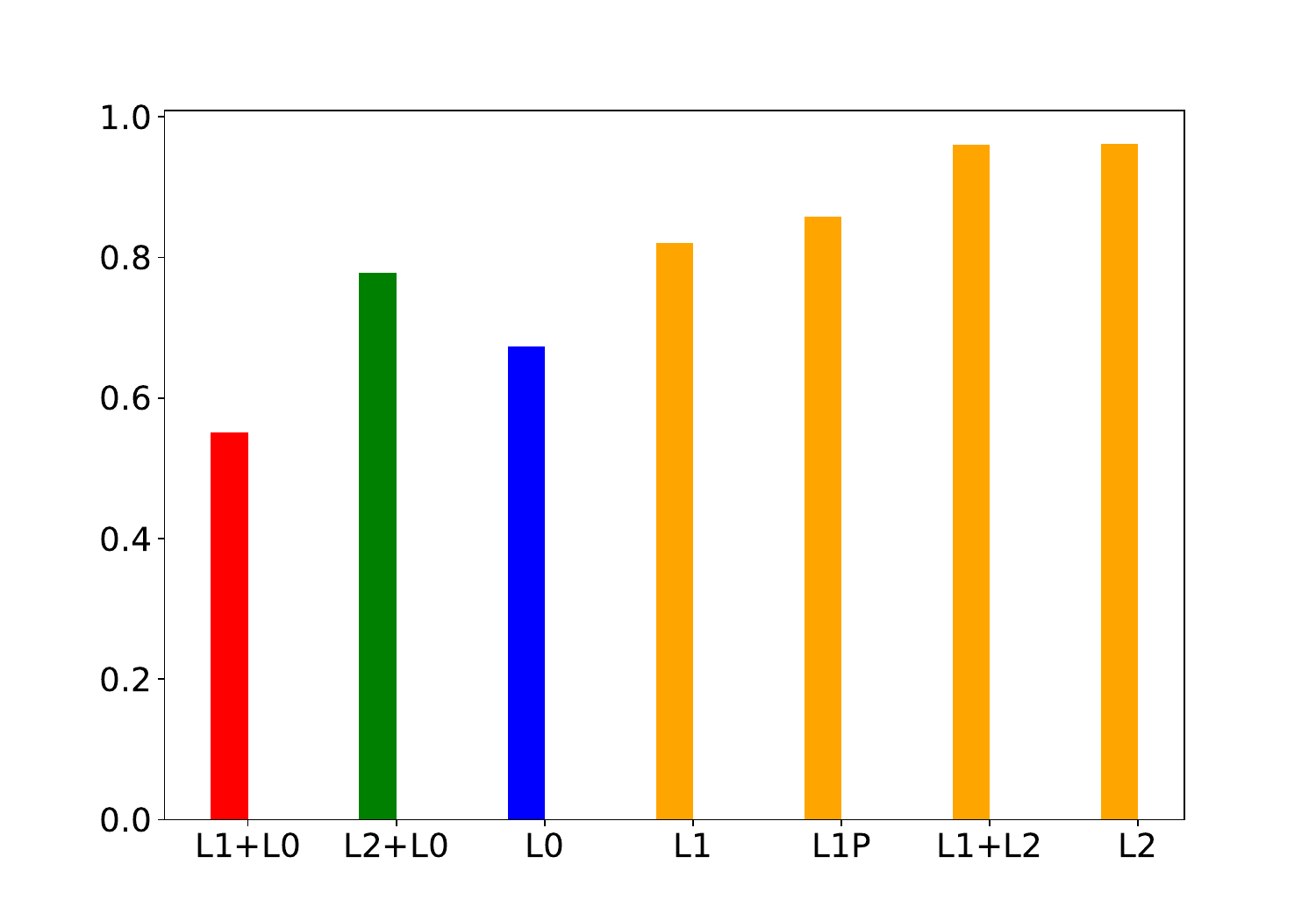}&
		\includegraphics[width=0.22\textwidth,height=0.18\textheight,  trim =.5cm 1cm 2.5cm 2cm, clip = true ]{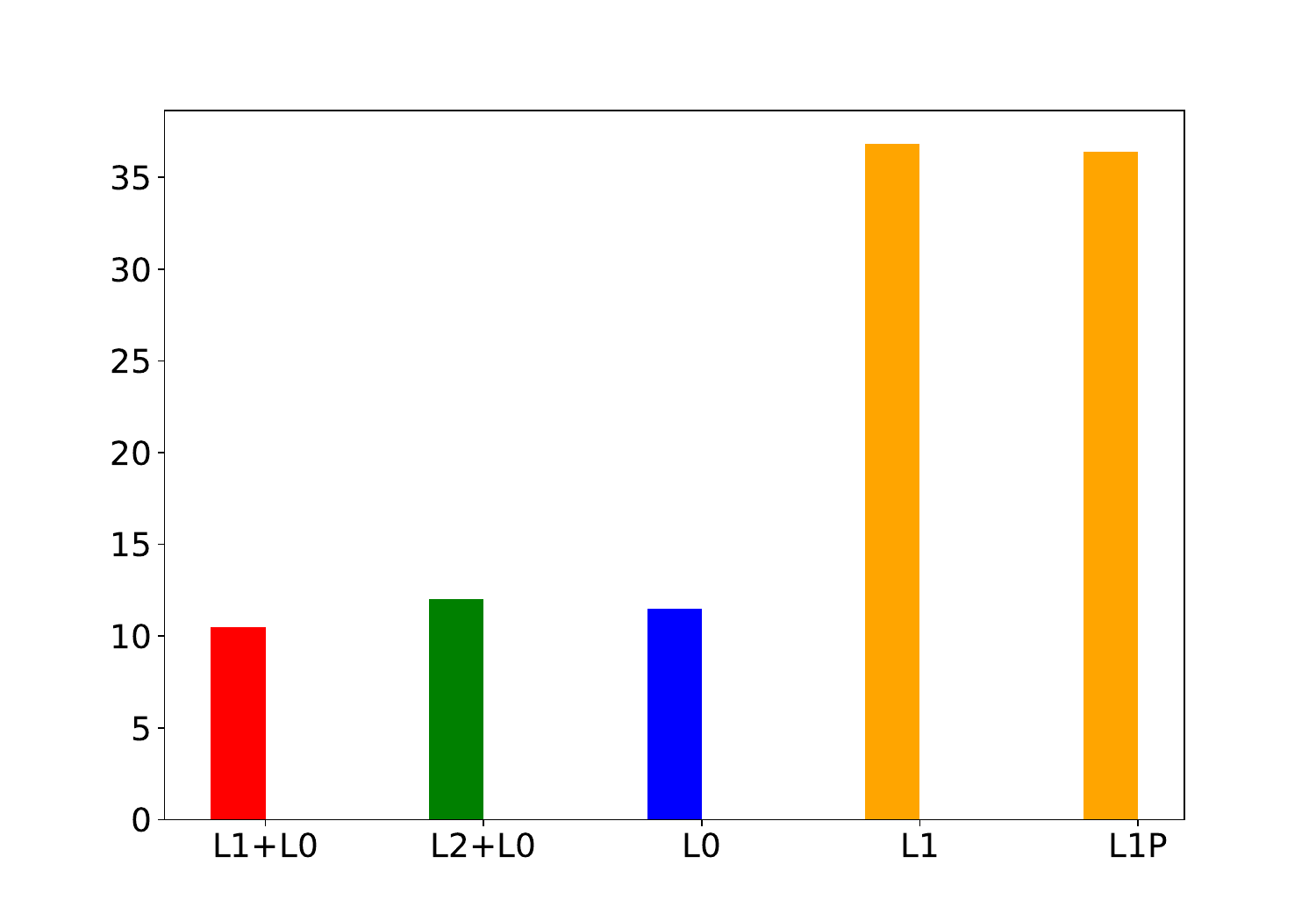}\\
	\end{tabular}
	\caption{ {\small{ Performance of the methods on four real datasets. We observe that pure L0 tended to underfit by selecting models that are overly sparse.  L1+L0/L2+L0 worked well both in terms of prediction and in terms of sparsity, when compared to the best available L1 models. L1 led to models with good predictive accuracy, but at the cost of a significant increase in density. } } }
	\label{fig: real}
\end{figure}

\vspace{-.8em}
\subsection{Real Datasets}

We now compare the performance of the methods on real datasets, as described below.

\noindent {\textbf{Triazine dataset}}  is taken from the \texttt{libsmv} website (\url{https://www.csie.ntu.edu.tw/~cjlin/libsvmtools/datasets/regression/triazines}). It contains $186$ observations and $60$ features, to which we added $500$ features generated as Gaussian noise.

\noindent {\textbf{Riboflavin dataset}}, taken from R package \texttt{hdi}, pertains to riboflavin production for $n=71$ observations of Bacillus subtilis.  Each observation contains $p=4088$ gene expression features.

\noindent {\textbf{Leukemia dataset}}, available at \url{http://cilab.ujn.edu.cn/datasets.htm}, is a classification dataset, with $72$ observations and $7129$ features.
We keep the top $2000$ features based on correlation screening and create a semi-synthetic response using $\mathbf{y} = \mathbf{\mathbf{X}}\B{\beta}^* +\B{\epsilon}$  with $\epsilon_{i} \stackrel{\text{iid}}{\sim}N(0,\sigma^2)$, where we set $\text{SNR}=4$ and let ${\beta}^*_{j} \in \{0,1\}$ with~10 randomly chosen coefficients set to~$1$.


\noindent {\textbf{Rat dataset}}. Using the same processing steps as~\cite{weng2013regularization}, we analyze the RNA from the eyes of $120$ twelve-week old male rats by considering $18,975$ probes expressed in the eye tissue. We thank Dr. Haolei Weng for providing the microarray dataset and the preprocessing code.

For each example, we standardize the features and the response.  We randomly split each dataset into new training and test sets, compute all the estimators and, for each method, keep the estimator with the best test accuracy.  Figure~\ref{fig: real} displays the results averaged over 10 random splits.

%% file: Supplement_July2021.tex
%
%

\section*{Supplementary Material  for ``Subset Selection with Shrinkage: Sparse Linear Modeling when the SNR is Low''}

\section{Computational details}

\subsection{Proof of Proposition~\ref{prop-conv-1}}\label{proof-prop-dfo-1}
\begin{enumerate}
\item[(a)] It follows from~\eqref{ubound-1} that for any $\B\beta$ satisfying $\|\B\beta\|_{0} \leq k$:
\begin{align}
F(\B\beta)&= Q_{L}(\B\beta, \B\beta) + \lambda \| \B\beta \|_{q} &  \nonumber \\
 &\geq \inf_{ \|\B\eta\|_{0} \leq {\mmk}  } \;\; \left(Q_{L}(\B\eta, \B\beta) + \lambda \| \B\eta \|_{q} \right) & \nonumber \\
 &= \inf_{  \|\B\eta\|_{0} \leq {\mmk}   } \;\; \left ( \frac{L}{2} \| \B\eta - \B\beta \|_2^2 + \langle \nabla f(\B\beta), \B\eta - \B\beta  \rangle + f(\B\beta)  + \lambda \| \B\eta \|_{q} \right ) & \nonumber \\
&=  \inf_{ \|\B\eta\|_{0} \leq {\mmk}  } \;\; \left (  \frac{L}{2} \left\| \B\eta - \left ( \B\beta - \frac{1}{L} \nabla f(\B\beta) \right) \right\|_2^2
- \frac{1}{2L} \|\nabla f(\B\beta) \|_2^2  + f(\B\beta)+   \lambda \| \B\eta \|_{q}  \right)  & \label{proof-1-1-0}  \\
&=  \;\; \left (  \frac{L}{2} \left\| \widehat{\B\eta} - \left ( \B\beta - \frac{1}{L} \nabla f(\B\beta) \right) \right\|_2^2
- \frac{1}{2L} \|\nabla f(\B\beta) \|_2^2  + f(\B\beta) \right)  + \lambda \| \hat{\B\eta} \|_{q}. &  \label{proof-1-1-1}
\end{align}
Note that in~\eqref{proof-1-1-1} above we use the notation
$\hat{\B\eta}$ to denote a minimizer of~\eqref{proof-1-1-0}. We now follow the proof in Proposition~6 in \cite{bertsimas2015best} to arrive at:
\begin{equation}\label{suff-decrease-1}
F(\B\beta) \geq  \frac{L - L_0}{2} \left \|\widehat{\B\eta} - \B\beta \right\|_2^2 + F(\widehat{\B\eta}).
\end{equation}
In particular, using $\hat{\B\eta} = \B\beta^{(m+1)}$, $\B\beta = \B\beta^{(m)}$ and $L \geq {L}_0$, we see that the sequence
$F(\B\beta^{(m)})$ is decreasing. Because $F(\B\beta) \geq 0$, we observe that the sequence $F(\B\beta^{(m)})$ converges to some $F^* \geq 0$.
%

\item[(b)] Summing inequalities ~\eqref{suff-decrease-1} for $ 1 \leq m \leq M,$
we obtain
\begin{equation}
\sum_{m=1}^{M} \left ( F ( \B\beta^{(m)}  ) - F( \B\beta^{(m+1)} )  \right )  \geq
 \frac{L - {L}_0}{2}\sum_{m=1}^{M} \| \B\beta^{(m+1)} - \B\beta^{(m)}\|_2^2,
 \end{equation}
leading to
$$ F( \B\beta^{(1)} ) - F( \B\beta^{(M+1)} )  \geq  \frac{M (L - {L}_0) }{2}  \min_{m = 1, \ldots, M}  \| \B\beta^{(m+1)} - \B\beta^{(m)}\|_2^2.$$
Because the decreasing sequence $F(\B\beta^{(m)})$ converges to  $F(\B\beta^*)=F^*$, say, we  arrive at the conclusion in Part (b).

\end{enumerate}

\subsection{Stronger formulations: adding implied inequalities}\label{strong-formu-1}
We use the following notation for the model matrix: $\M{X}=[\M{x}_{1}, \ldots, \M{x}_{p}]$.
We consider a structured version of Problem~\eqref{mio-form-2} with additional implied inequalities (cuts) for improved lower bounds:
\begin{subequations}\label{mio-form-21}
\begin{align}
\mini& ~~ \frac{u}{2}  + \lambda v \nonumber  \\
\sbt&~~  \| \M{y} - \M{X} \B\beta\|_{2}^2 \leq u \label{obj-con1-1}\\
&  \| \B\beta \|_{q} \leq v \label{obj-con11-1}\\
 & -{\mathcal M}_{j} z_{j} \leq  \beta_{j} \leq {\mathcal M}_{j} z_{j}, j \in [p]  \nonumber\\
 & z_{j} \in \{ 0 , 1 \}, j \in [p] \nonumber \\
 & \sum\limits_{j} z_{j} = k \nonumber\\
 &  - \mm_{i} \leq  \beta_{i} \leq \mm_{i}, i \in [p] \label{obj-con12-1} \\
& -\bar{\mm}^{-}_{i}  \leq  \langle \M{x}_{i}, \B\beta \rangle  \le \bar{\mm}^{+}_{i}, i \in [n] \label{obj-con13-1}\\
 & \|\B\beta \|_{1} \leq \mm_{\ell_1}, \label{obj-con14-1}
\end{align}
\end{subequations}
where (a) $\mm_{i}, i \in [p]$ denote bounds on $\beta_{i}$'s via  constraint~\eqref{obj-con12-1};
(b) $-\bar{\mm}^{-}_{i},\bar{\mm}^{+}_{i}$ denote bounds on the predicted values $\langle \M{x}_{i}, \B\beta\rangle$ for $i \in [n]$ via constraint~\eqref{obj-con13-1};
(c) $\mm_{\ell_1}$, in constraint~\eqref{obj-con14-1}, denotes an upper bound on the $\ell_{1}$-norm of the regression coefficients $\|\B\beta \|_{1}$.

The additional cuts in Problem~\eqref{mio-form-21} help the progress of the MIO solver -- the implied inequalities rule out several fractional solutions, thereby helping in obtaining superior lower bounds within a fixed computational budget.   The caveat, however, is that the resulting formulation has additional variables -- hence more work needs to be done within every node of the branch-and-bound tree.
Section~\ref{bigM-compute1} presents ways  to compute these bounds -- Section~\ref{sec-bigm-CO-1} describes ways to compute them via convex optimization -- these are bounds implied
by an optimal solution to Problem~\eqref{best-subset-penalized}. Section~\ref{sec-bigm-DFO} describes ways to compute these bounds based on good heuristic solutions.


\subsection{Computing problem specific parameters}\label{bigM-compute1}
\subsubsection{Computing parameters via convex optimization}\label{sec-bigm-CO-1}
Formulation~\eqref{mio-form-1} involves a BigM value $\mm$ -- tighter formulations can be obtained by using variable dependent BigM values for the $\beta_{i}$:
$$  - \mm_{i} z_{i}  \leq \beta_{i} \leq \mm_{i} z_{i}, ~~~~ i \in [p]. $$
In addition, implied constraints (or bounds) on $\langle \M{x}_{i}, \B\beta \rangle$'s can also be added:
$$ -  \bar{\mm}_{i} \leq  \langle \M{x}_{i}, \B\beta  \rangle \leq \bar{\mm}_{i}, ~~~ i \in [n].$$
We discuss how to compute these from data using convex optimization. Note that, because $\B\beta$ is $k$-sparse, we have
$|\langle \M{x}_{i}, \B\beta \rangle| \leq   \mm \| \M{x}_{i} \|_{k,1},$
where for a vector $\M{a} \in \mathbb{R}^{p}$ the quantity $\| \M{a} \|_{k,1}$ denotes the $\ell_{1}$-norm of the $k$-largest (in absolute value) entries of $\M{a}$.
We can set $\bar{\mm}_{i}  \leq  \mm \| \M{x}_{i} \|_{k,1}$. Note also that $ \| \B\beta \|_{1} \leq \mm k:=\mm_{\ell_1}$.
We now upper bound each coefficient $\beta_i$ by solving the quadratic optimization problems:
\begin{equation}\label{bound-bigM-1}
\begin{myarray}[1.3]{r c }
 \mm_i^+ =   \max&~~~ \beta_i   \\
 \sbt & \frac{1}{2} \|\mathbf{y} - \mathbf{X}\B{\beta} \|_2^2 + \lambda \| \B{\beta} \|_q \le \text{UB}\\
  & \| \B{\beta} \|_{\infty} \le \mathcal{M} \\
  & \| \B\beta \|_{1} \leq \mm_{\ell_1}\\
  & -\bar{\mm}^{-}_{i}  \leq  \langle \M{x}_{i}, \B\beta \rangle  \le \bar{\mm}^{+}_{i}, i \in [n]
\end{myarray}~~~~~~~~~
\begin{myarray}[1.3]{r c }
 \mm_i^- =   \max&~~~ -\beta_i   \\
 \sbt & \frac{1}{2} \|\mathbf{y} - \mathbf{X}\B{\beta} \|_2^2 + \lambda \| \B{\beta} \|_q \le \text{UB}\\
  & \| \B{\beta} \|_{\infty} \le \mathcal{M} \\
    & \| \B\beta \|_{1} \leq \mm_{\ell_1}\\
  &-\bar{\mm}^{-}_{i} \leq \langle \M{x}_{i}, \B\beta \rangle  \le \bar{\mm}^{+}_{i}, i \in [n]
\end{myarray}
\end{equation}
where $\text{UB}$ is an upper bound to Problem~\eqref{best-subset-penalized} obtained via Algorithm~1, for example.
Upon solving Problem~\eqref{bound-bigM-1}, we set $\mm_{i} = \max \{ \mm_i^+,  \mm_i^-   \}$ for all $i \in [p]$.
Consequently, we can update the bounds
$\mm = \| \mm_{i} \|_{\infty}$, $\bar{\mm}_{i}$ and $\mm_{\ell_{1}}$ -- such bound tightening methods have been proposed in~\cite{mazumder2015discrete}
in the context of the Discrete Dantzig Selector problem.

Similarly, we can also obtain bounds on $\langle \M{x}_{j}, \B\beta\rangle$ by solving the following pair of optimization problems for all $j \in [n]$.
 \begin{equation}\label{bound-bigM-2}
\begin{myarray}[1.3]{r c }
 \bar{\mm}_j^+ =   \max&~~~ \langle \M{x}_{j}, \B\beta \rangle   \\
 \sbt & \frac{1}{2} \|\mathbf{y} - \mathbf{X}\B{\beta} \|_2^2 + \lambda \| \B{\beta} \|_q \le \text{UB}\\
  &  - \mathcal{M}^{-}_{i} \leq {\beta}_{i}  \le \mathcal{M}^{+}_{i}, i \in [p] \\
  & \| \B\beta \|_{1} \leq \mm_{\ell_1}\\
& -\bar{\mm}^{-}_{i}  \leq  \langle \M{x}_{i}, \B\beta \rangle  \le \bar{\mm}^{+}_{i}, i \in [n]
\end{myarray}~~~~~~~~~
\begin{myarray}[1.3]{r c }
 \bar{\mm}_j^- =   \max&~~~ -\langle \M{x}_{j}, \B\beta \rangle \\
 \sbt & \frac{1}{2} \|\mathbf{y} - \mathbf{X}\B{\beta} \|_2^2 + \lambda \| \B{\beta} \|_q \le \text{UB}\\
  &  - \mathcal{M}^{-}_{i} \leq {\beta}_{i}  \le \mathcal{M}^{+}_{i}, i \in [p] \\
    & \| \B\beta \|_{1} \leq \mm_{\ell_1}\\
& -\bar{\mm}^{-}_{i}  \leq  \langle \M{x}_{i}, \B\beta \rangle  \le \bar{\mm}^{+}_{i}, i \in [n].
\end{myarray}
\end{equation}
Upon solving Problem~\eqref{bound-bigM-2}, we can set $\bar{\mm}_{i} = \max \{ | \bar{\mm}_j^+|, | \bar{\mm}_j^-|\}$.
The bounds thus obtained can be used to tighten the bounds used in Problems~\eqref{bound-bigM-1} and~\eqref{bound-bigM-2}.
New bounds on $\{\mm_{i}\}$ and $\{\bar{\mm}_{i}\}$ can be obtained by solving the new problems with the updated bounds.

\medskip

\begin{rem}
Problems~\eqref{bound-bigM-1},~\eqref{bound-bigM-2} drop the cardinality constraint on $\B\beta$ -- hence the derived bounds
need not be tight, i.e., $\mm_{i} > |\hat{{\beta}}_{i}(\lambda; k)|$, where $ \hat{\B\beta}(\lambda; k)$ denotes an optimal solution to Problem~\eqref{best-subset-penalized}.
\end{rem}


\subsubsection{Computing parameters via Algorithm~1}\label{sec-bigm-DFO}
We note that the BigM values $\mm_{i}, i \in [p]$ can also be based on the solutions obtained from the heuristic algorithms. For example, we can set
$\mm_{i} = \tau \| \hat{\B\beta}(\lambda; k)\|_{\infty}$ for all $i \in [p]$ for some multiplier $\tau \in \{ 1.5, 2\}$, for example. Similarly, the bounds $\bar{\mm}_{i}$ can be set to
$\tau |\langle \M{x}_{i}, \hat{\B\beta}(\lambda; k)\rangle|$ for all $i \in [n]$.  Such bounds are usually tighter and are obtained as a simple by-product of Algorithm~1.

\section{Proofs of the results in Section~\ref{sec:theory}}

\subsection{Proof of Theorem~\ref{probs.thm}}

We first note that the probability of event $\mathcal{F}$ is at least $1-\delta_0/2$ by Theorem~4.1 in \cite{bellec2018slope}.  Next, we establish the probability bound for~$\mathcal{E}_s$. 

Because the columns of~$\bX$ have unit Euclidean norm, we can write $\|\bX\bu\|\le\|\bu\|_1\le \sqrt{s}\|\bu\|$ for every $\bu\in B_0(s)$.  Hence, taking $\delta_0=s/(2ep)$, we derive
\begin{equation}
\label{G.ineq}
\sqrt{\log(1/\delta_0)}\|\bX\bu\|\le\sqrt{s\log(2ep/s)}\|\bu\|.
\end{equation}
It follows from Stirling's formula that $\log(s!)\ge s\log(s/e)$, and hence
\begin{equation*}
\sum_{j=1}^s\log(2p/j)=s\log(2p)-\log(s!)\le s\log(2ep/s).
\end{equation*}
Thus, using the Cauchy-Schwarz inequality and taking into account $\|\bu\|_0\le s$, we arrive at
\begin{equation}
\label{H.ineq}
\sum_{j=1}^p u_j^\sharp\sqrt{\log(2p/j)}\le \|\bu\|\sqrt{\sum_{j=1}^s\log(2p/j)}
\le \sqrt{s\log(2ep/s)}\|\bu\|.
\end{equation}
Inequalities~(\ref{H.ineq}) and~(\ref{G.ineq}) yield
\begin{equation*}
[4+\sqrt{2}]\sigma\max\Big(\sum_{j=1}^p u_j^\sharp\sqrt{\log(2p/j)},\sqrt{\log(1/\delta_0)}\|\bX\bu\|\Big) \le [4+\sqrt{2}]\sigma\sqrt{s\log(2ep/s)}\|\bu\|.
\end{equation*}
Consequently, when $\delta_0=s/(2ep)$, we have $\mathcal{F}\subseteq\mathcal{E}_s$.  Because the probability of event $\mathcal{F}$ is at least $1-s/(4ep)$, we have established the stated probability bound for~$\mathcal{E}_s$.

The result for~$\mathcal{H}$ follows from the standard tail probability bounds for maxima of Gaussian random variables (for example, those in \cite{buhlmann2011statistics}).  The result for~$\mathcal{G}_s$ follows from the argument in the proof of Lemma~8 in \cite{raskutti2011minimax}, with appropriate modifications in order to incorporate the uncertainty parameter~$\delta_0$.

\subsection{Proof of Theorem~\ref{slow.rate.thm}}

We consider an arbitrary~$\bbeta\in B_0(k)$ and note that
\begin{equation*}
\|\bY-\bX\hat\bbeta_2\|^2+\lambda\|\hat\bbeta_2\|\le \|\bY-\bX\bbeta\|^2+\lambda\|\bbeta\|,
\end{equation*}
which implies
\begin{equation}
\label{basic.ineq}
\|\bff^*-\bX\hat\bbeta_2\|^2+\lambda\|\hat\bbeta_2\|\le \|\bff^*-\bX\bbeta\|^2+2\bepsilon^\top\bX(\hat\bbeta_2-\bbeta)+\lambda\|\bbeta\|.
\end{equation}
We will derive prediction error bounds for~$\hat\bbeta_2$ by controlling the term ~$\bepsilon^\top\bX(\hat\bbeta_2-\bbeta)$. 

We first focus on establishing the slow rate. On the event~$\mathcal{E}_{2k}$ we have
\begin{equation}
\label{cross.prod.ineq}
  \bepsilon^\top\bX(\hat\bbeta_2-\bbeta)\le [4+\sqrt{2}]\sigma\sqrt{2k\log(ep/k)}\|\bbeta-\hat\bbeta_2\|.
\end{equation}
Combining this inequality with~(\ref{basic.ineq}) and using the lower bound imposed on~$\lambda$, we derive
\begin{equation}
\label{main.slow.rate.bnd}
\|\bff^*-\bX\hat\bbeta_2\|^2\le \|\bff^*-\bX\bbeta\|^2+2\lambda\|\bbeta\|.
\end{equation}
Thus, we have established the first slow rate prediction error bound.

Repeating the arguments in the proof of Theorem~\ref{probs.thm}, we see that on the event~$\mathcal{F}$ we have either (a) inequality~(\ref{cross.prod.ineq}), which implies~(\ref{main.slow.rate.bnd}), or (b) the following inequality:
\begin{equation}
\label{prelim.in.prod.bnd}
  \bepsilon^\top\bX(\hat\bbeta_2-\bbeta)\le [4+\sqrt{2}]\sigma\sqrt{\log(1/\delta_0)}\|\bX(\bbeta-\hat\bbeta_2)\|,  
\end{equation}
which implies
\begin{equation}
\label{2nd.slow.rate.bnd.prelim}
\|\bff^*-\bX\hat\bbeta_2\|^2\le \|\bff^*-\bX\bbeta\|^2+\lambda\|\bbeta\| + 2[4+\sqrt{2}]\sigma\sqrt{\log(1/\delta_0)}\big(\|\bff^*-\bX\hat\bbeta_2\|+\|\bff^*-\bX\bbeta\|\big).
\end{equation}
We bound the last term in the above display by two applications of the inequality
\begin{equation}
\label{ab.ineq}
2ab\le \alpha a^2+\alpha^{-1}b^2,
\end{equation}
which holds for every~$\alpha>0$ and $a,b\in\mathbb{R}$.  Setting~$\alpha=2$, we derive inequalities
\begin{equation*}
2[4+\sqrt{2}]\sigma\sqrt{\log(1/\delta_0)}\|\bff^*-\bX\hat\bbeta_2\|\le 2[4+\sqrt{2}]^2\sigma^2{\log(1/\delta_0)}+\|\bff^*-\bX\hat\bbeta_2\|^2/2
\end{equation*}
and
\begin{equation*}
\sigma\sqrt{\log(1/\delta_0)}\|\bff^*-\bX\bbeta\|\lesssim \sigma^2{\log(1/\delta_0)}+\|\bff^*-\bX\bbeta\|^2.
\end{equation*}
Taking into account inequality~(\ref{2nd.slow.rate.bnd.prelim}), we then arrive at the second slow rate prediction error bound:
\begin{equation*}
\|\bff^*-\bX\hat\bbeta_2\|^2\lesssim \|\bff^*-\bX\bbeta\|^2+\lambda\|\bbeta\| + \sigma^2\log(1/\delta_0).
\end{equation*}

We now establish the fast rate.  Starting with inequality~(\ref{basic.ineq}) and restricting our attention to event~$\mathcal{G}_{2k}$, we derive
\begin{equation*}
\|\bff^*-\bX\hat\bbeta_2\|^2\le \|\bff^*-\bX\bbeta\|^2+2\sigma\big[10k\log(ep/[2k])+\log(1/\delta_0)\big]^{1/2}\big(\|\bff^*-\bX\hat\bbeta_2\|+\|\bff^*-\bX\bbeta\|\big)+\lambda\|\bbeta-\hat\bbeta_2\|.
\end{equation*}
We bound the second term on the right-hand side by two applications of inequality~(\ref{ab.ineq}), in which we set $\alpha=4$ in order to have $\|\bff^*-\bX\hat\bbeta_2\|^2$ appear with the multiplier~$1/4$.  We bound the last term on the right-hand side using
\begin{equation*}
\lambda\|\bbeta-\hat\bbeta_2\|\le \gamma_{2k}^{-1}\lambda\|\bX(\bbeta-\hat\bbeta_2)\|\le \gamma_{2k}^{-1}\lambda\big(\|\bff^*-\bX\hat\bbeta_2\|+\|\bff^*-\bX\bbeta\|\big),
\end{equation*}
 and then apply~(\ref{ab.ineq}) with $\alpha=2$ again to derive
\begin{equation*}
\gamma_{2k}^{-1}\lambda\|\bff^*-\bX\hat\bbeta_2\|\le \gamma_{2k}^{-2}\lambda^2 + \|\bff^*-\bX\hat\bbeta_2\|^2/4
\end{equation*}
and
\begin{equation*}
\gamma_{2k}^{-1}\lambda\|\bff^*-\bX\bbeta\|\lesssim \gamma_{2k}^{-2}\lambda^2+\|\bff^*-\bX\bbeta\|^2.
\end{equation*}
Rearranging the resulting terms we arrive at the fast rate prediction error bound:
\begin{equation*}
\|\bff^*-\bX\hat\bbeta_2\|^2\lesssim \|\bff^*-\bX\bbeta\|^2+\sigma^2 k\log(ep/[2k])+\gamma_{2k}^{-2}\lambda^2+\sigma^2\log(1/\delta_0).
\end{equation*}

\subsection{Proof of Corollary~\ref{lin.mod.l2}}
The first prediction error bound is a direct consequence of Theorems~\ref{probs.thm} and~\ref{slow.rate.thm}. The last two prediction error bounds are derived from Theorems~\ref{probs.thm} and the corresponding bounds in Theorem~\ref{slow.rate.thm} by setting $\delta_0=1/p$ and $\delta_0=(k/p)^k$, respectively.  The estimation error bound follows from the inequality $\gamma_{2k}^{2}\|\hat\bbeta_2-\bbeta^*\|^2\le \|\bX(\hat\bbeta_2-\bbeta^*)\|^2$.

{
\subsection{Proof of Corollary~\ref{lin.mod.opt.gap}}
We let~$Q(\bbeta)$ denote the the objective function in~(\ref{opt.problem.theory}) when~$q=2$. Because $UB=Q(\widetilde{\bbeta}_2)$, $LB\le Q(\bbeta^*)$, and $UB=LB/(1-\tau)$, we derive
\begin{equation*}
Q(\widetilde{\bbeta}_2)\le Q(\bbeta^*)/(1-\tau).
\end{equation*}
Because $Q(\widetilde{\bbeta}_2)=\|\bY-\bX\widetilde{\bbeta}_2\|^2+\lambda\|\widetilde{\bbeta}\|$ and $Q(\bbeta^*)= \|\bepsilon\|^2+\lambda\|\bbeta^*\|$, we then have
\begin{equation*}
\|\bY-\bX\widetilde{\bbeta}_2\|^2+\lambda\|\widetilde{\bbeta}\le \|\bepsilon\|^2/(1-\tau)+\lambda\|\bbeta^*\|/(1-\tau).
\end{equation*}
Repeating the arguments in the proof of the second slow rate in Theorem~\ref{slow.rate.thm} while incorporating the optimality gap, we derive that
\begin{equation}
\label{prf.opt.gap.bnd1}
\|\bff^*-\bX\widetilde{\bbeta}_2\|^2\le 2\lambda\|\bbeta^*\|/(1-\tau) + 4[4+\sqrt{2}]^2\sigma^2{\log(1/\delta_0)}+2\|\bepsilon\|^2\tau/(1-\tau)
\end{equation}
on the event~$\mathcal{F}$. Standard chi-square tail bounds imply that, with an appropriate multiplicative constant, inequality $\|\bepsilon\|^2\lesssim \sigma^2[n\vee\log(p)]$ holds with probability at least $1-1/(2p)$. Letting $\delta_0=1/(2p)$, noting $\tau\le1$, and recalling that $1/(1-\tau)$ is upper-bounded by a universal constant, we then conclude that inequality
\begin{equation*}
\|\bff^*-\bX\widetilde{\bbeta}_2\|^2\lesssim \lambda\|\bbeta^*\| + \sigma^2[\log(p)+\tau n]
\end{equation*}
holds with probability at least $1-1/p$, establishing the first error bound in Corollary~\ref{lin.mod.opt.gap}.

Revisiting inequality~(\ref{prf.opt.gap.bnd1}) with $\delta_0=1/p$, we note that (as $n\rightarrow\infty$) the right-hand side is of the order
\begin{equation*}
2\lambda\|\bbeta^*\|\Big(1+\frac{\tau}{1-\tau}\Big) + 4[4+\sqrt{2}]^2\sigma^2{\log(p)}\left\{1+\frac{\tau n}{2[4+\sqrt{2}]^2(1-\tau)\log(p)}\right\}.
\end{equation*}
Thus, the multiplicative increase in the error bound relative to the case $\tau=0$ is at most
\begin{equation*}
1+\frac{\tau}{1-\tau}\left\{1\vee\frac{n}{58\log(p)}\right\}.
\end{equation*}

We now focus on the second error bound in Corollary~\ref{lin.mod.opt.gap}. Repeating the arguments in the proof of the fast rate in Theorem~\ref{slow.rate.thm}, incorporating the optimality gap, and keeping track of the constants, we arrive at the following error bound:
\begin{equation}
\label{prf.opt.gap.bnd2}
\|\bff^*-\bX\hat\bbeta_2\|^2\le 8\sigma^2\big[10k\log(ep/[2k])+\log(1/\delta_0)\big]+2\gamma_{2k}^{-2}\lambda^2\Big(1+\frac{\tau}{1-\tau}\Big)+2\|\bepsilon\|^2\tau/(1-\tau),
\end{equation}
which holds on the event~$\mathcal{G}_{2k}$. Letting $\delta_0=(k/p)^k/2$, and again using the chi-square tail bounds to control $\|\bepsilon\|^2$, we then conclude that inequality
\begin{equation*}
\|\bff^*-\bX\widetilde{\bbeta}_2\|^2\lesssim \sigma^2[k\log(ep/k)+\tau n]+\gamma_{2k}^{-2}\lambda^2
\end{equation*}
holds with probability at least $1-(k/p)^k$, establishing the second error bound in Corollary~\ref{lin.mod.opt.gap}.

Revisiting inequality~(\ref{prf.opt.gap.bnd2}) with $\delta_0=(k/p)^k$, we note that (as $n\rightarrow\infty$) the right-hand side is of the order
\begin{equation*}
88\sigma^2\log(ep/[2k])\left\{1+\frac{\tau n}{44(1-\tau)\log(ep/[2k])}\right\}
+2\gamma_{2k}^{-2}\lambda^2\Big(1+\frac{\tau}{1-\tau}\Big).
\end{equation*}
Thus, the multiplicative increase in the error bound relative to the case $\tau=0$ is at most
\begin{equation*}
1+\frac{\tau}{1-\tau}\left\{1\vee\frac{n}{43\log(ep/[2k])}\right\}.
\end{equation*}
}

\subsection{Proof of Corollary~\ref{Expect.slow.rate}}

Let~$c_0$ be the universal constant from the second slow rate error bound in Theorem~\ref{slow.rate.thm}. Take an arbitrary~$\bbeta\in B_0(k)$ and define
\begin{equation*}
W=\|\bff^*-\bX\hat\bbeta_2\|^2 - c_0\|\bff^*-\bX\bbeta\|^2 -c_0\lambda\|\bbeta\|.
\end{equation*}
By Theorems~\ref{probs.thm} and~\ref{slow.rate.thm} we have $W\le c_0 \sigma^2\log(1/\delta_0)$ with probability at least $1-\delta_0/2$.  Thus,
\begin{equation*}
2\mathbb{P}\big(W>w\big)\le e^{-w/[c_0\sigma^2]},
\end{equation*}
for every non-negative~$w$.  Consequently,
\begin{equation*}
\mathbb{E}W \le \int_0^{\infty}\mathbb{P}\big(W>w\big)dw \le \tfrac12\int_0^{\infty}e^{-w/[c_0\sigma^2]}dw \le \frac{c_0\sigma^2}{2},
\end{equation*}
and the first stated bound follows from the definition of~$W$.

The second stated bound follows by an analogous argument, together with an additional observation that $k\log (ep/[2k])$ is bounded away from zero by a positive universal constant.

\subsection{Proof of Theorem~\ref{L1.rate.thm}}

We consider an arbitrary~$\bbeta\in B_0(k)$. In the~$\ell_1$ setting, inequality~(\ref{basic.ineq}) becomes
\begin{equation}
\label{basic.ineq2}
\|\bff^*-\bX\hat\bbeta_1\|^2+\lambda\|\hat\bbeta_1\|_1\le \|\bff^*-\bX\bbeta\|^2+2\bepsilon^\top\bX(\hat\bbeta_1-\bbeta)+\lambda\|\bbeta\|_1.
\end{equation}
On the event~$\mathcal{H}$, we then have 
\begin{equation*}
2\bepsilon^\top\bX(\hat\bbeta_1-\bbeta) \le 2\|\bX^T\bepsilon\|_{\infty}\Big[\|\bbeta\|_1+\|\hat\bbeta_1\|_1\Big] \le \lambda\Big[ \|\hat\bbeta_1\|_1 + \|\bbeta\|_1\Big].
\end{equation*}
Consequently,
\begin{equation*}
\|\bff^*-\bX\hat\bbeta_1\|^2 \le  \|\bff^*-\bX\bbeta\|^2 + 2\lambda\|\bbeta\|_1,
\end{equation*}
which completes the proof of the first slow rate error bound.

We now restrict our attention to the event $\mathcal{F}$.
Note that, because
\begin{equation*}
\sum_{j=1}^p u_j^\sharp\sqrt{\log(2p/j)}\le \sqrt{\log(2p)}\|\bu\|_1,
\end{equation*}
we must have either (a) inequality
\begin{equation*}
  \bepsilon^\top\bX(\hat\bbeta_1-\bbeta)\le [4+\sqrt{2}]\sigma\sqrt{\log(2p)}\|\bbeta-\hat\bbeta_1\|_1,
\end{equation*}
which implies
\begin{equation*}
\|\bff^*-\bX\hat\bbeta_1\|^2\le \|\bff^*-\bX\bbeta\|^2+2\lambda\|\bbeta\|_1;
\end{equation*}
or (b) the following inequality:
\begin{equation*}
  \bepsilon^\top\bX(\hat\bbeta_1-\bbeta)\le [4+\sqrt{2}]\sigma\sqrt{\log(1/\delta_0)}\|\bX(\bbeta-\hat\bbeta_2)\|,
\end{equation*}
which implies
\begin{equation*}
\|\bff^*-\bX\hat\bbeta_1\|^2\le \|\bff^*-\bX\bbeta\|^2+\lambda\|\bbeta\|_1 + [8+2\sqrt{2}]\sigma\sqrt{\log(1/\delta_0)}\Big(\|\bff^*-\bX\hat\bbeta_1\|+\|\bff^*-\bX\bbeta\|\Big).
\end{equation*}
Bounding the last term in the above display by two applications of~(\ref{ab.ineq}) with $\alpha=2$ yields
\begin{equation}
\label{2nd.slowrate.bnd.L1.prf}
\|\bff^*-\bX\hat\bbeta_1\|^2\lesssim \|\bff^*-\bX\bbeta\|^2+\lambda\|\bbeta\|_1+ \sigma^2\log(1/\delta_0),
\end{equation}
which establishes the second slow rate error bound.

We now move to the fast rate.  Starting with~(\ref{basic.ineq2}), using inequalities
\begin{equation*}
\lambda\|\bbeta\|_1- \lambda\|\hat\bbeta_1\|_1 \le \lambda\|\bbeta-\hat\bbeta_1\|_1 \le \lambda\sqrt{2k}\|\bbeta-\hat\bbeta_1\|,
\end{equation*}
and restricting our attention to the event~$\mathcal{G}_{2k}$, we derive
\begin{equation*}
\|\bff^*-\bX\hat\bbeta_1\|^2\le \|\bff^*-\bX\bbeta\|^2+\sigma\Big[10k\log(ep/[2k])+\log(1/\delta_0)\Big]^{1/2}\|\bX(\bbeta-\hat\bbeta_1)\|+\lambda\sqrt{k}\|\bbeta-\hat\bbeta_1\|.
\end{equation*}
Repeating the argument used to establish the fast rate part of Theorem~\ref{slow.rate.thm}, we arrive at
\begin{equation*}
\|\bff^*-\bX\hat\bbeta_1\|^2\lesssim \|\bff^*-\bX\bbeta\|^2+\sigma^2 k\log(ep/[2k])+\gamma_{2k}^{-2}\lambda^2k+\sigma^2\log(1/\delta_0).
\end{equation*}

\subsection{Proof of Corollary~\ref{L1.Expect.slow.rate}}

This result follows by an argument analogous to the one used in the proof of Corollary~\ref{Expect.slow.rate}.

\subsection{Proof of Corollary~\ref{cor.comp.L1.L2}}
This result follows directly from the slow rate parts of Theorems~\ref{slow.rate.thm} and~\ref{L1.rate.thm}.

\subsection{Proof of Theorem~\ref{best.sub.lower.bnd}}
The following result will allow us to lower-bound the magnitude of the cross-product term in the sum of squares function.
\begin{lem}
\label{lemma.sup}
Let~$S\subset\{1,...,p\}$ have cardinality~$q$, and let~$s$ be an integer in~$[1,q]$. There exists a positive universal constant~$\tilde{c}$, such that
\begin{equation*}
\max_{\text{supp}(\bv)\subset S,\, \|\bv\|_0\le s,\, \|\bX\bv\|=1}|\bepsilon^\top\bX\bv|\gtrsim \sigma \gamma_{2s} \sqrt{s\log(eq/s)}
\end{equation*}
with probability at least $1-2(eq/s)^{-\tilde{c}\gamma_{2s}^2 s}$.
\end{lem}
Lemma~\ref{lemma.sup} is proved in the next subsection.

Using Maurey's argument \citep{pisier1980remarques}, we can bound the error in approximating $\bX\bbeta^*$ with $\bX\bbeta$, when~$\bbeta$ is restricted to an~$\ell_0$ ball.  More specifically, by Lemma A.1 in \cite{rigollet2011exponential}, there exists a vector $\tilde\bbeta^*\in B_0(k/2)$ such that $\|\bX\bbeta^*-\bX\tilde\bbeta^*\|^2\le 2\|\bbeta^*\|^2_1/k$.  For convenience, we define
$\bD^*=\bX\bbeta^*-\bX\tilde\bbeta^*$. Minimizing the sum of squares is equivalent to minimizing the function
\begin{equation*}
G(\bbeta) = \|\bY-\bX\bbeta\|^2 - \|\bY-\bX\tilde\bbeta^*\|^2 = \|\bX\bbeta-\bX\tilde\bbeta^*\|^2+2(\bD^*+\bepsilon)^\top(\bX\tilde\bbeta^*-\bX\bbeta).
\end{equation*}
Given a vector~$\bu\in\mathbb{R}^p$, we define
\begin{equation*}
H(\bu)=\|\bX\bu\|^2-2(\bD^*+\bepsilon)^\top\bX\bu.
\end{equation*}
Given an index set~$\mathcal{I}$ and a vector~$\bbeta$, we will write~$\bbeta_{\mathcal{I}}$ for the vector that (a) matches~$\bbeta$ element by element on the index set~$\mathcal{I}$; and (b) has its support contained in~$\mathcal{I}$.
Let~$\tilde S$ denote the support of $\tilde\bbeta^*$. Note that if $\bbeta_{\tilde S}=\tilde\bbeta^*$ and~$\|\bbeta\|_0\le k$, then
\begin{equation*}
G(\bbeta)=\|\bX\bbeta_{\tilde{S}^c}\|^2-2(\bD^*+\bepsilon)^\top\bX\bbeta_{\tilde{S}^c}=H(\bbeta_{\tilde{S}^c}).
\end{equation*}
Note that $|\tilde S|\le k/2$, and hence
\begin{equation}
\label{LB.thm.seq1}
\min_{\|\bbeta\|_0\le k}G(\bbeta) \le \min_{\bbeta_{\tilde S}=\tilde\bbeta^*,\,\|\bbeta\|_0\le k}G(\bbeta)
\le \min_{\bbeta_{\tilde S}=\tilde\bbeta^*,\,\|\bbeta\|_0\le k}H(\bbeta_{\tilde{S}^c})
\le \min_{\text{supp}(\bu)\subseteq \tilde{S}^c,\, \|\bu\|_0\le k/2}H(\bu).
\end{equation}
To simplify the notation, we define $\mathcal{V}_k=\{\bv\in\mathbb{R}^p,\;\text{s.t. supp}(\bv)\subseteq \tilde{S}^c,\,\|\bv\|_0\le k/2,\,\|\bX\bv\|=1\}$ and $c_{\bv}=(\bD^*+\bepsilon)^\top\bX\bv$.  In addition to the inequalities in~(\ref{LB.thm.seq1}) we also have
\begin{equation*}
\min_{\text{supp}(\bu)\subseteq \tilde{S}^c,\, \|\bu\|_0\le k/2}H(\bu)\le \min_{\mathcal{V}_k}H(c_{\bv}\bv)=\min_{\mathcal{V}_k}\big[-c_{\bv}^2\big]
= -\max_{\mathcal{V}_k}|(\bD^*+\bepsilon)^\top\bX\bv|^2.
\end{equation*}
Consequently,
\begin{equation}
\label{G.bnd}
\min_{\|\bbeta\|_0\le k}G(\bbeta) \le  -\max_{\mathcal{V}_k}|(\bD^*+\bepsilon)^\top\bX\bv|^2.
\end{equation}

Note that if $\|\bX\bv\|=1$, then $|(\bD^*+\bepsilon)^\top\bX\bv| \ge |\bepsilon^\top\bX\bv| - \|\bD^*\|$.  Also note that
\begin{equation}
\label{Delta-star.UB}
\|\bD^*\|\le \|\bbeta^*\|_1/\sqrt{k/2}\lesssim \sigma \gamma_k\sqrt{k\log(ep/k)},
\end{equation}
with a sufficiently small multiplicative constant due to the assumption on~$\|\bbeta^*\|_1$.  Note that the cardinality of~$\tilde{S}^c$ is at least~$p/2$.  Thus, applying Lemma~\ref{lemma.sup}, with $s=k/2$ and $q=p/2$, to lower bound $\max_{\mathcal{V}_k}|\bepsilon^\top\bX\bv|$, we derive that
\begin{equation*}
\max_{\mathcal{V}_k}|(\bD^*+\bepsilon)^\top\bX\bv|\gtrsim \sigma\gamma_k\sqrt{k\log(ep/k)},
\end{equation*}
with probability at least $1-2(ep/k)^{-\tilde{c}\gamma_{k}^2 k/2}$.  Thus, inequality~(\ref{G.bnd}) and the definition of $\hat\bbeta_{\ell_0}$ yield
\begin{equation*}
G(\hat\bbeta_{\ell_0})\le\min_{\bbeta:\;\|\bbeta\|_0\le k}G(\bbeta)\lesssim -\sigma^2\gamma_k^2k\log(ep/k).
\end{equation*}
Because $2(\bD^*+\bepsilon)^\top(\bX\tilde\bbeta^*-\bX\bbeta)\le G(\bbeta)$ for each~$\bbeta$, we derive
\begin{equation}
\label{prf.seq1}
|(\bD^*+\bepsilon)^\top(\bX\hat\bbeta_{\ell_0}-\bX\tilde\bbeta^*)|\gtrsim \sigma^2\gamma_k^2k\log(ep/k).
\end{equation}
Taking into account~(\ref{Delta-star.UB}), which holds with a sufficiently small multiplicative constant, we derive
\begin{equation}
\label{prf.seq2}
|{\bD^*}^\top(\bX\hat\bbeta_{\ell_0}-\bX\tilde\bbeta^*)|\le\|{\bD^*}\|\|\bX\hat\bbeta_{\ell_0}-\bX\tilde\bbeta^*\|\lesssim \sigma\gamma_k\sqrt{k\log(ep/k)}\|\bX\hat\bbeta_{\ell_0}-\bX\tilde\bbeta^*\|.
\end{equation}
Furthermore, on the event $\mathcal{G}_{2k}$ with $\delta_0=(ep/k)^{-k}$, which holds with probability at least $1-\delta_0$ by Theorem~\ref{probs.thm}, we have
\begin{equation}
\label{prf.seq3}
|{\bepsilon}^\top(\bX\hat\bbeta_{\ell_0}-\bX\tilde\bbeta^*)|\lesssim \sigma\sqrt{k\log(ep/k)}\|\bX\hat\bbeta_{\ell_0}-\bX\tilde\bbeta^*\|.
\end{equation}
Combining inequalities~(\ref{prf.seq1}), (\ref{prf.seq2}) and~(\ref{prf.seq3}), we arrive at
\begin{equation*}
\|\bX\hat\bbeta_{\ell_0}-\bX\bbeta^*\|+\|\bD^*\|\ge\|\bX\hat\bbeta_{\ell_0}-\bX\tilde\bbeta^*\|\gtrsim \sigma\gamma_k\sqrt{k\log(ep/k)}.
\end{equation*}
Note that $\|\bX\hat\bbeta_{\ell_0}-\bX\tilde\bbeta^*\|\le\|\bX\hat\bbeta_{\ell_0}-\bX\bbeta^*\|+\|\bD^*\|$, by the triangle inequality.  Let $\c=\tilde{c}/2$.
Applying~(\ref{Delta-star.UB}), which holds with a sufficiently small multiplicative constant, we conclude that
\begin{equation*}
\|\bX\hat\bbeta_{\ell_0}-\bX\bbeta^*\|\gtrsim \sigma\gamma_k\sqrt{k\log(ep/k)},
\end{equation*}
with probability at least $1-2(ep/k)^{-c\gamma_{k}^2 k}-(ep/k)^{-k}$.

\subsection{Proof of Lemma~\ref{lemma.sup}}
\label{prf.lem}
Note that if~$s>q/2$, then we can establish the bound for $s=\lfloor q/2\rfloor$ and use
\begin{equation*}
\max_{\|\bv\|_0\le s,\, \|\bX\bv\|=1}|\bepsilon^\top\bX\bv|\ge \max_{\|\bv\|_0\le \lfloor q/2\rfloor,\, \|\bX\bv\|=1}|\bepsilon^\top\bX\bv|.
\end{equation*}
Hence, we will focus on the case $s\le q/2$.

We write~$|\cdot|$ for the cardinality of a set. Applying Lemma~F.1 in \cite{bellec2018slope}, which is closely related to the results in \cite{verzelen2012minimax}, we deduce that there exists a subset~$\mathcal{H}$ of the set~$\{-1,0,1\}^p$,
with
\begin{equation*}
\log\big(|\mathcal{H}|\big)\gtrsim s\log(eq/s),
\end{equation*}
such that $\text{supp}(\bv)\subset S$, $\|\bv\|_0\le s$, $\|\bX\bv\|^2\le s$ and $\|\bv_1-\bv_2\|^2\ge s/4$, for all $\bv,\bv_1,\bv_2\in\mathcal{H}$. Note that the last inequality implies
\begin{equation*}
\|\bX\bv_1-\bX\bv_2\|^2\ge \gamma_{2s}^2s/4.
\end{equation*}
Consequently, by Sudakov's minoration \citep[for example, Proposition~3.15 in][]{massart1},
\begin{equation*}
E\max_{\bv\in\mathcal{H}}\bepsilon^\top \bX\bv \gtrsim \sigma \gamma_{2s}\sqrt{s\log(|\mathcal{H}|)} \gtrsim \sigma \gamma_{2s} s \sqrt{\log(eq/s)}.
\end{equation*}
Define $W=\max_{\bv\in\mathcal{H}}\bepsilon^\top \bX\bv$ and $v=\max_{\bv\in\mathcal{H}}SD(\bepsilon^\top \bX\bv)$ by~$v$.
By the concentration inequality for the supremum of a Gaussian process \citep[for example, Theorem~3.12 in][]{massart1}, we have, for all~$t\ge0$,
\begin{equation*}
P\left(W\le EW-vt\right)\le2\exp(-t^2/2).
\end{equation*}
Note that
\begin{equation*}
v\le \sigma\max_{\bv\in\mathcal{H}}\|\bX\bv\|\le \sigma\sqrt{s}.
\end{equation*}
Consequently, if $t\le\gamma_{2s}\sqrt{\tilde{c}s\log(eq/s)}$ with a sufficiently small positive universal constant~$\tilde{c}$, then $EW-vt\gtrsim \sigma \gamma_{2s} s \sqrt{\log(eq/s)}$, and hence
\begin{equation*}
\max_{\bv\in\mathcal{H}}\bepsilon^\top \bX\bv \gtrsim \sigma \gamma_{2s} s \sqrt{\log(eq/s)},
\end{equation*}
with probability at least $1-2\exp(-\tilde{c}\gamma_{2s}^2 s\log(eq/s))$.  We complete the proof by noting that
\begin{equation*}
\max_{\|\bv\|_0\le s,\, \|\bX\bv\|=1}\bepsilon^\top \bX\bv \ge s^{-1/2}\max_{\bv\in\mathcal{H}}\bepsilon^\top \bX\bv.
\end{equation*}

\subsection{Proof of Proposition~\ref{prop1}}

Note that the assumptions imposed on~$b$ imply
\begin{equation}
\label{b.bnds}
b\gtrsim1 \qquad  \text{and}  \qquad  b\lesssim1,
\end{equation}
where the universal constant in the second bound can be chosen to be sufficiently small.  Also note that
\begin{equation}
\label{b.norm.bnds}
\|\bbeta^*\|=b\sigma \sqrt{[\log(ep)]/k^*} \qquad  \text{and}  \qquad  \|\bbeta^*\|_1=b\sigma \sqrt{\log(ep)}.
\end{equation}
Thus, to establish the result of Proposition~\ref{prop1}, we only need to demonstrate that
\begin{equation}
\label{Ex.rest.LB}
\min_{k\in[0,p]}\|\bX\bbeta^*-\bX\widehat\bbeta_{\ell_0}\|^2\gtrsim \sigma^2 \log(ep),
\end{equation}
with high probability.

Let~$\mathbf{1}$ denote a $p$-dimensional vector of ones, and note that
\begin{equation*}
\label{Ex.sig.LB}
\|\bX\bbeta^*\|^2\ge  \rho_l{\bbeta^*}^\top\mathbf{1}\mathbf{1}^\top\bbeta^*  = \rho_l\|\bbeta^*\|^2_1 =  \rho_l b^2\sigma^2\log(ep) \gtrsim \sigma^2\log(ep).
\end{equation*}
We conclude that for $k=0$ bound~(\ref{Ex.rest.LB}) holds with probability one.
For the remainder of the proof we focus on the case of $k\in[p]$.

Minimizing the sum of squares is equivalent to minimizing the function
\begin{equation*}
L(\bbeta) = \|\bX\bbeta\|^2-2\bY^\top\bX\bbeta.
\end{equation*}
Define $c_j=\bY^\top \bX_j$ and let $\be_j$ denote the $j$-th coordinate vector in $\mathbb{R}^p$.  Because $\bX\be_j=\bX_j$ and $\|\bX_j\|=1$, we have
\begin{equation*}
\min_{\|\bbeta\|_0= 1}L(\bbeta) \le \min_{j}L(c_j\be_j)=\min_{j}-c_j^2 = -\max_{j}|\bY^\top\bX_j|^2.
\end{equation*}
We also have
\begin{eqnarray*}
\max_{j}|\bY^\top\bX_j|^2 &=& \max_{j}\Big(|\bepsilon^\top\bX_j|^2+2(\bepsilon^\top\bX_j)(\bX_j^\top\bX\bbeta^*)+|\bX_j^\top\bX\bbeta^*|^2\Big)\\
&\ge& \max_{j}\Big(|\bepsilon^\top\bX_j|^2-2|\bepsilon^\top\bX_j|\|\bX\bbeta^*\|\Big)\\
&\ge& \max_{j}\Big(|\bepsilon^\top\bX_j|^2/2-2\|\bX\bbeta^*\|^2\Big),
\end{eqnarray*}
where we used bound~(\ref{ab.ineq}) with $a=|\bepsilon^\top\bX_j|$, $b=\|\bX\bbeta^*\|$ and $\alpha=1/2$ to get the last inequality.
Applying Lemma~\ref{lemma.sup}, we derive that
\begin{equation*}
\max_{j}|\bepsilon^\top\bX_j|\gtrsim (1-\rho_u)\sigma^2\log(ep),
\end{equation*}
with probability at least $1-2(ep)^{-\tilde{c}(1-\rho_u)}$, for some positive universal constant~$\tilde{c}$.

Inequalities~(\ref{b.norm.bnds}) and~(\ref{b.bnds}), together with the fact that columns of~$\bX$ have unit norm, yield
\begin{equation}
\label{Ex.signal.UB}
\|\bX\bbeta^*\|^2\le \|\bbeta^*\|_1^2 \le  b^2\sigma^2\log(ep) \lesssim \sigma^2\log(ep),
\end{equation}
with a sufficiently small universal constant.  Consequently,
\begin{equation}
\label{Ex.L.UB}
\min_{\|\bbeta\|_0\le k}L(\bbeta)\le\min_{\|\bbeta\|_0= 1}L(\bbeta)\lesssim -\sigma^2\log(ep),
\end{equation}
uniformly over~$k\in[p]$ and with probability at least $1-2(ep)^{-a}$, for some positive universal constant~$a$.

We conduct the rest of the argument on the high-probability event where~(\ref{Ex.L.UB}) holds.  On this event we have the bound
\begin{equation}
\label{L.min.value.bnd}
L(\hat\bbeta_{\ell_0}) = \|\bX\hat\bbeta_{\ell_0}\|^2-2\bY^\top\bX\hat\bbeta_{\ell_0}\lesssim -\sigma^2\log(ep),
\end{equation}
in which the universal constant does not depend on~$k$.  Given a set $S\subseteq\{1,...,p\}$, we define~$\hat\bbeta_S=\arg\min_{\text{supp}(\bbeta)\subseteq S}L(\bbeta)$ and note that $\|\bX\hat\bbeta_S\|^2=\bY^\top\bX\hat\bbeta_S$.  Consequently, $\|\bX\hat\bbeta_{\ell_0}\|^2=\bY^\top\bX\hat\bbeta_{\ell_0}$, and hence bound~(\ref{L.min.value.bnd}) implies
\begin{equation*}
 \|\bX\hat\bbeta_{\ell_0}\|^2\gtrsim \sigma^2\log(ep).
\end{equation*}
Bound~(\ref{Ex.rest.LB}) then follows from the inequality $\|\bX\bbeta^*-\bX\hat\bbeta_{\ell_0}\|\ge \|\bX\hat\bbeta_{\ell_0}\|-\|\bX\bbeta^*\|$ and bound~(\ref{Ex.signal.UB}), applied with a sufficiently small universal constant.

\subsection{Proof of Proposition~\ref{prop2}}

Let~$\mathbf{1}$ denote a $p$-dimensional vector of ones, and note that
\begin{equation*}
\bX^\top\bX=(1-\rho)\mathbf{I}+\rho\mathbf{1}\mathbf{1}^\top.
\end{equation*}
Hence, for every $\bu\in\mathbb{R}^p$,
\begin{equation*}
\|\bX\bu\|^2 = (1-\rho)\|u\|^2 + \rho(\mathbf{1}^\top\bu)^2 \ge (1-\rho)\|u\|^2,
\end{equation*}
which implies $\gamma_k^2\ge 1-\rho$.  Also note that, by~(\ref{b.norm.bnds}), $\|\bbeta^*\|=b\sigma\sqrt{[\log(ep)]/k^*}$, which can be made smaller than any given multiple of $\sigma\sqrt{k\log(ep/k)}$ under the assumptions imposed on~$b$, $k$ and~$k^*$ in Proposition~\ref{prop2}. Under this scenario, we can apply Theorem~\ref{best.sub.lower.bnd}, which leads to
 \begin{equation*}
\|\bX\hat\bbeta_{\ell_0}-\bX\bbeta^*\|^2\gtrsim\sigma^2 k{\log (ep/k)}.
\end{equation*}

\subsection{Proof of Theorem~\ref{pen2.thm}}

We first establish the bound for~$\hat\bbeta_2^B$, and then establish the one for~$\hat\bbeta_1^B$. We note that throughout the proof the positive multiplicative factors in inequalities~$\lesssim$ and~$\gtrsim$ are universal constants, which are independent from all other parameters such as $n$, $p$, $\sigma$, $\bbeta^*$, $\bbeta$ and~$s$.

\textbf{Expected prediction error bound for~$\hat\bbeta_2^B$}.\\
To simplify the expressions, we drop the subscript and the superscript in~$\hat\bbeta_2^B$ and simply write~$\hat\bbeta$.

Taking an arbitrary~$\bbeta\in \mathbb{R}^p$, we note that
\begin{equation}
\label{basic.ineq.pen2}
\|\bff^*-\bX\hat\bbeta\|^2+\lambda_{\hat\bbeta}\|\hat\bbeta\|+\mu_{\hat\bbeta}|\|\hat\bbeta\|_0\le \|\bff^*-\bX\bbeta\|^2+2\bepsilon^\top\bX(\hat\bbeta-\bbeta)+\lambda_{\bbeta}\|\bbeta\|+\mu_{\bbeta}\|\bbeta\|_0.
\end{equation}
In the setting where~$a\gtrsim \sigma$, with a sufficiently large multiplicative constant, we will bound the term $2\bepsilon^\top\bX(\hat\bbeta-\bbeta)-\lambda_{\hat\bbeta}\|\hat\bbeta\|$.  Similarly, in the case~$b\gtrsim \sigma^2$ we will bound $2\bepsilon^\top\bX(\hat\bbeta-\bbeta)-\mu_{\hat\bbeta}|\|\hat\bbeta\|_0$.

We first consider the case~$a\gtrsim \sigma$, which implies $\lambda_{\bbeta}\gtrsim \sigma\sqrt{\|\bbeta\|_0\log (ep/\|\bbeta\|_0)}$.
We let $\hat\bu=\hat\bbeta-\bbeta$ and restrict our attention to event~$\mathcal{F}$, defined in Section~\ref{sec.prelims}, which holds with probability at least $1-\delta_0/2$.  On event~$\mathcal{F}$, we have either $\bepsilon^\top\bX\hat\bu \lesssim \sigma \sum_{j=1}^p \hat u_j^\sharp\sqrt{\log(2p/j)}$ or
$\bepsilon^\top\bX\hat\bu \lesssim \sigma \sqrt{\log(1/\delta_0)}\|\bX\hat\bu\|$.  In the latter scenario, repeating the argument after inequality~(\ref{prelim.in.prod.bnd}) in the proof of Theorem~\ref{slow.rate.thm} yields
\begin{equation}
\label{or.bnd2.pen2}
\|\bff^*-\bX\hat\bbeta\|^2+\mu_{\hat\bbeta_2}|\|\hat\bbeta\|_0\lesssim \|\bff^*-\bX\bbeta\|^2+\lambda_{\bbeta}\|\bbeta\|+\sigma^2\log(1/\delta_0)+\mu_{\bbeta}\|\bbeta\|_0.
\end{equation}
We now focus on the event $\bepsilon^\top\bX\hat\bu \lesssim \sigma \sum_{j=1}^p \hat u_j^\sharp\sqrt{\log(2p/j)}$.  In view of~(\ref{H.ineq}), we have
\begin{eqnarray*}
\sigma \sum_{j=1}^p \hat u_j^\sharp\sqrt{\log(2p/j)} &\le& \sigma \sum_{j=1}^p \hat \beta_j^\sharp\sqrt{\log(2p/j)}  +  \sigma \sum_{j=1}^p {\beta_j}^\sharp\sqrt{\log(2p/j)}\\
&\le & \sigma\sqrt{\hat k\log(2ep/\hat k)}\|\hat\bbeta\|  +  \sigma\sqrt{\|\bbeta\|_0\log(2ep/\|\bbeta\|_0)}\|\bbeta\|.
\end{eqnarray*}
Thus, if $a\gtrsim \sigma$ with a sufficiently large universal constant, then $2\bepsilon^\top\bX\hat\bu  \le \lambda_{\hat\bbeta}\|\hat\bbeta\|  + \lambda_{\bbeta}\|\bbeta\|$. Combining this bound with inequality~(\ref{basic.ineq.pen2}), we derive
\begin{equation}
\label{or.bnd1.pen2}
\|\bff^*-\bX\hat\bbeta\|^2+\mu_{\hat\bbeta_2}|\|\hat\bbeta\|_0\lesssim \|\bff^*-\bX\bbeta\|^2+2\lambda_{\bbeta}\|\bbeta\|+\mu_{\bbeta}\|\bbeta\|_0.
\end{equation}

Bounds~(\ref{or.bnd2.pen2}) and~(\ref{or.bnd1.pen2}) imply that, for each $\delta_0\in(0,1)$,
\begin{equation}
\label{or.bnd12.pen2}
\|\bff^*-\bX\hat\bbeta\|^2+\mu_{\hat\bbeta_2}|\|\hat\bbeta\|_0\lesssim \|\bff^*-\bX\bbeta\|^2+\lambda_{\bbeta}\|\bbeta\|+\sigma^2\log(1/\delta_0)+\mu_{\bbeta}\|\bbeta\|_0
\end{equation}
with probability at least $1-\delta_0/2$.

We now focus on the case~$b\gtrsim \sigma^2$, which implies $\mu_{\bbeta}\gtrsim \sigma^2\log (ep/\|\bbeta\|_0)$.  Given an $s\in[p]$, we consider the event $\mathcal{G}_s$, defined in Section~\ref{sec.prelims}, where we take $\delta_0=(s/[ep])^s\epsilon_0$.  Here, $\epsilon_0\in(0,1)$ is an arbitrary value that does not depend on~$s$.
On the event $\mathcal{G}=\cap_{s=1}^p\mathcal{G}_s$ we have
\begin{eqnarray*}
\bepsilon^\top\bX\hat\bu &\lesssim& \sigma\sqrt{\|\hat\bu\|_0\log(ep/\|\hat\bu\|_0)+\log(1/\epsilon_0)}\|\bX\hat\bu\| \\
&\lesssim&  \sigma\sqrt{\|\hat\bbeta\|_0\log(ep/\|\hat\bbeta\|_0)+\|\bbeta\|_0\log(ep/\|\bbeta\|_0)+\log(1/\epsilon_0)}\|\bX\hat\bu\|.
\end{eqnarray*}
Noting that $\|\bX\hat\bu\|\le \|\bff^*-\bX\hat\bbeta\|+\|\bff^*-\bX\bbeta\|$ and applying inequality~(\ref{ab.ineq}) twice, we derive
\begin{equation*}
2\bepsilon^\top\bX\hat\bu \le c \Big[ \sigma^2\|\hat\bbeta\|_0\log(ep/\|\hat\bbeta\|_0)+\sigma^2\|\bbeta\|_0\log(ep/\|\bbeta\|_0)+\sigma^2\log(1/\epsilon_0)+\|\bff^*-\bX\bbeta\|^2\Big]+\|\bff^*-\bX\hat\bbeta\|^2/2,
\end{equation*}
for some universal constant~$c $.  Taking into account inequality~(\ref{basic.ineq.pen2}), we deduce that
\begin{equation*}
\|\bff^*-\bX\hat\bbeta\|^2 +2[b-c \sigma^2]\|\hat\bbeta\|_0\log(ep/\|\hat\bbeta\|_0)\lesssim \mu_{\bbeta}\|\bbeta\|_0+\sigma^2\log(1/\epsilon_0)+\|\bff^*-\bX\bbeta\|^2+\lambda_{\bbeta}\|\bbeta\|
\end{equation*}
on the event $\mathcal{G}$. Note that
\begin{equation*}
\mathbb{P}(\mathcal{G}^c)\le\sum_{s=1}^p\mathcal{P}(\mathcal{G}_s^c)\le\sum_{s=1}^p(s/[ep])^s\epsilon_0\le\sum_{s=1}^p e^{-s}\epsilon_0\le \epsilon_0.
\end{equation*}
Consequently, if we let $b\ge2c \sigma^2$, then
\begin{equation}
\label{or.bnd3.pen2}
\|\bff^*-\bX\hat\bbeta\|^2+\mu_{\hat\bbeta}|\|\hat\bbeta\|_0\lesssim \|\bff^*-\bX\bbeta\|^2+\sigma^2\log(1/\epsilon_0)+\lambda_{\bbeta}\|\bbeta\|+\mu_{\bbeta}\|\bbeta\|_0
\end{equation}
with probability at least $1-\epsilon_0$.

Combining bounds~(\ref{or.bnd12.pen2}) and~(\ref{or.bnd3.pen2}), we conclude that, for each $\delta_0\in(0,1)$,
\begin{equation}
\label{or.bnd.fin.pen2}
\|\bff^*-\bX\hat\bbeta\|^2+\mu_{\hat\bbeta}|\|\hat\bbeta\|_0\lesssim \inf_{\bbeta\in\mathbb{R}^p}\Big[ \|\bff^*-\bX\bbeta\|^2+\lambda_{\bbeta}\|\bbeta\|+\mu_{\bbeta}\|\bbeta\|_0\Big] +\sigma^2\log(1/\delta_0)
\end{equation}
with probability at least $1-\delta_0$.
Repeating the argument in the proof of Corollary~\ref{Expect.slow.rate}, we derive
\begin{equation}
\label{exp.bnd.pen2}
\mathbb{E}\|\bff^*-\bX\hat\bbeta\|^2\lesssim  \inf_{\bbeta\in\mathbb{R}^p}\Big[ \|\bff^*-\bX\bbeta\|^2+\lambda_{\bbeta}\|\bbeta\|+\mu_{\bbeta}\|\bbeta\|_0\Big]+\sigma^2,
\end{equation}
which establishes the first bound in the statement of Theorem~\ref{pen2.thm}.

\textbf{Expected prediction error bound for~$\hat\bbeta_1^B$}.\\
To simplify the expressions, we drop the subscript and the superscript in~$\hat\bbeta_1^B$ and simply write~$\hat\bbeta$.

In the case~$b\gtrsim \sigma^2$ we repeat the argument in the corresponding part of the proof for~$\hat\bbeta_2^B$ to derive a counterpart of inequality~(\ref{or.bnd.fin.pen2}).  We deduce that, for each $\delta_0\in(0,1)$,
\begin{equation}
\label{or.bnd1.pen1}
\|\bff^*-\bX\hat\bbeta\|^2+\mu_{\hat\bbeta}|\|\hat\bbeta\|_0\lesssim \inf_{\bbeta\in\mathbb{R}^p}\Big[ \|\bff^*-\bX\bbeta\|^2+\lambda\|\bbeta\|_1+\mu_{\bbeta}\|\bbeta\|_0\Big] +\sigma^2\log(1/\delta_0)
\end{equation}
with probability at least $1-\delta_0$.

In the case $\lambda\ge c_0\sigma\sqrt{\log(ep)}$, we repeat the argument in the second slow rate part of the proof of Theorem~\ref{L1.rate.thm} to derive a slight modification of inequality~(\ref{2nd.slowrate.bnd.L1.prf}), containing the additional $\ell_0$ penalty terms.  Thus, we again deduce that inequality~(\ref{or.bnd1.pen1}) holds for each $\delta_0\in(0,1)$ with probability at least $1-\delta_0$.

As before, starting with probability bound~(\ref{or.bnd1.pen1}) and repeating the argument in the proof of Corollary~\ref{Expect.slow.rate} we conclude that
\begin{equation}
\label{exp.bnd.pen1}
\mathbb{E}\|\bff^*-\bX\hat\bbeta\|^2\lesssim  \inf_{\bbeta\in\mathbb{R}^p}\Big[ \|\bff^*-\bX\bbeta\|^2+\lambda\|\bbeta\|_1+\mu_{\bbeta}\|\bbeta\|_0\Big]+\sigma^2,
\end{equation}
which establishes the second bound in the statement of Theorem~\ref{pen2.thm}.

{
\subsection{Proof of Corollary~\ref{data.driven.cor}}

The error rates in Corollary~\ref{data.driven.cor} follow directly from inequalities~(\ref{exp.bnd.pen2}) and~(\ref{exp.bnd.pen1}).

To control the sparsity of~$\hat\bbeta_2^B$, we first establish that inequality $\|\hat\bbeta_2^B\|_0 \lesssim  (k^*\vee1)\{1+[\log(1/\delta_0)]^2\}$ holds with probability at least $1-\delta_0$, and then use this fact to bound $\mathbb{E}\|\hat\bbeta_2^B\|_0$.

We define $g(x)=\log(ep/x)x$ for $x\in[0,p]$, with $g(0)=0$.  We note that $g(x)\ge x$, and $g(x)$ is monotone increasing and continuous.  We also note that if $C>0$, $x_1\in[0,p]$ and $x_2\in[0,p]$, then
\begin{equation}
\label{g.prop}
g(x_1) \le Cg(x_2) \quad\Rightarrow \quad x_1 \le 2C[1\vee\log(2C)]x_2.
\end{equation}
To establish the above relationship we note that
\begin{equation*}
g(x_1) \le Cg(x_2) \le g(2Cx_2)/2 + Cx_2\log(2C)\le\max\{g(2Cx_2),2Cx_2\log(2C)\}
\end{equation*}
and consider two cases. If $g(x_1) \le g(2Cx_2)$, then $x_1 \le 2Cx_2$.  Alternatively, if $g(x_1) \le 2Cx_2\log(2C)$, then $x_1 \le 2C\log(2C)x_2$.

We write $\hat k = \|\hat\bbeta_2^B\|_0$ and note that under each corresponding set of assumptions on the tuning parameters in Corollary~\ref{data.driven.cor}, inequality~(\ref{or.bnd.fin.pen2}) yields $(\mu_{\hat\bbeta_2^B})\hat k\lesssim b\log(ep/[k^*\vee 1])[k^*\vee 1]\{1+\log(1/\delta_0)\}$.  Taking into account $(\mu_{\hat\bbeta_2^B})\hat k=bg(\hat k)$, we rewrite the last inequality as $g(\hat k) \lesssim g(k^*\vee1)\{1+\log(1/\delta_0)\}$. Hence, by property~\ref{g.prop}, we have $\hat k \lesssim \{1+ \log(1/\delta_0)\}[1\vee\log(1+2\log(1/\delta_0))](k^*\vee1)$.  Consequently,
$\hat k / [k^*\vee1] \lesssim  1+[\log(1/\delta_0)]^2$ with probability at least $1-\delta_0$.  Finally, we bound $\mathbb{E}\hat k / [k^*\vee1]$ using an argument analogous to the one in the proof of Corollary~\ref{Expect.slow.rate}:
$\mathbb{E}\hat k / [k^*\vee1] \lesssim 1 + \int_0^{\infty}e^{-w^{1/2}}dw \lesssim 1$.

To establish the sparsity bound for~$\hat\bbeta_1^B$, we note that for all of corresponding tuning parameter settings in Corollary~\ref{data.driven.cor}, inequality~(\ref{or.bnd1.pen1}) yields $(\mu_{\hat\bbeta_1^B})\|\hat\bbeta_1^B\|_0\lesssim b\log(ep/[k^*\vee 1])[k^*\vee 1]\{1+\log(1/\delta_0)\}$, with probability at least $1-\delta_0$.  The sparsity bound then follows by repeating the argument in the last paragraph of the proof for~$\hat\bbeta_2^B$.
}

\section{Additional experiments}\label{sec-exp-suppl}

\subsection{Ultra-high dimensional examples}
\begin{figure}[h!]
\renewcommand{\baselinestretch}{1.25}
	\centering
	\begin{tabular}{l c c c}
		\multicolumn{4}{c} { \sf {Example~1: Large settings: $n=100, p=1000$, $k^*=15$} }\\
		&\sf {\small{$\rho=0.2, \text{SNR}=1$}} &  \sf {\small{$\rho=0.2, \text{SNR}=2$}} & \sf {\small{$\rho=0.8, \text{SNR}=3$}}\\
		\rotatebox{90}{\sf {\small{~~~~~~~~~Prediction Error}}} &
		\includegraphics[width=0.3\textwidth,height=0.18\textheight,  trim =1.8cm 1cm 2cm 1.3cm, clip = true ]{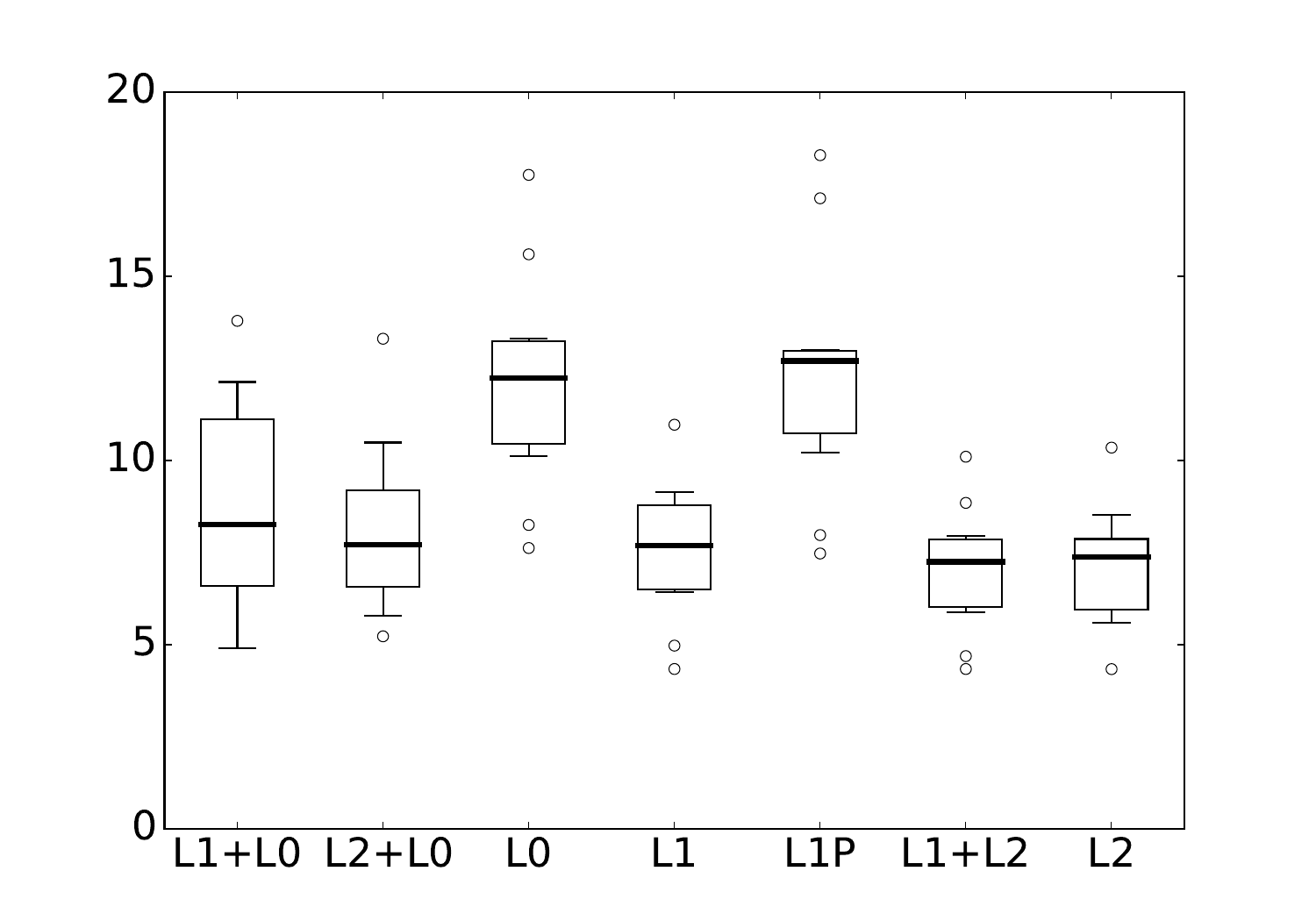}&
		\includegraphics[width=0.3\textwidth,height=0.18\textheight,  trim =1.2cm 1cm 2cm 1.3cm, clip = true ]{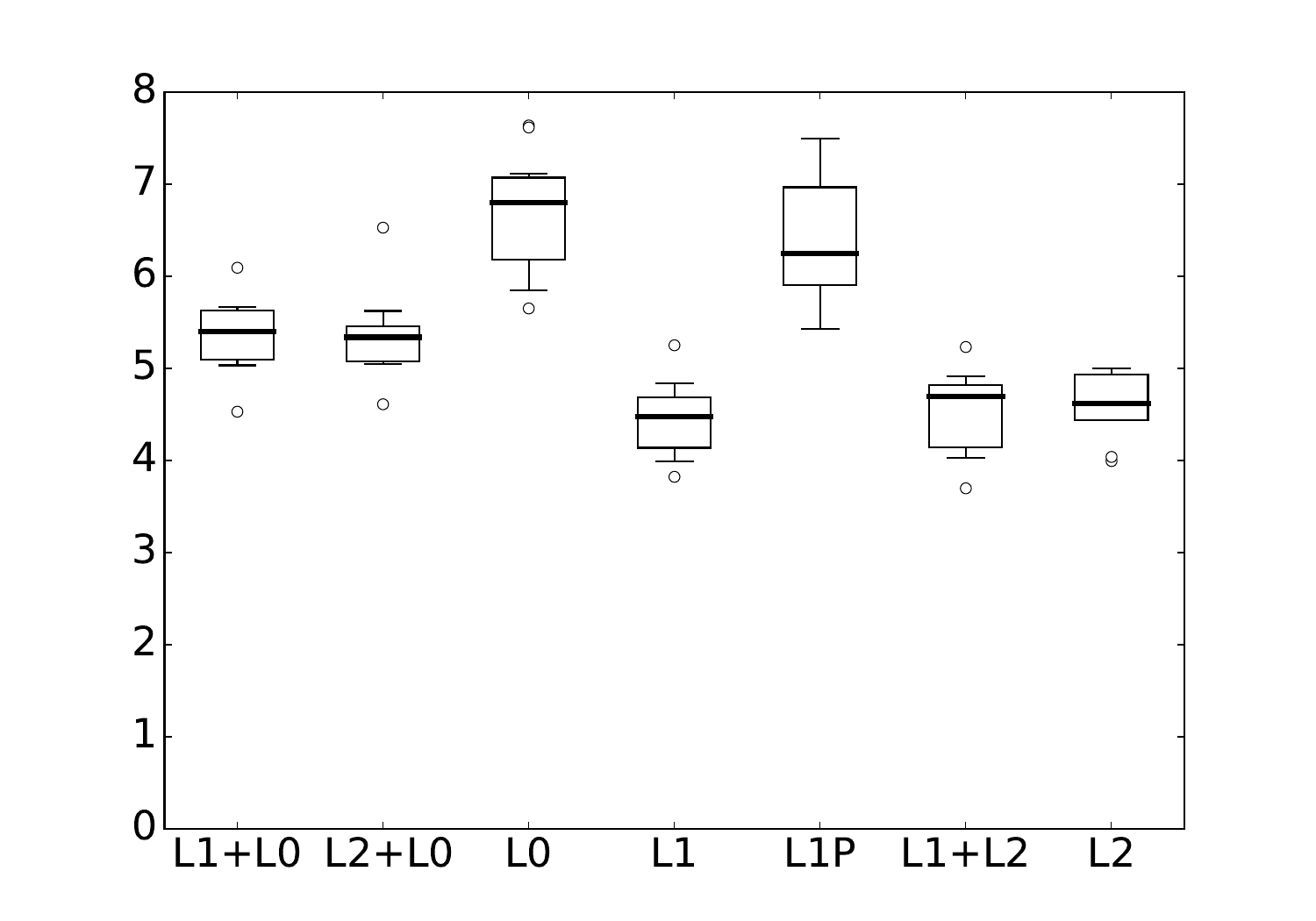}&
		\includegraphics[width=0.3\textwidth,height=0.18\textheight,  trim =1.2cm 1cm 2cm 1.3cm, clip = true ]{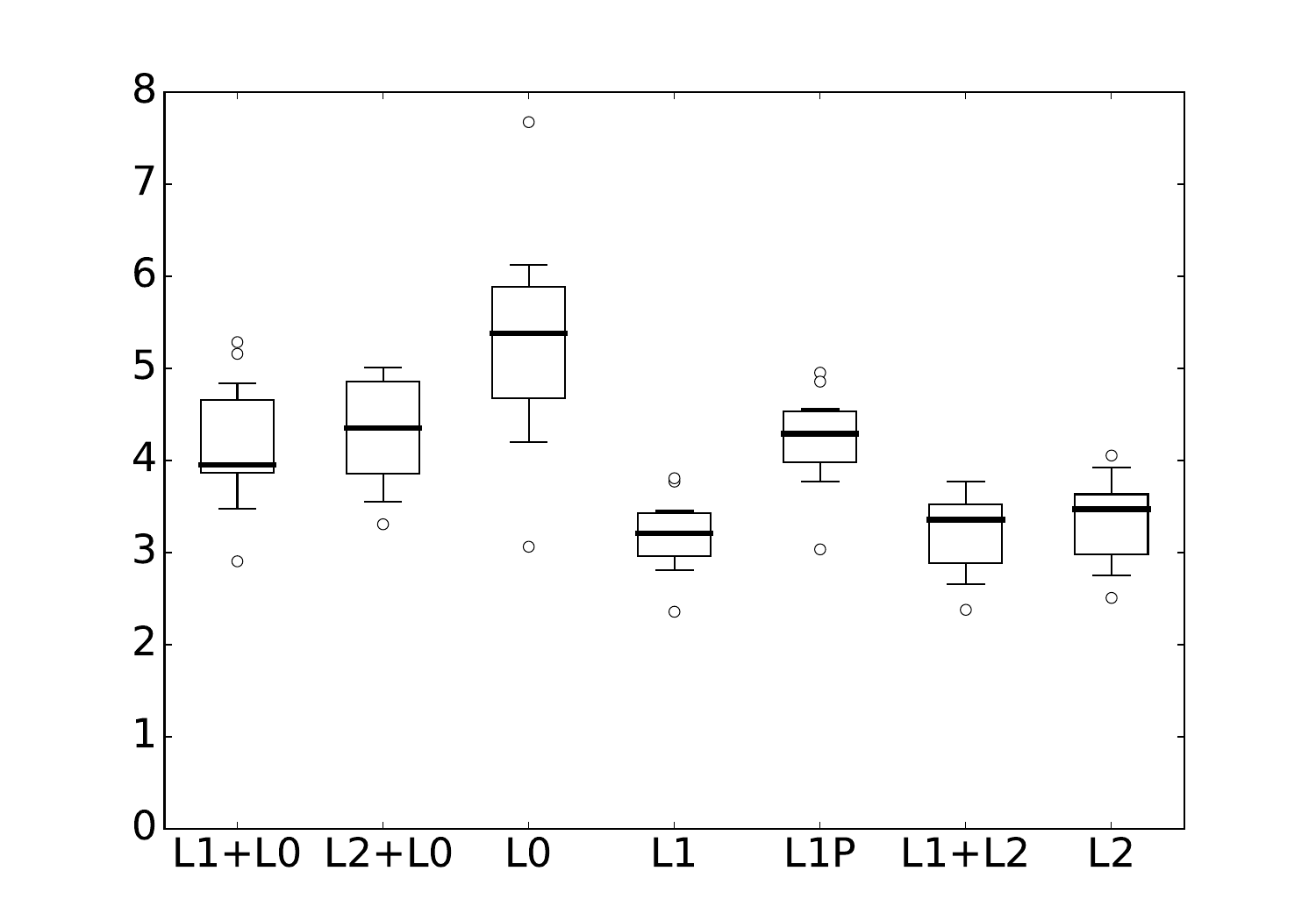}\\
		
		\rotatebox{90}{\sf {\small{~~~~~~~\# nonzeros}}}&
		\includegraphics[width=0.3\textwidth,height=0.18\textheight,  trim =1.8cm 1cm 2cm 1cm, clip = true ]{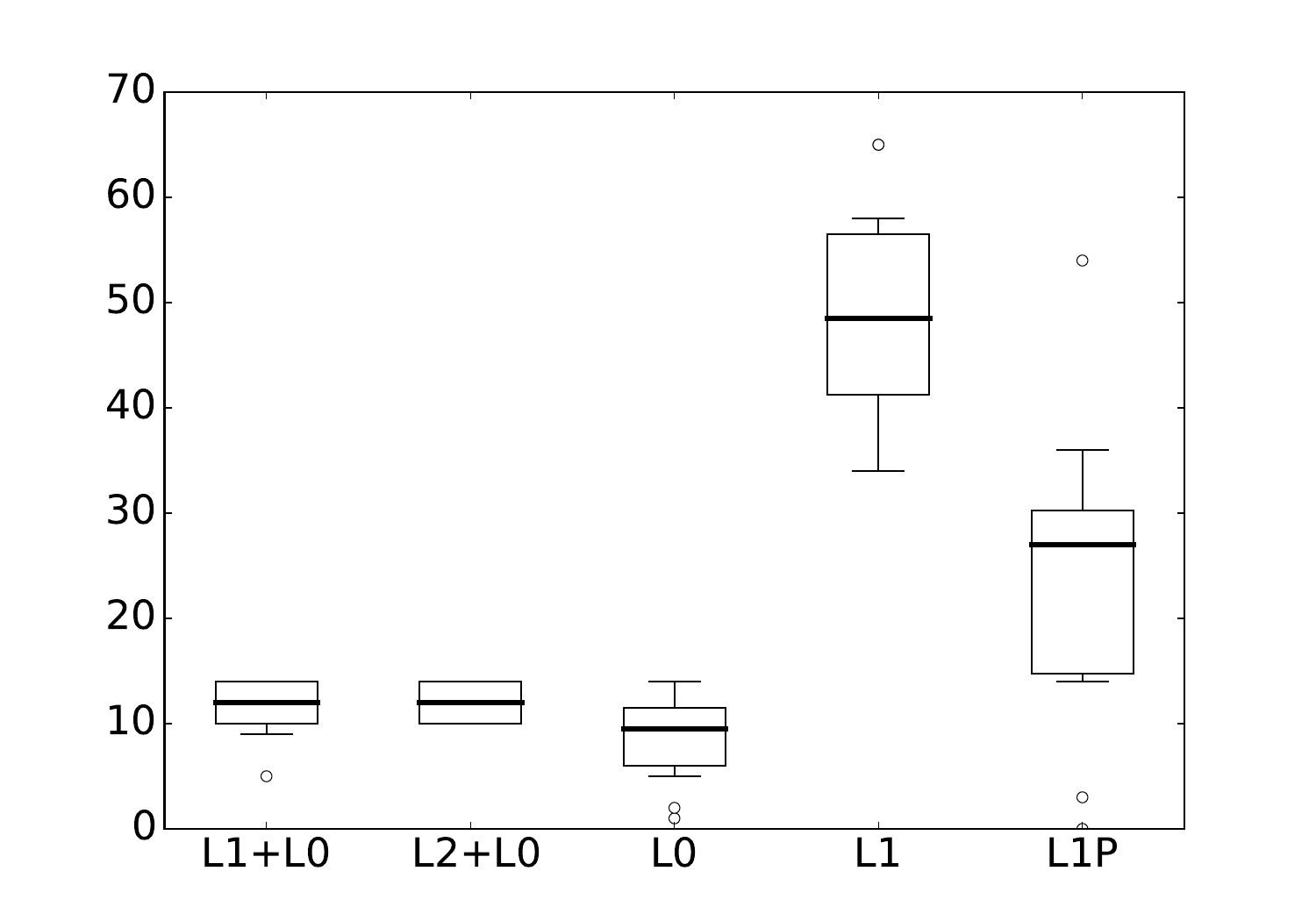}&
		\includegraphics[width=0.3\textwidth,height=0.18\textheight,  trim =1.2cm 1cm 2cm 1cm, clip = true ]{\plotA/N100_P1000_k015_rho0.2_SNR1.0_Sigma1/sparsity_boxplot_fixed_design_averaged.pdf}&
		\includegraphics[width=0.3\textwidth,height=0.18\textheight,  trim =1.2cm 1cm 2cm 1cm, clip = true ]{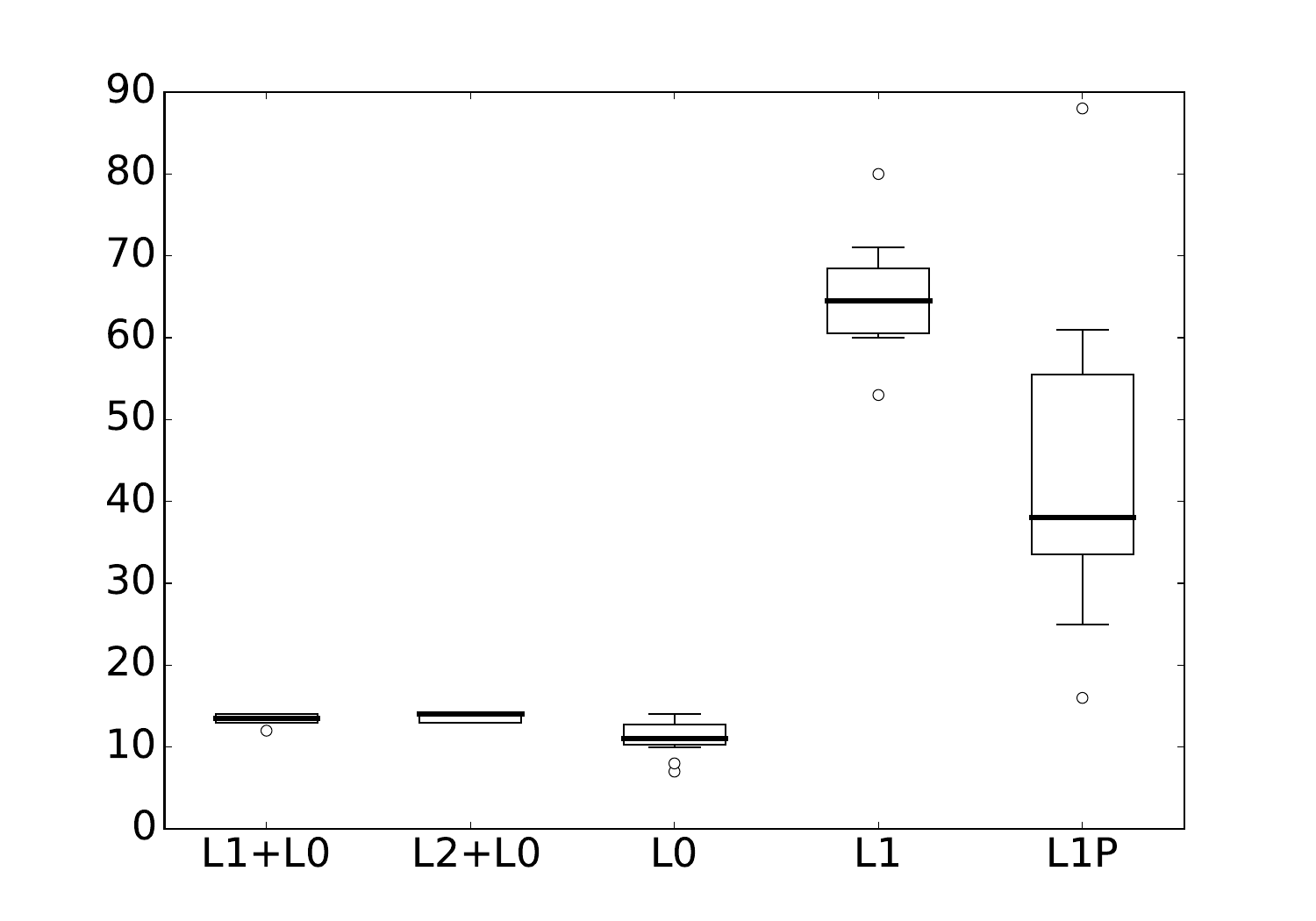}
			\end{tabular}
	\caption{{\small{Example 1 simulations for different values of $n$, $p$, $\rho$, and SNR.
}}}
	\label{fig: ultrahigh.exp1}
\end{figure}

\begin{figure}[h!]
\renewcommand{\baselinestretch}{1.25}
	\centering
	\begin{tabular}{l c c c}
		\multicolumn{4}{c} { \sf {Example~1: Large settings: $n=50, p=1000$, $k^*=10$} }\\
		&\sf {\small{$\rho=0.2, \text{SNR}=1$}} &  \sf {\small{$\rho=0.2, \text{SNR}=2$}} & \sf {\small{$\rho=0.8, \text{SNR}=3$}}\\
		\rotatebox{90}{\sf {\small{~~~~~~~~~Prediction Error}}} &
		\includegraphics[width=0.3\textwidth,height=0.18\textheight,  trim =1.8cm 1cm 2cm 1.3cm, clip = true ]{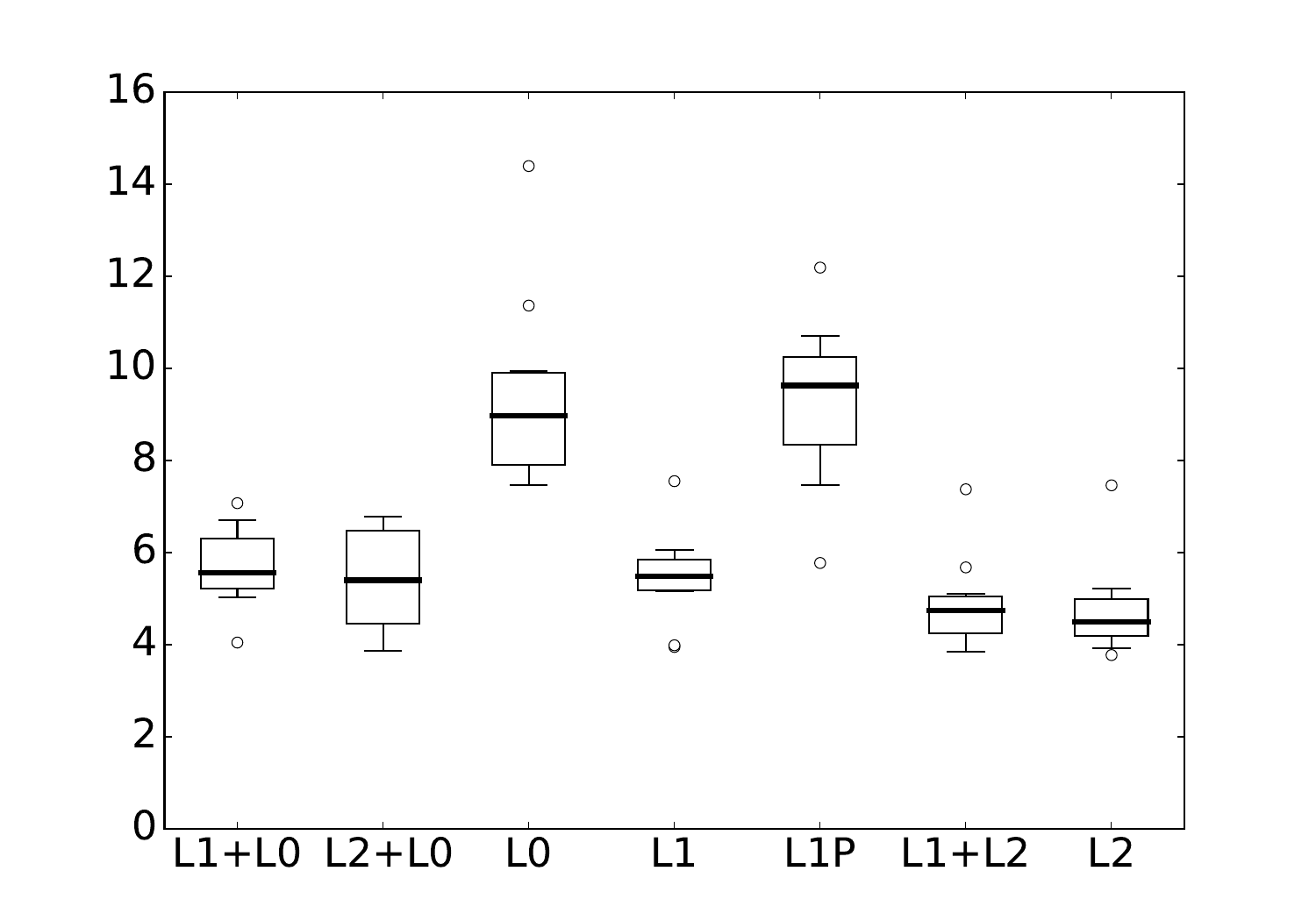}&
		\includegraphics[width=0.3\textwidth,height=0.18\textheight,  trim =1.2cm 1cm 2cm 1.3cm, clip = true ]{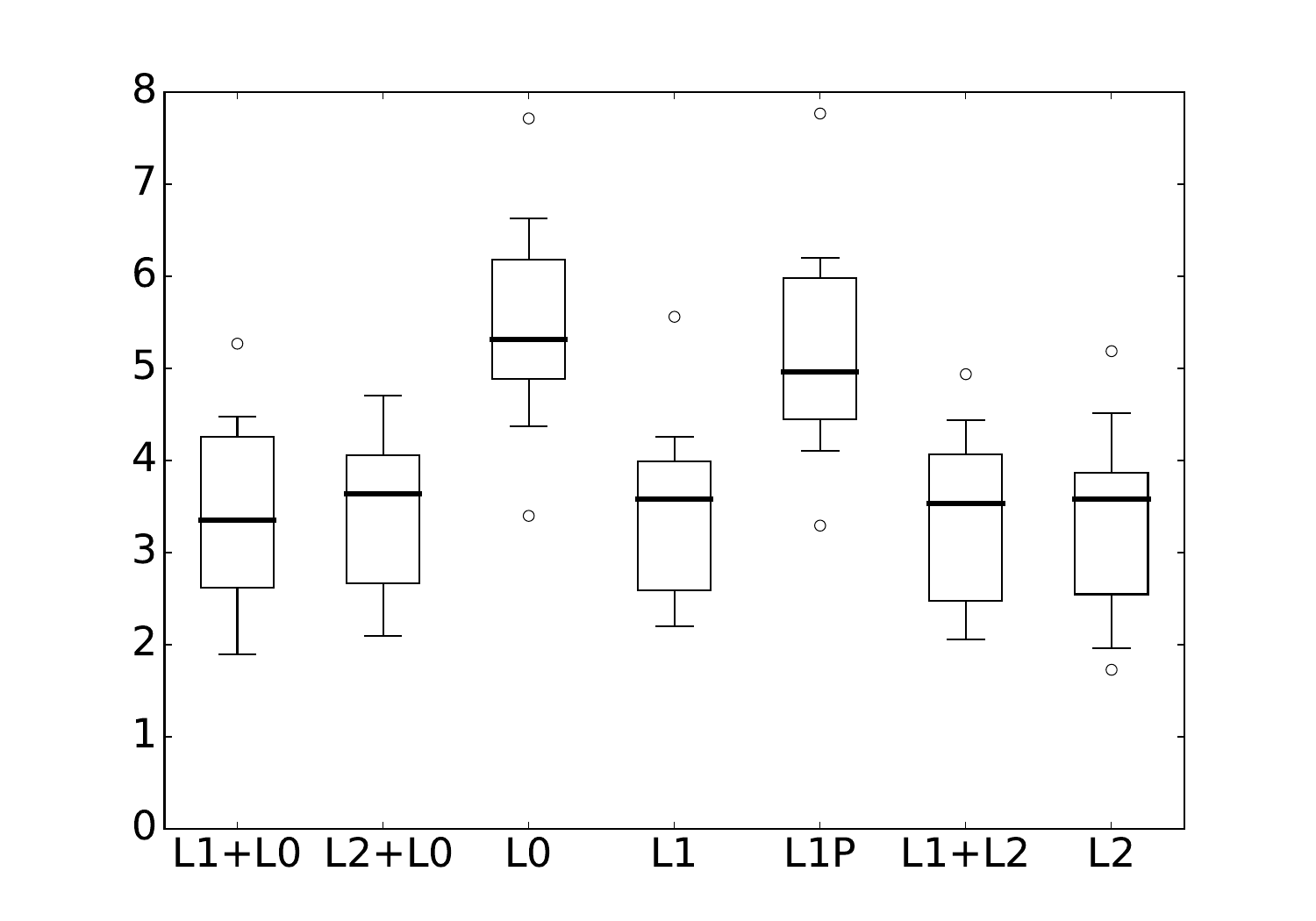}&
		\includegraphics[width=0.3\textwidth,height=0.18\textheight,  trim =1.2cm 1cm 2cm 1.3cm, clip = true ]{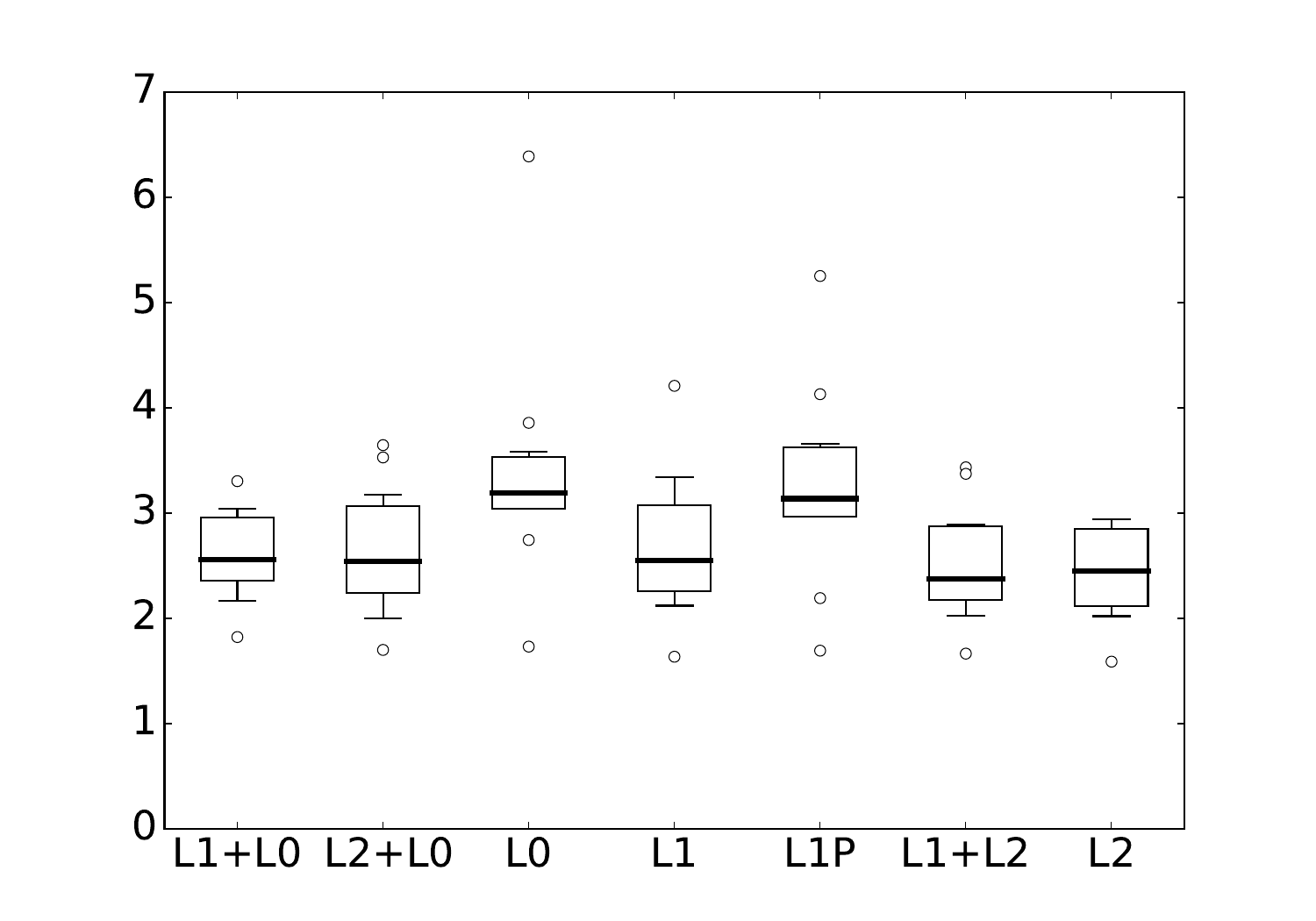}\\
		
		\rotatebox{90}{\sf {\small{~~~~~~~\# nonzeros}}}&
		\includegraphics[width=0.3\textwidth,height=0.18\textheight,  trim =1.8cm 1cm 2cm 1cm, clip = true ]{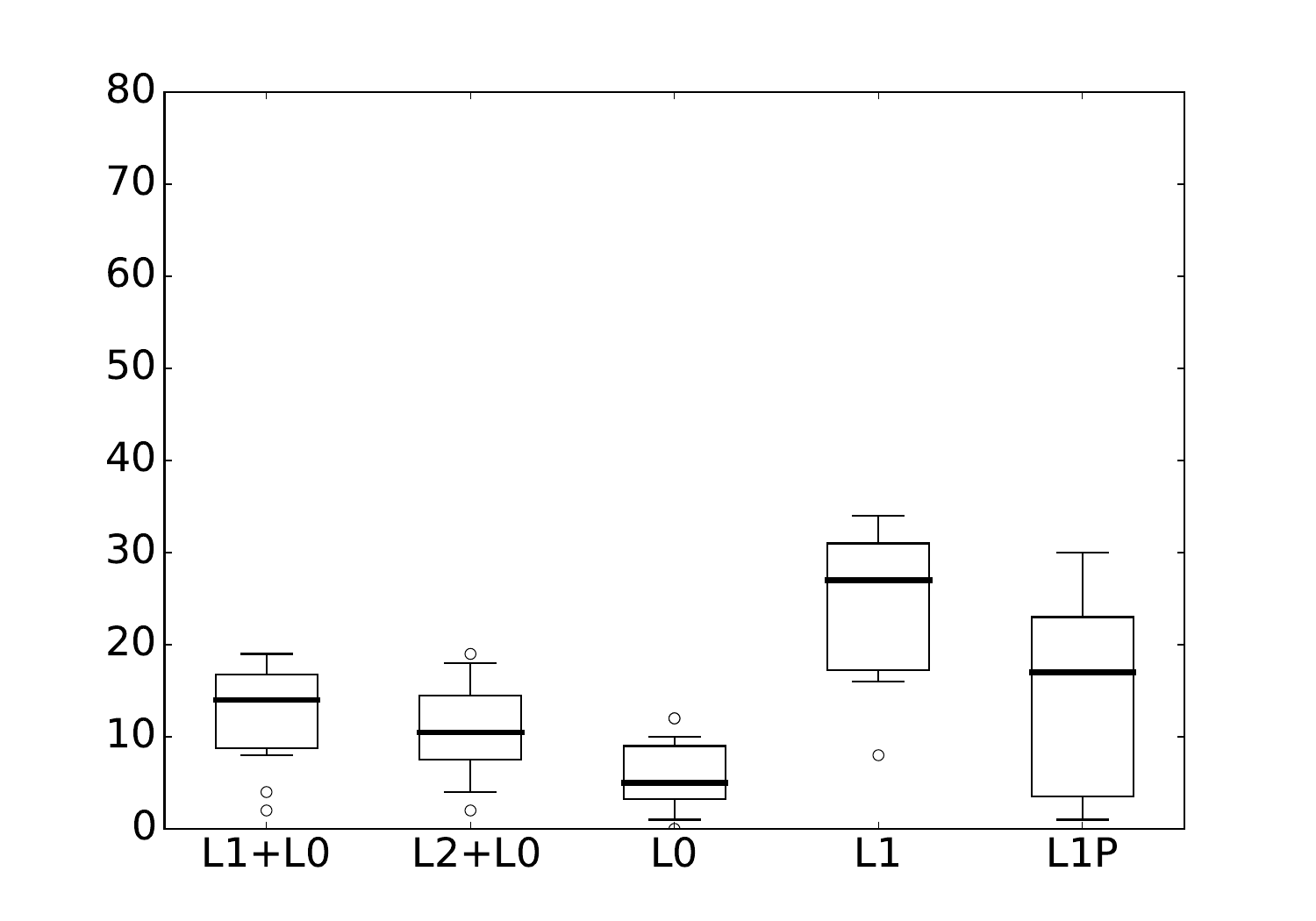}&
		\includegraphics[width=0.3\textwidth,height=0.18\textheight,  trim =1.2cm 1cm 2cm 1cm, clip = true ]{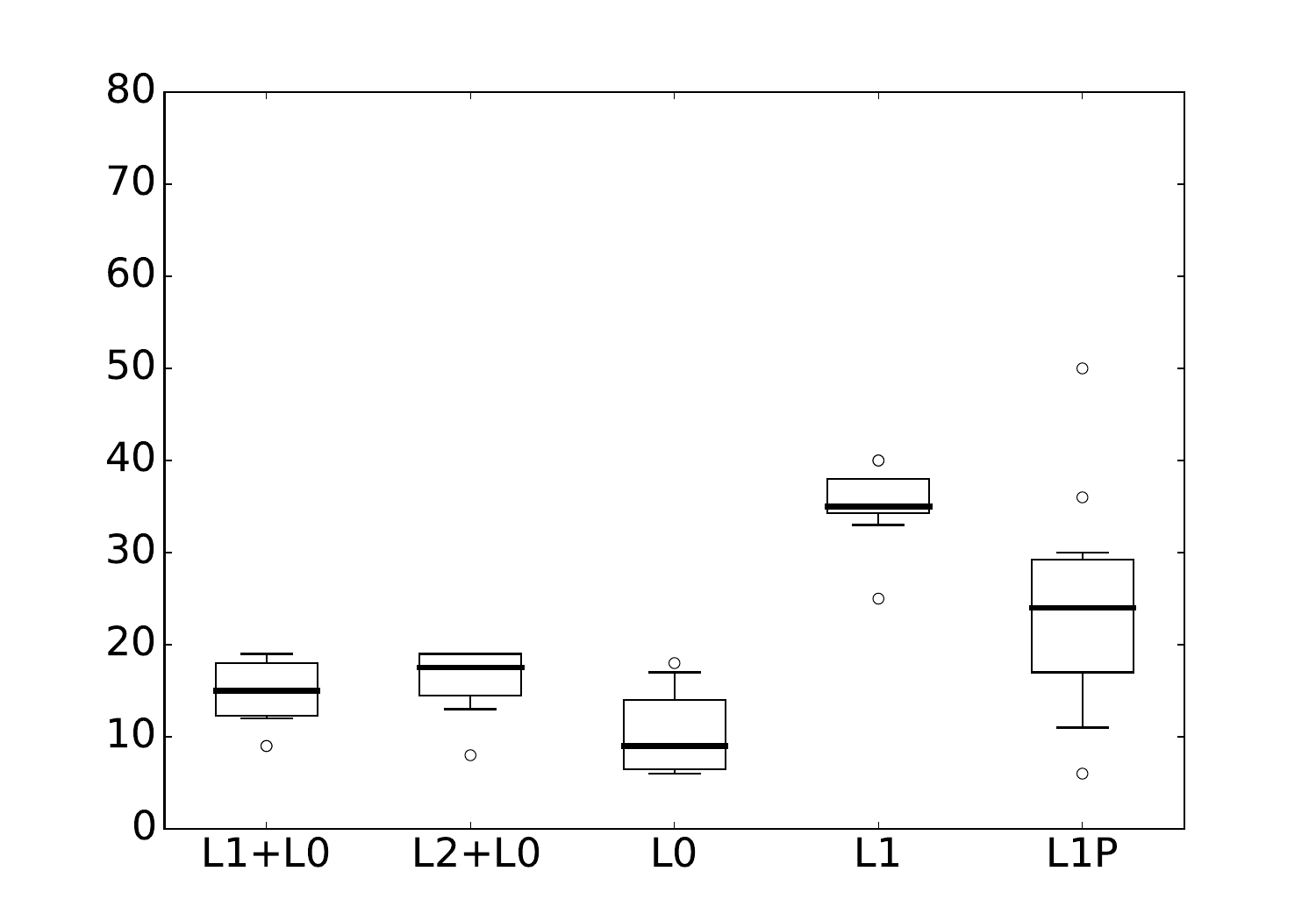}&
		\includegraphics[width=0.3\textwidth,height=0.18\textheight,  trim =1.2cm 1cm 2cm 1cm, clip = true ]{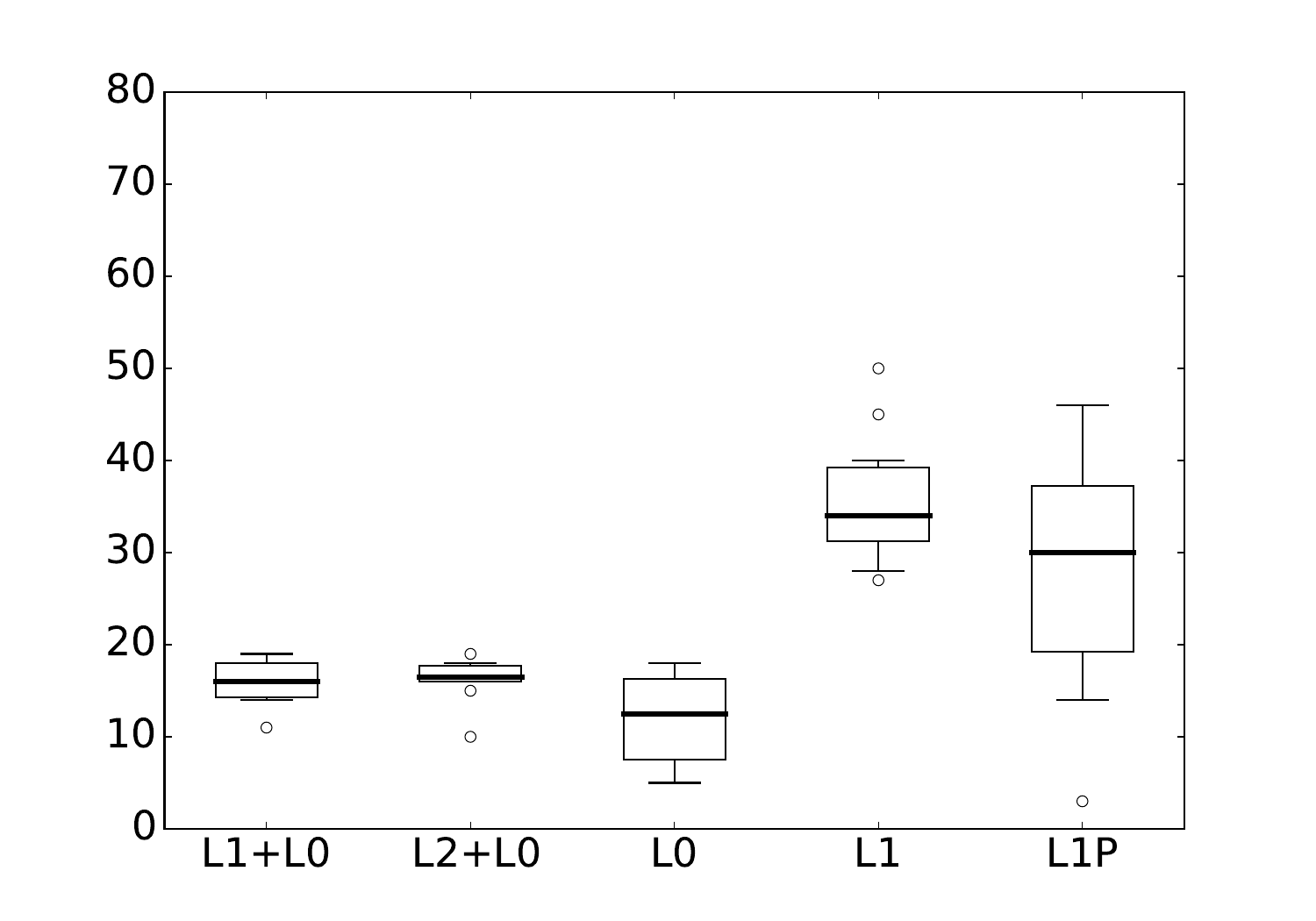}
			\end{tabular}
	\caption{{Example 1 simulations for different values of $n$, $p$, $\rho$, and SNR.}}
	\label{fig: ultrahigh.exp2}
\end{figure}

{Figures~\ref{fig: ultrahigh.exp1} and~\ref{fig: ultrahigh.exp2} summarize the results of additional experiments corresponding to the challenging ultra-high dimensional setting \citep{verzelen2012minimax} with $k^*\log(p/k^*)>n/2$. These experiments complement the ones that are reported in Figure~\ref{fig: example1}. Qualitatively, the results are overall similar to those in Figure~\ref{fig: example1}, especially with respect to the effect of adding~$\ell_1$ or~$\ell_2$ regularization to best subset selection. However, the predictive performance of all the methods in the challenging ultra-high dimensional setting is significantly worse than before, while the corresponding relative standing of dense models such as Ridge, Elastic net, and Lasso is improved.}

\newpage

\subsection{Comparisons with Bayesian methods}

\begin{figure}[h!]
\renewcommand{\baselinestretch}{1.25}
	\centering
	\begin{tabular}{c c c c}
		\multicolumn{2}{c}{Example~1: $n=100, p=1000$, $\rho=0.2$, SNR=2.} \\
		\sf {\scriptsize{Prediction error}} &\sf {\scriptsize{\# nonzeros }} \\
		\includegraphics[width=0.35\textwidth,height=0.16\textheight,  trim = 0.2cm 1cm 1.5cm 1cm, clip = true ]{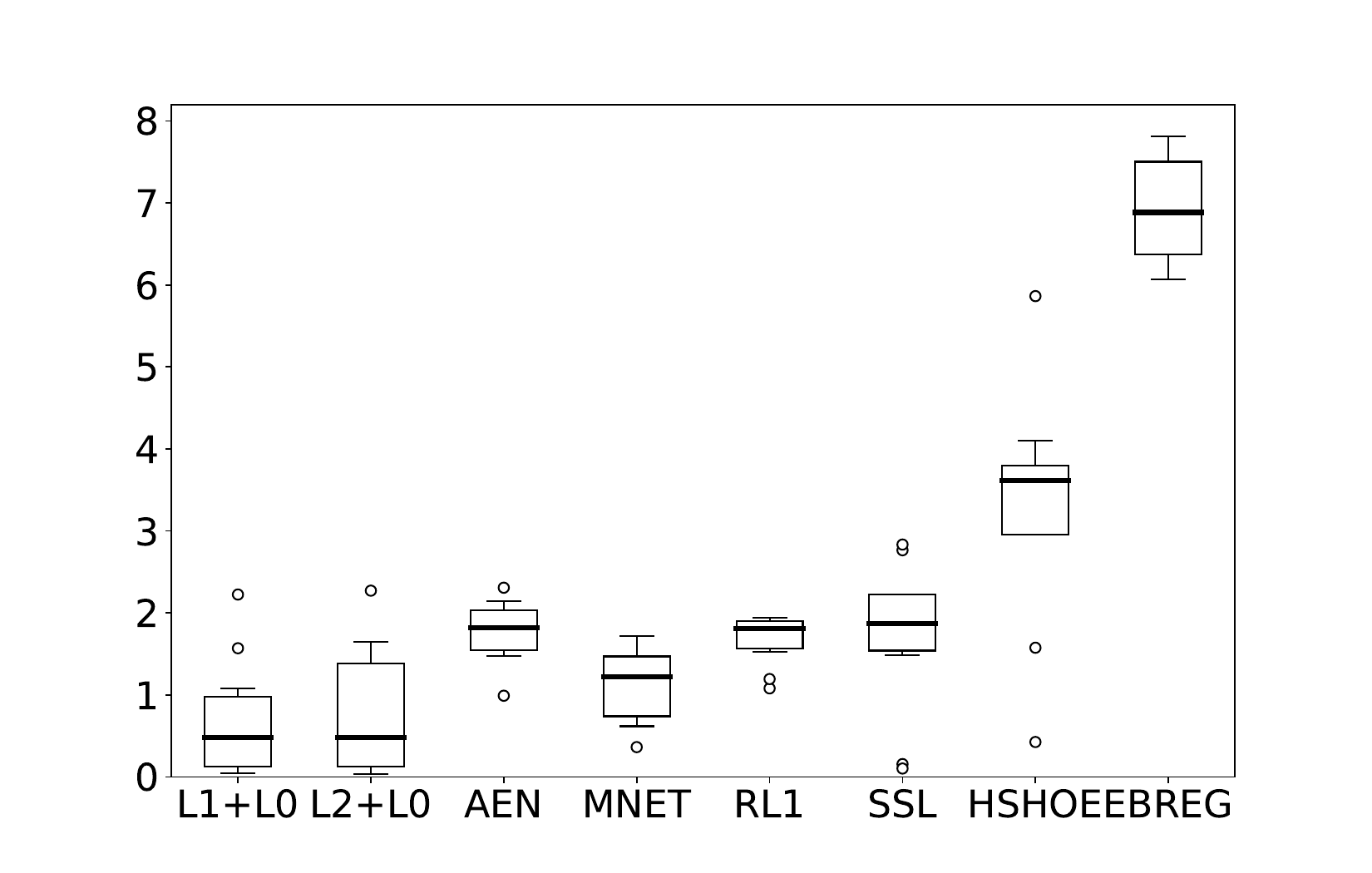}&
		\includegraphics[width=0.33\textwidth,height=0.16\textheight,  trim =.2cm 1cm 1.5cm 1cm, clip = true ]{\plotB/bayesian/N100_P1000_k07_rho0.2_SNR2.0_Sigma1/sparsity_boxplot_fixed_design_averaged_alt.pdf}
		\\
		\multicolumn{2}{c}{Example~2: $n=100, p=1000$, $\rho=0.1$, SNR=3.} \\
		\sf {\scriptsize{Prediction error}} &\sf {\scriptsize{\# nonzeros}}\\
		\includegraphics[width=0.35\textwidth,height=0.16\textheight,  trim =.2cm 1cm 1.5cm 1cm, clip = true ]{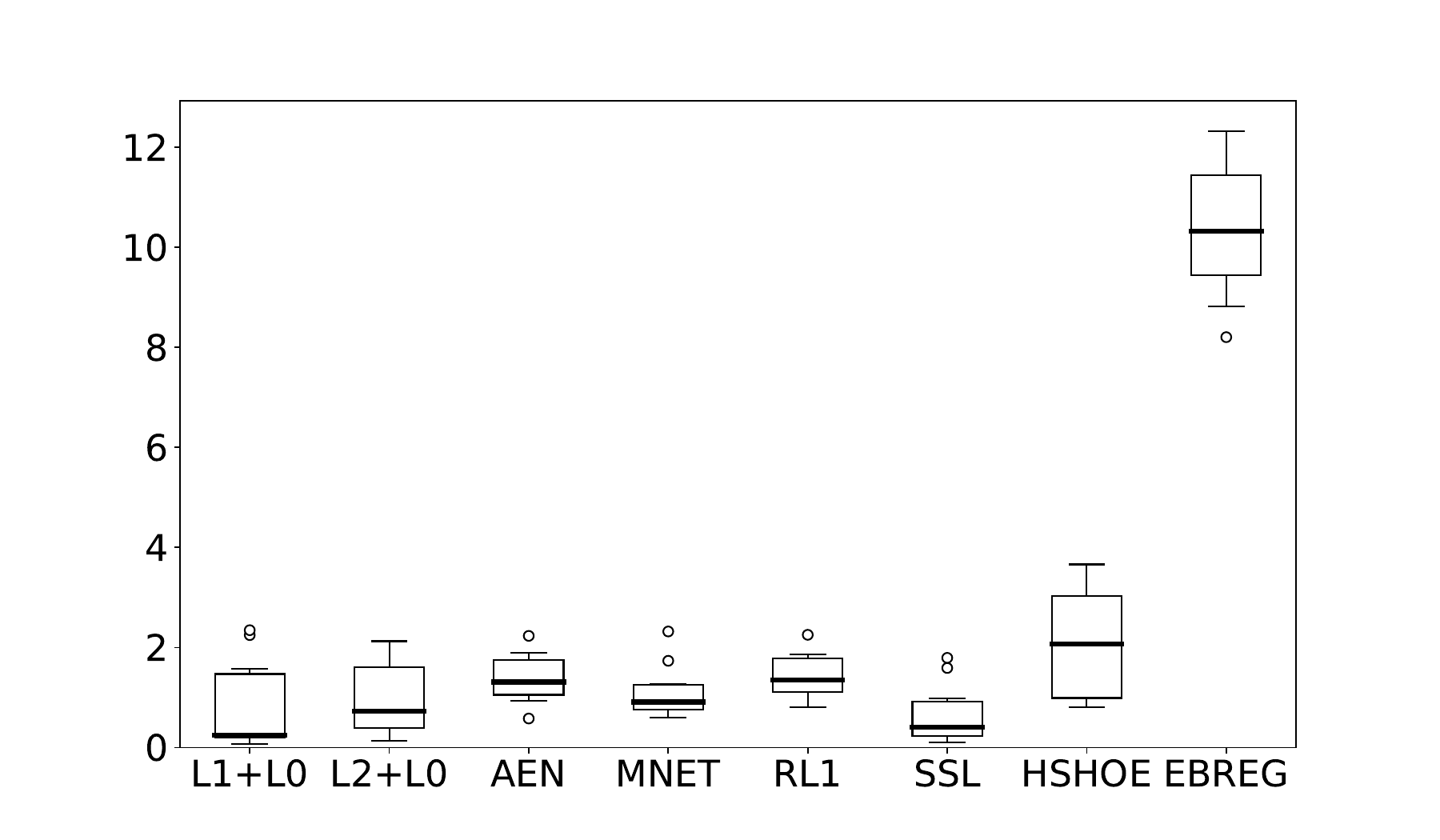}
		&
		\includegraphics[width=0.33\textwidth,height=0.16\textheight,  trim =.2cm 1cm 1.5cm 1cm, clip = true ]{\plotB/bayesian/N100_P1000_k07_rho0.1_SNR3.0_Sigma2/sparsity_boxplot_fixed_design_averaged_alt.pdf}\\
				\multicolumn{2}{c}{Example~1: $n=100, p=100$, $\rho=0.2$, SNR=2.} \\
		\sf {\scriptsize{Prediction error}} &\sf {\scriptsize{\# nonzeros }} \\
		\includegraphics[width=0.35\textwidth,height=0.16\textheight,  trim = 0.2cm 1cm 1.5cm 1cm, clip = true ]{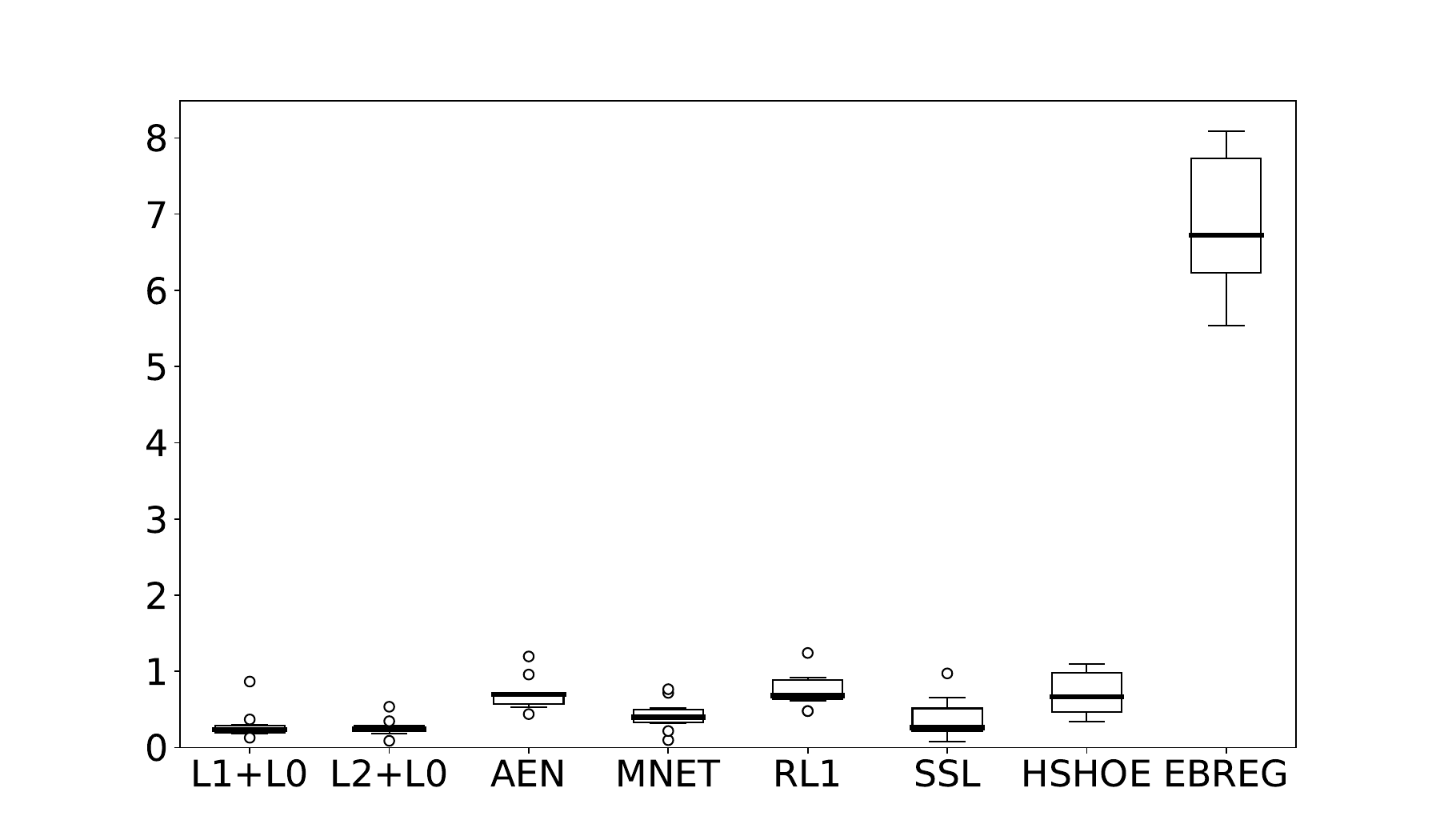}&
		\includegraphics[width=0.33\textwidth,height=0.16\textheight,  trim =.2cm 1cm 1.5cm 1cm, clip = true ]{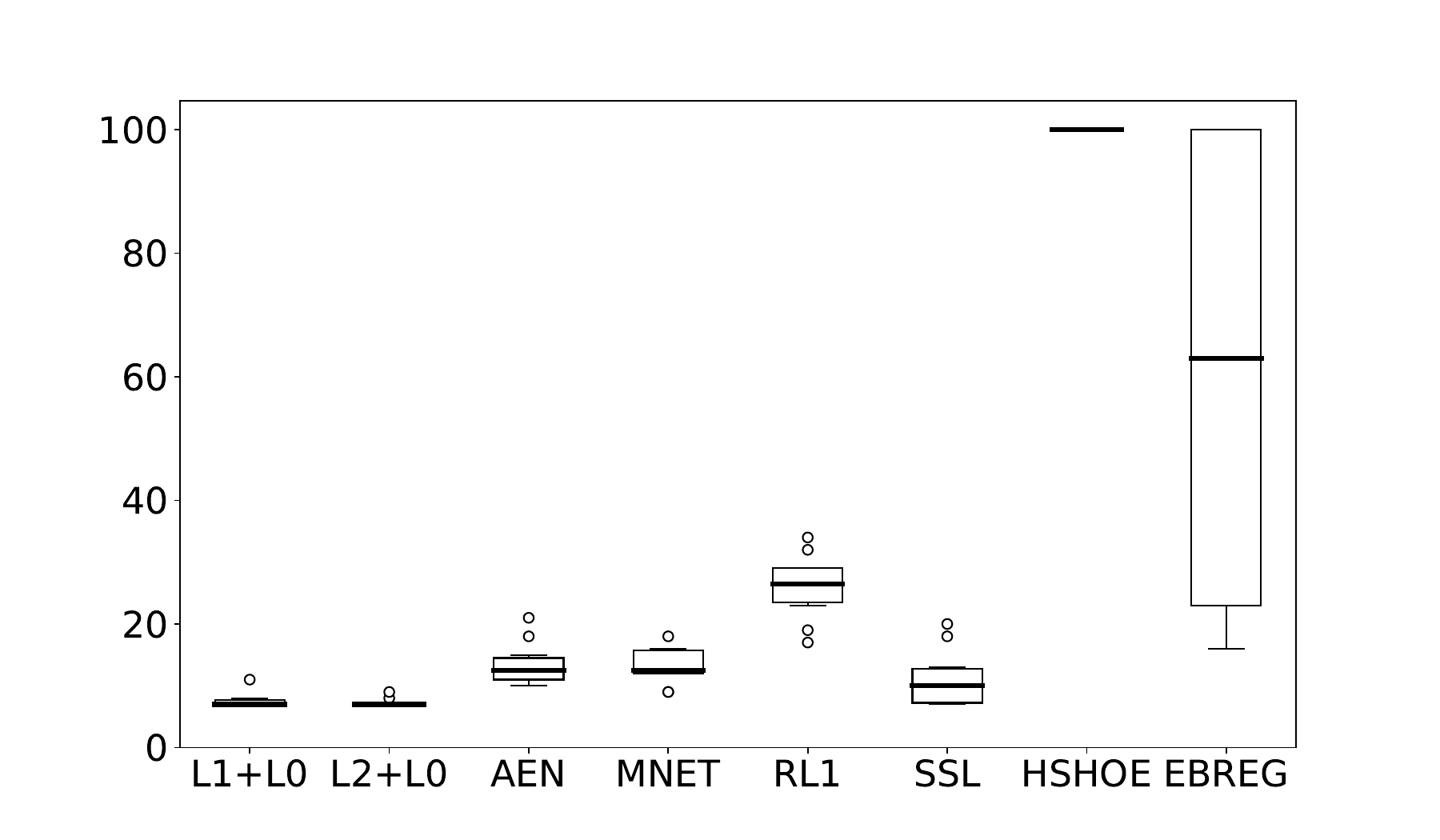}
		\\
		\multicolumn{2}{c}{Example~2: $n=100, p=100$, $\rho=0.1$, SNR=3.} \\
		\sf {\scriptsize{Prediction error}} &\sf {\scriptsize{\# nonzeros}}\\
		\includegraphics[width=0.35\textwidth,height=0.16\textheight,  trim =.2cm 1cm 1.5cm 1cm, clip = true ]{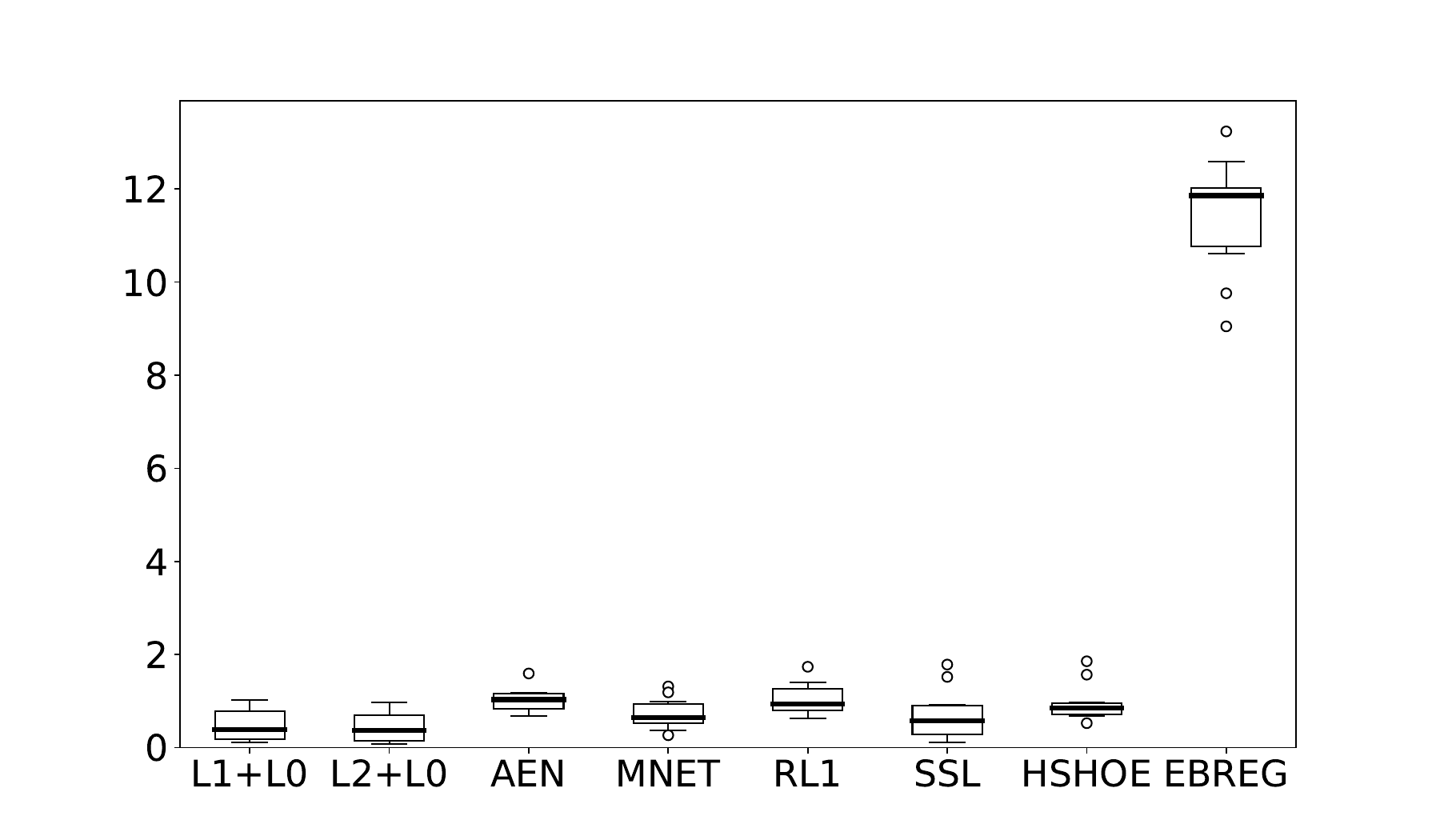}
		&
		\includegraphics[width=0.33\textwidth,height=0.16\textheight,  trim =.2cm 1cm 1.5cm 1cm, clip = true ]{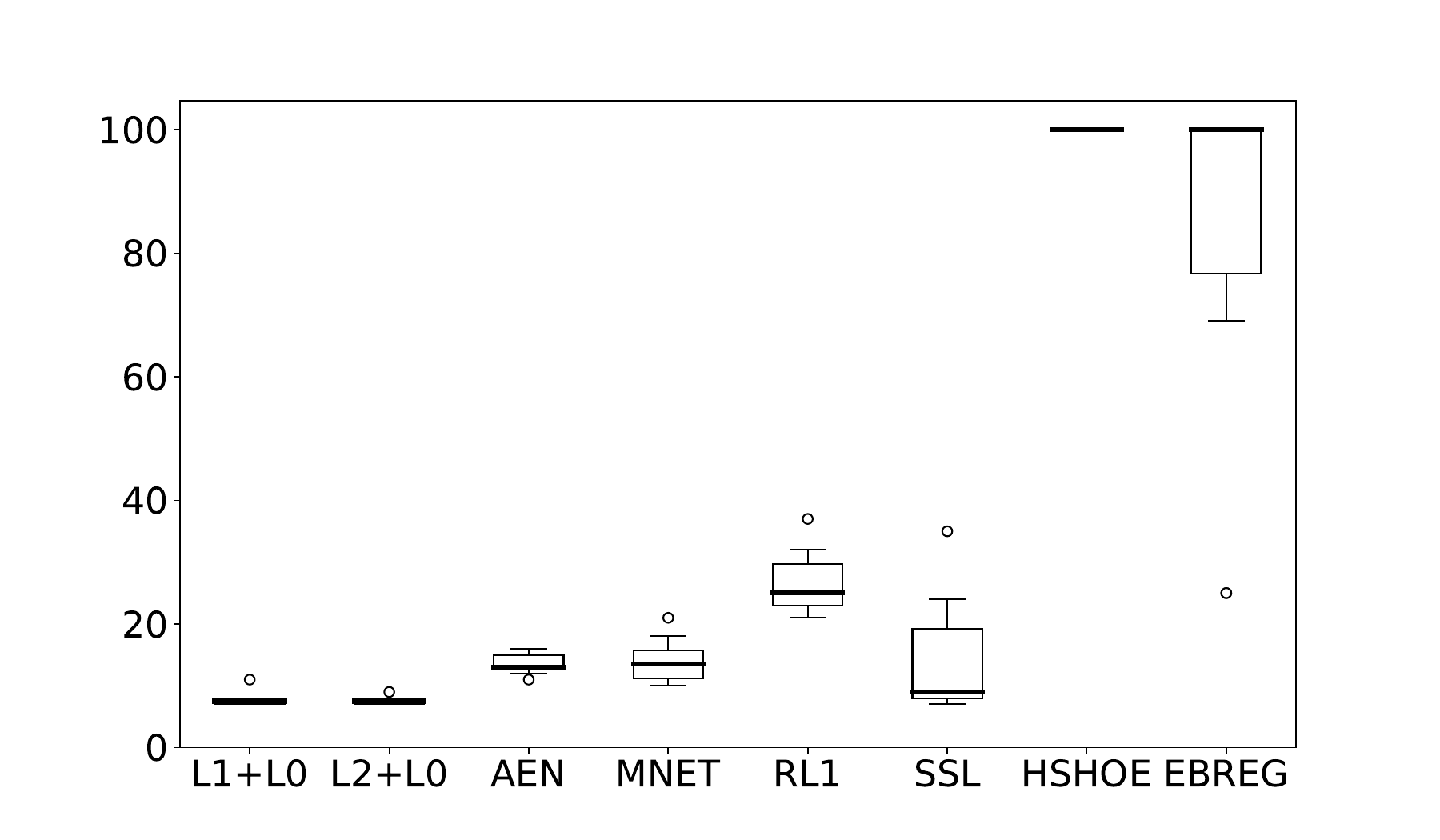}\\
	\end{tabular}
	\caption{ {Experimental results for the proposed methods, L0+L1 and L0+L2, as well as adaptive elastic net (AEN), Mnet, relaxed Lasso (RL1), \texttt{SSLASSO} (SSL), \texttt{horseshoe} (HSHOE), and \texttt{ebreg} (EBREG) methods. Due to the density of the corresponding solutions, we do not report the sparsity for HSHOE and EBREG in the top two panels.}}
	\label{fig: bayesian.full}
\end{figure}

{Figure~\ref{fig: bayesian.full} summarizes the results of additional experiments that include three state-of-the-art Bayesian approaches: the spike-and-slab Lasso method \citep{rockova2018spike}, implemented using R package {\texttt{SSLASSO}}; the empirical Bayes method of \cite{martin2017empirical}, implemented using R package \texttt{ebreg}; and the horseshoe regression \citep{carvalho2010horseshoe}, implemented using R package \texttt{horseshoe}. In the experiments that we consider, and with the default settings for the tuning parameters, the predictive performance of the last two methods is not quite as good as that of the competitors. The predictive performance of spike-and-slab Lasso is on par with the best performing methods (but somewhat worse overall than that of the proposed approach); however, their models are denser than those of the proposed approach. Overall, the proposed approach performed favorably in terms of both the model sparsity and the prediction accuracy. We note that the experiments in the top two panels of Figure~\ref{fig: bayesian.full} are the same as those in Figure~\ref{fig: exampleA}; however, Figure~\ref{fig: bayesian.full} also includes the results for \texttt{horseshoe} and \texttt{ebreg}.}

%% file: OR_FINAL.bbl
\begin{thebibliography}{64}
\providecommand{\natexlab}[1]{#1}
\providecommand{\url}[1]{\texttt{#1}}
\expandafter\ifx\csname urlstyle\endcsname\relax
  \providecommand{\doi}[1]{doi: #1}\else
  \providecommand{\doi}{doi: \begingroup \urlstyle{rm}\Url}\fi

\bibitem[Aarts and Lenstra(1997)]{aarts1997local}
E.~H. Aarts and J.~K. Lenstra.
\newblock \emph{Local search in combinatorial optimization}.
\newblock Princeton University Press, 1997.

\bibitem[Amini et~al.(2012)Amini, Kamilov, and Unser]{amini2012analog}
A.~Amini, U.~S. Kamilov, and M.~Unser.
\newblock The analog formulation of sparsity implies infinite divisibility and
  rules out bernoulli-gaussian priors.
\newblock In \emph{2012 IEEE Information Theory Workshop}, pages 682--686.
  Ieee, 2012.

\bibitem[Atamturk and Gomez(2019)]{atamturk2019rank}
A.~Atamturk and A.~Gomez.
\newblock Rank-one convexification for sparse regression.
\newblock \emph{arXiv preprint arXiv:1901.10334}, 2019.

\bibitem[Bartlett et~al.(2012)Bartlett, Mendelson, and Neeman]{bartlett2012}
P.~L. Bartlett, S.~Mendelson, and J.~Neeman.
\newblock L1-regularized linear regression: persistence and oracle
  inequalities.
\newblock \emph{Probability theory and related fields}, 154\penalty0
  (1-2):\penalty0 193--224, 2012.

\bibitem[Bellec et~al.(2018)Bellec, Lecu{\'e}, and Tsybakov]{bellec2018slope}
P.~C. Bellec, G.~Lecu{\'e}, and A.~B. Tsybakov.
\newblock Slope meets lasso: improved oracle bounds and optimality.
\newblock \emph{The Annals of Statistics}, 46\penalty0 (6B):\penalty0
  3603--3642, 2018.

\bibitem[Belloni and Chernozhukov(2013)]{belloni2013least}
A.~Belloni and V.~Chernozhukov.
\newblock Least squares after model selection in high-dimensional sparse
  models.
\newblock \emph{Bernoulli}, 19\penalty0 (2):\penalty0 521--547, 2013.

\bibitem[Belloni et~al.(2014)Belloni, Chernozhukov, and
  Wang]{belloni2014pivotal}
A.~Belloni, V.~Chernozhukov, and L.~Wang.
\newblock Pivotal estimation via square-root lasso in nonparametric regression.
\newblock \emph{The Annals of Statistics}, 42\penalty0 (2):\penalty0 757--788,
  2014.

\bibitem[Bertsimas and Weismantel(2005)]{bertsimas2005optimization_new}
D.~Bertsimas and R.~Weismantel.
\newblock \emph{Optimization over integers}.
\newblock Dynamic Ideas Belmont, 2005.

\bibitem[Bertsimas et~al.(2016)Bertsimas, King, and
  Mazumder]{bertsimas2015best}
D.~Bertsimas, A.~King, and R.~Mazumder.
\newblock Best subset selection via a modern optimization lens.
\newblock \emph{Annals of Statistics}, 44\penalty0 (2):\penalty0 813--852,
  2016.

\bibitem[Bhadra et~al.(2019)Bhadra, Datta, Li, Polson, and
  Willard]{bhadra2019prediction}
A.~Bhadra, J.~Datta, Y.~Li, N.~G. Polson, and B.~Willard.
\newblock Prediction risk for the horseshoe regression.
\newblock \emph{The Journal of Machine Learning Research}, 20\penalty0
  (1):\penalty0 2882--2920, 2019.

\bibitem[Bickel et~al.(2009)Bickel, Ritov, and Tsybakov]{bickel1}
P.~Bickel, Y.~Ritov, and A.~Tsybakov.
\newblock Simultaneous analysis of lasso and dantzig selector.
\newblock \emph{The Annals of Statistics}, 37:\penalty0 1705--1732, 2009.

\bibitem[Boyd and Vandenberghe(2004)]{BV2004}
S.~Boyd and L.~Vandenberghe.
\newblock \emph{Convex Optimization}.
\newblock Cambridge University Press, Cambridge, 2004.

\bibitem[Breiman(1996)]{breiman1996heuristics}
L.~Breiman.
\newblock Heuristics of instability and stabilization in model selection.
\newblock \emph{The annals of statistics}, 24\penalty0 (6):\penalty0
  2350--2383, 1996.

\bibitem[B{\"u}hlmann and {van-de-Geer}(2011)]{buhlmann2011statistics}
P.~B{\"u}hlmann and S.~{van-de-Geer}.
\newblock \emph{Statistics for high-dimensional data}.
\newblock Springer, 2011.

\bibitem[Carvalho et~al.(2010)Carvalho, Polson, and
  Scott]{carvalho2010horseshoe}
C.~M. Carvalho, N.~G. Polson, and J.~G. Scott.
\newblock The horseshoe estimator for sparse signals.
\newblock \emph{Biometrika}, 97\penalty0 (2):\penalty0 465--480, 2010.

\bibitem[Comminges et~al.(2012)Comminges, Dalalyan, et~al.]{comminges2012tight}
L.~Comminges, A.~S. Dalalyan, et~al.
\newblock Tight conditions for consistency of variable selection in the context
  of high dimensionality.
\newblock \emph{The Annals of Statistics}, 40\penalty0 (5):\penalty0
  2667--2696, 2012.

\bibitem[Dalalyan et~al.(2017)Dalalyan, Hebiri, and
  Lederer]{dalalyan2017prediction}
A.~S. Dalalyan, M.~Hebiri, and J.~Lederer.
\newblock On the prediction performance of the lasso.
\newblock \emph{Bernoulli}, 23\penalty0 (1):\penalty0 552--581, 2017.

\bibitem[Fan and Lv(2013)]{fan2013asymptotic}
Y.~Fan and J.~Lv.
\newblock Asymptotic properties for combined {L}1 and concave regularization.
\newblock \emph{Biometrika}, 101\penalty0 (1):\penalty0 57--70, 2013.

\bibitem[Frank and Friedman(1993)]{FF93}
I.~Frank and J.~Friedman.
\newblock A statistical view of some chemometrics regression tools (with
  discussion).
\newblock \emph{Technometrics}, 35\penalty0 (2):\penalty0 109--148, 1993.

\bibitem[Gamarnik and Zadik(2017)]{david2017high}
D.~Gamarnik and I.~Zadik.
\newblock High dimensional regression with binary coefficients. estimating
  squared error and a phase transtition.
\newblock In \emph{Conference on Learning Theory}, pages 948--953, 2017.

\bibitem[Greenshtein and Ritov(2004)]{GR2004}
E.~Greenshtein and Y.~Ritov.
\newblock Persistence in high-dimensional linear predictor selection and the
  virtue of overparametrization.
\newblock \emph{Bernoulli}, 10:\penalty0 971--988, 2004.

\bibitem[Greenshtein(2006)]{greenshtein2006best}
E.~Greenshtein.
\newblock Best subset selection, persistence in high-dimensional statistical
  learning and optimization under $\ell_{1}$ constraint.
\newblock \emph{The Annals of Statistics}, 34\penalty0 (5):\penalty0
  2367--2386, 2006.

\bibitem[Hastie et~al.(2015)Hastie, Mazumder, Lee, and Zadeh]{hastie2015matrix}
T.~Hastie, R.~Mazumder, J.~D. Lee, and R.~Zadeh.
\newblock Matrix completion and low-rank svd via fast alternating least
  squares.
\newblock \emph{Journal of Machine Learning Research}, 16:\penalty0 3367--3402,
  2015.

\bibitem[Hastie et~al.(2020)Hastie, Tibshirani, and Tibshirani]{hastie2020best}
T.~Hastie, R.~Tibshirani, and R.~Tibshirani.
\newblock Best subset, forward stepwise or lasso? {A}nalysis and
  recommendations based on extensive comparisons.
\newblock \emph{Statistical Science}, 35\penalty0 (4):\penalty0 579--592, 2020.

\bibitem[Hazimeh and Mazumder(2018)]{hazimeh2018fast}
H.~Hazimeh and R.~Mazumder.
\newblock Fast best subset selection: Coordinate descent and local
  combinatorial optimization algorithms.
\newblock \emph{arXiv preprint arXiv:1803.01454}, 2018.

\bibitem[Hazimeh et~al.(2020)Hazimeh, Mazumder, and Saab]{hazimeh2020sparse}
H.~Hazimeh, R.~Mazumder, and A.~Saab.
\newblock Sparse regression at scale: Branch-and-bound rooted in first-order
  optimization.
\newblock \emph{arXiv preprint arXiv:2004.06152}, 2020.

\bibitem[Hazimeh et~al.(2021)Hazimeh, Mazumder, and
  Radchenko]{hazimeh2021grouped}
H.~Hazimeh, R.~Mazumder, and P.~Radchenko.
\newblock Grouped variable selection with discrete optimization: Computational
  and statistical perspectives.
\newblock \emph{arXiv preprint arXiv:2104.07084}, 2021.

\bibitem[Hoerl and Kennard(1970)]{HK70}
A.~E. Hoerl and R.~Kennard.
\newblock Ridge regression: biased estimation for nonorthogonal problems.
\newblock \emph{Technometrics}, 12:\penalty0 55--67, 1970.

\bibitem[Huang et~al.(2016)Huang, Breheny, Lee, Ma, and Zhang]{huang2016mnet}
J.~Huang, P.~Breheny, S.~Lee, S.~Ma, and C.~Zhang.
\newblock The mnet method for variable selection.
\newblock \emph{Statistica Sinica}, 26:\penalty0 903?923, 2016.

\bibitem[James and Stein(1961)]{james1961estimation}
W.~James and C.~Stein.
\newblock Estimation with quadratic loss.
\newblock In \emph{Proceedings of the fourth Berkeley symposium on mathematical
  statistics and probability}, volume~1, pages 361--379, 1961.

\bibitem[Koltchinskii et~al.(2011)Koltchinskii, Lounici, and
  Tsybakov]{koltchinskii2011nuclear}
V.~Koltchinskii, K.~Lounici, and A.~Tsybakov.
\newblock Nuclear-norm penalization and optimal rates for noisy low-rank matrix
  completion.
\newblock \emph{The Annals of Statistics}, 39\penalty0 (5):\penalty0
  2302--2329, 2011.

\bibitem[Koren et~al.(2009)Koren, Bell, and Volinsky]{koren2009matrix}
Y.~Koren, R.~Bell, and C.~Volinsky.
\newblock Matrix factorization techniques for recommender systems.
\newblock \emph{IEEE Computer}, 42\penalty0 (8), 2009.

\bibitem[Lecu{\'e} and Mendelson(2017)]{lecue2017sparse}
G.~Lecu{\'e} and S.~Mendelson.
\newblock Sparse recovery under weak moment assumptions.
\newblock \emph{Journal of the European Mathematical Society}, 19\penalty0
  (3):\penalty0 881--904, 2017.

\bibitem[Linderoth and Lodi(2010)]{linderoth2010milp}
J.~T. Linderoth and A.~Lodi.
\newblock {MILP} software.
\newblock \emph{Wiley encyclopedia of operations research and management
  science}, 2010.

\bibitem[Liu and Wu(2007)]{liu2007variable}
Y.~Liu and Y.~Wu.
\newblock Variable selection via a combination of the l0 and l1 penalties.
\newblock \emph{Journal of Computational and Graphical Statistics}, 16\penalty0
  (4):\penalty0 782--798, 2007.

\bibitem[Lounici et~al.(2011)Lounici, Pontil, Tsybakov, and van~de
  Geer]{lounici1}
K.~Lounici, M.~Pontil, A.~Tsybakov, and S.~Geer.
\newblock Oracle inequalities and optimal inference under group sparsity.
\newblock \emph{The Annals of Statistics}, 39\penalty0 (4):\penalty0
  2164--2204, 2011.

\bibitem[Martin and Tang(2020)]{martin2020empirical}
R.~Martin and Y.~Tang.
\newblock Empirical priors for prediction in sparse high-dimensional linear
  regression.
\newblock \emph{Journal of Machine Learning Research}, 21\penalty0
  (144):\penalty0 1--30, 2020.

\bibitem[Martin et~al.(2017)Martin, Mess, and Walker]{martin2017empirical}
R.~Martin, R.~Mess, and S.~G. Walker.
\newblock Empirical bayes posterior concentration in sparse high-dimensional
  linear models.
\newblock \emph{Bernoulli}, 23\penalty0 (3):\penalty0 1822--1847, 2017.

\bibitem[Massart(2007)]{massart1}
P.~Massart.
\newblock \emph{Concentration inequalities and model selection}, volume~6.
\newblock Springer, 2007.

\bibitem[Mazumder and Radchenko(2017)]{mazumder2015discrete}
R.~Mazumder and P.~Radchenko.
\newblock {The Discrete Dantzig Selector}: Estimating sparse linear models via
  mixed integer linear optimization.
\newblock \emph{IEEE Transactions on Information Theory}, 63 (5):\penalty0 3053
  -- 3075, 2017.

\bibitem[Meinshausen(2007)]{meinshausen2007relaxed}
N.~Meinshausen.
\newblock Relaxed lasso.
\newblock \emph{Computational Statistics \& Data Analysis}, 52\penalty0
  (1):\penalty0 374--393, 2007.

\bibitem[Miller(2002)]{miller2002subset}
A.~Miller.
\newblock \emph{Subset selection in regression}.
\newblock CRC Press Washington, 2002.

\bibitem[Mitchell and Beauchamp(1988)]{mitchell1988bayesian}
T.~J. Mitchell and J.~J. Beauchamp.
\newblock Bayesian variable selection in linear regression.
\newblock \emph{Journal of the american statistical association}, 83\penalty0
  (404):\penalty0 1023--1032, 1988.

\bibitem[Mladenovi{\'c} and Hansen(1997)]{mladenovic1997variable}
N.~Mladenovi{\'c} and P.~Hansen.
\newblock Variable neighborhood search.
\newblock \emph{Computers \& operations research}, 24\penalty0 (11):\penalty0
  1097--1100, 1997.

\bibitem[Natarajan(1995)]{natarajan1995sparse}
B.~Natarajan.
\newblock Sparse approximate solutions to linear systems.
\newblock \emph{SIAM journal on computing}, 24\penalty0 (2):\penalty0 227--234,
  1995.

\bibitem[Nemhauser and Wolsey(1988)]{nemhauser1988integer}
G.~L. Nemhauser and L.~A. Wolsey.
\newblock Integer programming and combinatorial optimization.
\newblock \emph{Wiley, Chichester. GL Nemhauser, MWP Savelsbergh, GS Sigismondi
  (1992). Constraint Classification for Mixed Integer Programming Formulations.
  COAL Bulletin}, 20:\penalty0 8--12, 1988.

\bibitem[Nesterov(2013)]{nesterov2013gradient}
Y.~Nesterov.
\newblock Gradient methods for minimizing composite functions.
\newblock \emph{Mathematical Programming}, 140\penalty0 (1):\penalty0 125--161,
  2013.

\bibitem[Nesterov(2004)]{nesterov2004introductorynew}
Y.~Nesterov.
\newblock \emph{Introductory Lectures on Convex Optimization: A Basic Course}.
\newblock Kluwer, Norwell, 2004.

\bibitem[Pisier(1980)]{pisier1980remarques}
G.~Pisier.
\newblock Remarques sur un r{\'e}sultat non publi{\'e} de b. maurey.
\newblock \emph{S{\'e}minaire Analyse fonctionnelle (dit" Maurey-Schwartz")},
  pages 1--12, 1980.

\bibitem[Polson and Sun(2019)]{polson2019bayesian}
N.~G. Polson and L.~Sun.
\newblock Bayesian $\ell_0$-regularized least squares.
\newblock \emph{Applied Stochastic Models in Business and Industry},
  35\penalty0 (3):\penalty0 717--731, 2019.

\bibitem[Raskutti et~al.(2011)Raskutti, Wainwright, and
  Yu]{raskutti2011minimax}
G.~Raskutti, M.~Wainwright, and B.~Yu.
\newblock Minimax rates of estimation for high-dimensional linear regression
  over-balls.
\newblock \emph{Information Theory, IEEE Transactions on}, 57\penalty0
  (10):\penalty0 6976--6994, 2011.

\bibitem[Rigollet and Tsybakov(2011)]{rigollet2011exponential}
P.~Rigollet and A.~Tsybakov.
\newblock Exponential screening and optimal rates of sparse estimation.
\newblock \emph{The Annals of Statistics}, 39\penalty0 (2):\penalty0 731--771,
  2011.

\bibitem[Rockov{\'a} and George(2018)]{rockova2018spike}
V.~Rockov{\'a} and E.~I. George.
\newblock The spike-and-slab lasso.
\newblock \emph{Journal of the American Statistical Association}, 113\penalty0
  (521):\penalty0 431--444, 2018.

\bibitem[Soussen et~al.(2011)Soussen, Idier, Brie, and
  Duan]{soussen2011bernoulli}
C.~Soussen, J.~Idier, D.~Brie, and J.~Duan.
\newblock From bernoulli--gaussian deconvolution to sparse signal restoration.
\newblock \emph{IEEE Transactions on Signal Processing}, 59\penalty0
  (10):\penalty0 4572--4584, 2011.

\bibitem[Sun and Zhang(2012)]{sun2012scaled}
T.~Sun and C.-H. Zhang.
\newblock Scaled sparse linear regression.
\newblock \emph{Biometrika}, 99\penalty0 (4):\penalty0 879--898, 2012.

\bibitem[Tibshirani(1996)]{Ti96}
R.~Tibshirani.
\newblock Regression shrinkage and selection via the lasso.
\newblock \emph{Journal of the Royal Statistical Society, Series B},
  58:\penalty0 267--288, 1996.

\bibitem[Verzelen(2012)]{verzelen2012minimax}
N.~Verzelen.
\newblock Minimax risks for sparse regressions: Ultra-high dimensional
  phenomenons.
\newblock \emph{Electronic Journal of Statistics}, 6:\penalty0 38--90, 2012.

\bibitem[Vielma et~al.(2016)Vielma, Dunning, Huchette, and
  Lubin]{vielma2016extended}
J.~P. Vielma, I.~Dunning, J.~Huchette, and M.~Lubin.
\newblock Extended formulations in mixed integer conic quadratic programming.
\newblock \emph{Mathematical Programming Computation}, pages 1--50, 2016.

\bibitem[Wainwright(2009)]{wainwright2009sharp}
M.~J. Wainwright.
\newblock Sharp thresholds for high-dimensional and noisy recovery of sparsity
  using l1-constrained quadratic programming.
\newblock \emph{IEEE Transactions on Information Theory}, 2009.

\bibitem[Weng et~al.(2013)Weng, Feng, and Qiao]{weng2013regularization}
H.~Weng, Y.~Feng, and X.~Qiao.
\newblock Regularization after retention in ultrahigh dimensional linear
  regression models.
\newblock \emph{arXiv preprint arXiv:1311.5625}, 2013.

\bibitem[Zhang and Zhang(2012)]{zhang2012general}
C.-H. Zhang and T.~Zhang.
\newblock A general theory of concave regularization for high-dimensional
  sparse estimation problems.
\newblock \emph{Statistical Science}, 27\penalty0 (4):\penalty0 576--593, 2012.

\bibitem[Zhang et~al.(2017)Zhang, Wainwright, and Jordan]{zhang2017optimal}
Y.~Zhang, M.~J. Wainwright, and M.~I. Jordan.
\newblock Optimal prediction for sparse linear models? {L}ower bounds for
  coordinate-separable {M}-estimators.
\newblock \emph{Electronic Journal of Statistics}, 11\penalty0 (1):\penalty0
  752--799, 2017.

\bibitem[Zou and Hastie(2005)]{ZH2005}
H.~Zou and T.~Hastie.
\newblock Regularization and variable selection via the elastic net.
\newblock \emph{Journal of the Royal Statistical Society Series B.},
  67\penalty0 (2):\penalty0 301--320, 2005.

\bibitem[Zou and Zhang(2009)]{zou2009adaptive}
H.~Zou and H.~H. Zhang.
\newblock On the adaptive elastic-net with a diverging number of parameters.
\newblock \emph{Annals of statistics}, 37\penalty0 (4):\penalty0 1733, 2009.

\end{thebibliography}
